%% file: ANA-MUON-2018-03-PAPER.tex
\documentclass[preprint,atlasdraft=true,texlive=2016,UKenglish,texmf]{atlasdoc}
 
\usepackage{atlaspackage}
\usepackage{atlasbiblatex}
 
\usepackage{atlasphysics}
\usepackage{multirow}
 
\graphicspath{{logos/}{figures/}}
 
\usepackage{ANA-MUON-2018-03-PAPER-defs}

\AtlasTitle{Muon reconstruction and identification efficiency in ATLAS using the full Run 2 $pp$ collision data set at $\sqrt{s}=13$ TeV}
 
\PreprintIdNumber{CERN-EP-2020-199}
\AtlasAbstract{
This article documents the muon reconstruction and identification efficiency obtained by the ATLAS experiment for 139 fb$^{-1}$ of $pp$ collision data at $\sqrt{s}=13$ TeV collected
between 2015 and 2018 during Run 2 of the LHC. The increased instantaneous luminosity delivered by the LHC over this period required a reoptimisation of the criteria for the identification of prompt muons.
Improved and newly developed algorithms were deployed to preserve high muon identification efficiency with a low misidentification rate and good momentum resolution.
The availability of large samples of $Z\to\mu\mu$ and $J/\psi\to\mu\mu$ decays, and the minimisation of systematic uncertainties, allows the efficiencies of criteria for muon identification, primary vertex association, and isolation to be measured with an accuracy at the per-mille level in the bulk of the phase space, and up to the percent level in complex kinematic configurations.
Excellent performance is achieved over a range of transverse momenta from 3 GeV to several hundred GeV, and across the full muon detector acceptance of $|\eta|<2.7$. }
 
\author{The ATLAS Collaboration}
 
\AtlasRefCode{MUON-2018-03}
 
\AtlasDOI{DOI:10.1140/epjc/s10052-021-09233-2}
\AtlasJournalRef{Eur. Phys. J. C 81 (2021) 578}
\hypersetup{pdftitle={ATLAS document},pdfauthor={The ATLAS Collaboration}}
 
\begin{document}
 
\maketitle

\section{Introduction}
 
In the years from 2015 to 2018, Run~2 of the Large Hadron Collider (LHC) at CERN provided an unprecedented number of $pp$ collision events at a centre-of-mass energy of 13~\TeV{}.
The identification and accurate measurement of processes with muons in the final state is one of the main features of the ATLAS experiment~\cite{PERF-2007-01} at the LHC,
and a key element for a successful physics programme.
For example, Standard Model (SM) predictions can be tested by studying the leptonic decays of the $W$ or $Z/\gamma^*$ vector bosons, heavy-flavour hadrons undergoing weak decays into leptons can be identified with high signal-to-background ratio, and beyond-the-SM (BSM) resonances may be found in leptonic decay channels.
Highlight analyses where optimal muon identification performance has been fundamental are, for example, the measurement of Higgs boson properties~\cite{ATLAS:2020wny},
the precise determination of SM parameters in the quark-mixing sector~\cite{BPHY-2018-01},
and searches for BSM physics in extreme regions of phase space~\cite{Aad:2019qnd, EXOT-2018-30}.
Analyses targeting these and similar processes profit from the structure of the ATLAS muon reconstruction and identification systems~\cite{ATLAS-TDR-10}, which combine information from several subdetectors to reach almost 100\% reconstruction and identification efficiency over a wide range of transverse momenta (\pt{}), with background contamination at the per-mille level and good momentum resolution, even in challenging data-taking conditions characterised by a large number of interactions per LHC bunch crossing.
 
Compared to a previous publication~\cite{PERF-2015-10}, which reported on the muon identification performance on early $\sqrt{s}=13$~\TeV{} data,
this article describes refined and newly developed techniques that improved muon identification performance over a wide region in phase space,
and reduced the uncertainties related to the data-driven efficiency measurements by roughly a factor of five.
Specific care is dedicated to the improvement of muon identification algorithms and of the efficiency measurement in extreme regions of the phase space, such as \pt{} of a few~\GeV{} or a few~\TeV{},
the forward region of the detector where instrumentation coverage is poorer, or an environment polluted by a large number of $pp$ interactions.
Muon reconstruction and identification efficiencies in the bulk of the phase space are measured using the tag-and-probe method, applied to $Z\to\mu\mu$ data collected during the full Run~2. The available data set is about 40 times larger than that used in the previous publication, and revised algorithms for the efficiency extraction and for the modelling of background contamination are adopted. A similar approach is used for the measurements of vertex association and isolation selection efficiencies, while the measurements of muon reconstruction and identification efficiency at low \pt{} or in forward regions of the detector rely on the tag-and-probe method applied to a $J/\psi\to\mu\mu$ data set, and on a double-ratio method applied to the $Z\to\mu\mu$ data set, respectively.

This article is structured as follows. Section~\ref{sec:detector} briefly describes the experimental apparatus, Section~\ref{sec:datamc} provides details of the analysed data set and simulated samples,
Section~\ref{sec:reconstruction} summarises the muon candidate reconstruction process, and
Section~\ref{sec:identification} describes the algorithms developed for optimal muon identification. Section~\ref{sec:efficiency} and Section~\ref{sec:results} are the core of the article: the former describes the measurements of muon identification, vertex association and isolation selection efficiencies using several data-driven techniques, while the latter details the results obtained. Conclusions are given in Section~\ref{sec:conclusions}.

\section{ATLAS detector}
\label{sec:detector}
 
\newcommand{\AtlasCoordFootnote}{
ATLAS uses a right-handed coordinate system with its origin at the nominal interaction point (IP)
in the centre of the detector and the \(z\)-axis along the beam pipe.
The \(x\)-axis points from the IP to the centre of the LHC ring,
and the \(y\)-axis points upwards.
Cylindrical coordinates \((r,\phi)\) are used in the transverse plane,
\(\phi\) being the azimuthal angle around the \(z\)-axis.
The pseudorapidity is defined in terms of the polar angle \(\theta\) as \(\eta = -\ln \tan(\theta/2)\).
Angular distance is measured in units of \(\Delta R = \sqrt{(\Delta\eta)^{2} + (\Delta\phi)^{2}}\).}

The ATLAS detector~\cite{PERF-2007-01} at the LHC covers nearly the entire solid angle around the collision point.\footnote{\AtlasCoordFootnote}
ATLAS consists of an inner tracking detector surrounded by a thin superconducting solenoid, electromagnetic and hadronic calorimeters,
and a muon spectrometer incorporating three large superconducting toroidal magnets.
 
The inner-detector system (ID) is immersed in a \SI{2}{\tesla} axial magnetic field which bends
charged particles in the $r$--$\phi$ plane and provides tracking capabilities in the range \(|\eta| < 2.5\). The high-granularity silicon pixel detector covers the vertex region and typically provides four position measurements (hits) per track,
the first hit normally being in the insertable B-layer installed before Run~2~\cite{ATLAS-TDR-2010-19,PIX-2018-001}.
It is followed by the silicon microstrip tracker (SCT), which usually provides eight measurements per track.
These silicon detectors are complemented by the transition radiation tracker (TRT),
which enables radially extended track reconstruction up to \(|\eta| = 2.0\).
 
The calorimeter system covers the pseudorapidity range \(|\eta| < 4.9\).
Within the region \(|\eta|< 3.2\), electromagnetic calorimetry is provided by barrel and
endcap high-granularity lead/liquid-argon (LAr) calorimeters,
with an additional thin LAr presampler covering \(|\eta| < 1.8\)
to correct for energy loss in material in front of the calorimeters.
Hadronic calorimetry is provided by the steel/scintillator-tile calorimeter,
segmented into three barrel structures within \(|\eta| < 1.7\), and two copper/LAr hadronic endcap calorimeters.
The solid angle coverage is completed with forward copper/LAr and tungsten/LAr calorimeter modules
optimised for electromagnetic and hadronic measurements, respectively.
 
The muon spectrometer~\cite{ATLAS-TDR-10} (MS) comprises separate trigger and
high-precision tracking chambers measuring the deflection of muons in the $r$--$z$ plane due to a magnetic field generated by the superconducting air-core toroids.
The field integral of the toroids ranges between \num{2.0} and \SI{6.0}{\tesla\metre}
across most of the detector. A set of precision chambers covers the region \(|\eta| < 2.7\) with three stations of monitored drift tube (MDT) chambers. The innermost MDT station is replaced with cathode-strip chambers (CSCs) in the $|\eta|>2.0$ region, where the background is higher. Each MDT chamber provides six to eight $\eta$ measurements along the muon track, while the CSCs provide four simultaneous measurements of $\eta$ and $\phi$. The nominal single-hit resolution of the MDTs and CSCs is about \SI{80}{\micro\metre} and \SI{60}{\micro\metre}, respectively, in the bending plane.
The chambers are precisely aligned with a system based on optical sensors~\cite{ATLAS-TDR-10} designed to obtain a 10\% transverse momentum resolution for 1~\TeV\ muons.
The muon trigger system covers the range \(|\eta| < 2.4\) with resistive-plate chambers (RPCs, three doublet stations for $|\eta|<1.05$) in the barrel, and thin-gap chambers (TGCs, one triplet station followed by two doublets for $1.0<|\eta|<2.4$) in the endcap regions.
The RPCs and TGCs also provide tracking information complementary to the precision chambers, in particular improving the determination of
the track coordinate in the non-bending direction, referred to as the second coordinate. The typical spatial resolution for the position measurements in the RPCs and TGCs is 5--10~mm
in both the bending plane and in the non-bending direction.
 
Interesting events are selected by the first-level trigger system implemented in custom hardware,
followed by selections made by algorithms implemented in software in the high-level trigger~\cite{TRIG-2016-01}.
The first-level trigger accepts events from the \SI{40}{\MHz} bunch crossings at a rate below \SI{100}{\kHz},
which the high-level trigger reduces in order to record events to disk at about \SI{1}{\kHz}.
 
\section{Data and Monte Carlo samples}
\label{sec:datamc}
 
\subsection{Data set description}
The results presented in this article are obtained from an analysis of $pp$ collision events
collected by the ATLAS detector in the years from 2015 to 2018, with proton bunches colliding every 25~ns at a centre-of-mass energy of $\sqrt{s}=13$~\TeV{}.
The data set corresponds to an integrated luminosity of 139 fb$^{-1}$, with an average number of $pp$ collisions per bunch crossing of $\langle\mu\rangle=34$, and
maximum instantaneous luminosity of \mbox{$2.1\times 10^{34}$~cm$^{-2}$\,s$^{-1}$}.
The average number of interactions per bunch crossing varied during the data-taking, with values of $\langle\mu\rangle=13$, $\langle\mu\rangle=25$, $\langle\mu\rangle = 38$,
$\langle\mu\rangle=36$ during 2015, 2016, 2017, and 2018, respectively.

Events are accepted for the analysis only if both the solenoid and toroid magnets were on during data taking and if the ID,  MS, and calorimeter detectors were in good operating condition\footnote{Overall, 95.6\% of the recorded proton--proton collision data collected at $\sqrt{s}=13$~\TeV{} is certified for physics analysis.}~\cite{DAPR-2018-01}.
The criteria used to define the good operating condition of the RPC subsystem were reoptimised for data taking in 2017 and 2018,
allowing the use of about 1\% more integrated luminosity with no visible impact on muon reconstruction performance.
 
Events were selected online using dedicated muon trigger algorithms~\cite{Aad:2020uyd} that identified signatures consistent with the prompt decays of $Z$ and $J/\psi$ resonances into two muons, with the $J/\psi\to\mu\mu$ sample used to measure the low-\pT muon reconstruction and identification efficiency.

The  online selection of $Z\to\mu\mu$ candidates was based on single-muon trigger algorithms, to avoid any bias in the reconstruction and identification of the other muon from the decay.
The trigger algorithms imposed requirements on the muon candidate's minimum \pT\ and isolation with respect to nearby tracks in the ID. These requirements varied according to the LHC running conditions: in 2015 the muon \pT threshold was 20~\GeV{}, and a loose isolation selection was applied;
starting from 2016 the muon \pT threshold was increased to 26~\GeV{}, and a more restrictive isolation requirement was imposed.
 
The $J/\psi\to\mu\mu$ candidates events were selected online using several triggers,
all based on the identification of one muon candidate plus one MS track or one ID track.
The use of MS tracking information was needed for an unbiased measurement of the ID track reconstruction efficiency.
Similarly, the  use of ID tracking information was needed for unbiased measurements of the remaining components of the muon offline reconstruction and identification efficiency.
The muon-plus-track pair was required to form an invariant mass in the range 2.5--4.3~\GeV.
The muon candidate was required to have a minimum \pT of 4~\GeV{}, or 6~\GeV{}, depending on the data-taking year and on the specific trigger.
For one type of trigger the track was reconstructed using only the MS information and had to satisfy the requirement of \pT> 4~\GeV{} (as measured by the MS).
For two other types of triggers, active during different data-taking years, the track was reconstructed using only ID information: loose track requirements and a \pT threshold of 3.5~\GeV{} were imposed during 2015 and 2016;
whereas during 2017 and 2018 a newly deployed trigger algorithm based on partial event building (PEB) in a region of interest~\cite{Aad:2020uyd} allowed a lower \pT threshold of 3~\GeV{}.
A large number of events satisfied  the requirement for the online  $J/\psi\to\mu\mu$ selection, and therefore only a fraction of them were saved to disk.
This fraction varied depending on the instantaneous luminosity, on the data-taking year, and on the trigger type,
with the triggers based on the PEB technique allowing a larger fraction of events to be collected than in the previous years.

\subsection{Simulated event samples description}
 
The results presented in this article rely primarily on a comparison of selected $Z\to\mu\mu$ and prompt $J/\psi\to\mu\mu$ decays in data, referred to as signal, with the corresponding Monte Carlo (MC) simulated events.
 
The \mbox{$Z\rightarrow\mu\mu$}  signal process was simulated using the
\POWHEGBOXV{v2}~\cite{Alioli:2010xd} generator at next-to-leading order (NLO) in QCD with the CT10 parton distribution function (PDF) set~\cite{CT10} for the hard-scatter process.
Events were generated with a dimuon invariant mass above 40~\GeV{}.
The parton showering was simulated using \PYTHIAV{8.186}~\cite{Sjostrand:2014zea} with the
\textsc{CTEQ6L1} PDF set~\cite{Pumplin:2002vw} and the \textsc{AZNLO}~\cite{STDM-2012-23} set of tuned parameters (tune) for the underlying event.
About 210 million events were simulated for this process.
A set of signal samples of $Z^{*}/\gamma^{*}\rightarrow\mu\mu$ events with dimuon invariant mass generated above 120~\GeV{} and the same settings as described above was also used.
For comparisons to assess systematic uncertainties, additional  $Z\to\mu\mu$ events were generated using the \textsc{Sherpa}~v2.2.1 generator with a set-up described in detail in Ref.~\cite{ATL-PHYS-PUB-2016-003}.
 
The $J/\psi\to\mu\mu$ signal process was simulated using the \PYTHIAV{8.186}~\cite{Sjostrand:2014zea} leading-order generator,
with the \textsc{CTEQ6L1} PDF set and A14~\cite{ATL-PHYS-PUB-2014-021} as the underlying-event tune. In addition, \textsc{Photos}++~v3.52~\cite{Golonka:2005pn,Davidson:2010ew} was used to simulate the effect of final-state radiation.
To increase the effective number of events in the regions of phase space relevant to this analysis,
the events were generated in a reduced phase space, requiring at least one of the two muons to have $\pt>6$~\GeV{} and both muons to have $|\eta|<2.5$.
About 420 million events were simulated using this configuration.
 
Other MC simulated processes were used to study additional contributions from prompt muons, non-prompt muons, or hadrons misidentified as muon candidates.
The diboson, $Z\rightarrow\tau\tau$, and $W\rightarrow\mu\nu$ processes were simulated using the same generator and parton showering algorithm as the $Z\rightarrow\mu\mu$ signal sample.
The contribution from $\ttbar$ production was simulated at NLO using the \POWHEGBOXV{v2}
generator~\cite{Alioli:2010xd}, with the \textsc{NNPDF3.0 NLO} PDF set~\cite{Ball:2014uwa} and parton showering performed using \PYTHIAV{8.186} with the \textsc{NNPDF2.3 LO} PDF
set~\cite{Ball:2012cx} and A14 tune.
Multi-jet events involving heavy-flavour jets, namely $\bbbar$ and $\ccbar$ production, were simulated using \PYTHIAV{8B}~\cite{Sjostrand:2014zea} with the
\textsc{NNPDF2.3 LO} PDF set and A14 as the underlying-event tune.
 
All the generated events were passed through the simulation of the ATLAS detector based on $\GEANT4$~\cite{SOFT-2010-01,Agostinelli:2002hh} and reconstructed with the same algorithms as used for data.
 
The simulation of multiple proton--proton interactions in each bunch crossing, i.e.\ pile-up interactions, was done by adding the detector response simulation
of minimum-bias interactions, generated using \PYTHIAV{8.186} with the A3 min-bias tune~\cite{ATL-PHYS-PUB-2016-017}, on top of the hard-scattering process in amounts corresponding to the pile-up profile observed during the data-taking.
 
 
\section{Reconstruction}
\label{sec:reconstruction}
 
The main signature exploited for muon identification
in ATLAS is that of a minimum-ionising particle, as revealed by the presence of a track in the MS
or characteristic energy deposits in the calorimeters.
The muon reconstruction is based primarily on information from the ID and MS tracking detectors.
Information from the calorimeters is also used: in the determination of track parameters, to account for cases of large energy loss in the calorimeters, and for MS-independent tagging of ID tracks as muon candidates.
The reconstruction of charged particles in the ID is described in Refs.~\cite{Cornelissen:1020106,ATLAS-CONF-2010-072}.
In the following, the MS track reconstruction as well as different muon identification algorithms
based on the complete detector information are described. Additional details are available in Ref.~\cite{ATLAS-TDR-10}.
 
\subsection{Muon spectrometer stand-alone track reconstruction}
 
The reconstruction of tracks in the MS starts with the identification of short
straight-line local track segments reconstructed from hits in an individual MS station. Segments are
identified in the individual stations by means of a Hough transform~\cite{Illingworth198887}.
Segments in the different stations are combined into preliminary track candidates using a loose pointing
constraint based on the IP and a parabolic trajectory that constitutes a first-order approximation
to the muon bending in the magnetic field. Information from precision measurements in the bending plane
is combined with measurements of the second coordinate from the trigger detectors to create
three-dimensional track candidates. Finally, a global $\chi^2$ fit of the muon trajectory through the magnetic field is performed, taking into account the effects
of possible interactions in the detector material as well as the effects of possible misalignments between
the different detector chambers.
 
Using the muon trajectory as obtained from the global $\chi^2$ fit, outlier hits are removed
and hits along the trajectory that were not assigned to the original track candidate are
added. The track fit is then performed again using the updated hit information. Ambiguities
are resolved by removing tracks that share a large fraction of hits with higher-quality tracks; an exception is made in the case of tracks that are identical in two stations but share no hits in
a third station, to ensure a high efficiency for boosted low-mass dimuon systems. The final
set of tracks is re-fitted with a loose IP constraint and taking into account the energy
loss in the calorimeters, and back-extrapolated to the beam line. The $\pt$ of the extrapolated track is then expressed at the IP.
 
\subsection{Muon reconstruction based on complete detector information}
 
Global muon reconstruction is performed using information from the ID and MS
detectors as well as the calorimeters. The reconstruction proceeds according to five main
reconstruction strategies, leading to the corresponding muon \emph{types}: combined (CB), inside-out combined (IO), muon-spectrometer extrapolated (ME), segment-tagged (ST), and calorimeter-tagged (CT).
 
Combined muons are identified by matching MS tracks to ID tracks and performing a combined track
fit based on the ID and MS hits, taking into account the energy loss in the calorimeters.
Based on the particle trajectory from the combined fit, the muon spectrometer hits associated with the track
may again be updated and the track fit repeated. For $|\eta|>2.5$, MS tracks may be
combined with short track segments reconstructed from hits in the pixel and SCT detectors, leading
to a subset of CB muons referred to as silicon-associated forward (SiF) muons.
 
IO muons are reconstructed using a complementary inside-out algorithm, which extrapolates ID tracks to the MS and searches for at least three loosely-aligned MS hits. The ID track, the energy loss in the calorimeters and the MS hits are then used in a combined track fit.
This algorithm does not rely on an independently reconstructed MS track, and
therefore recovers some efficiency, for example in regions of limited MS coverage and
for low-$\pt$ muons which may not reach the middle MS station.
 
If an MS track cannot be matched to an ID track, its parameters are extrapolated to the beamline and used to define an ME muon. Such muons are used to extend the acceptance outside that of the ID, thus fully exploiting
the full MS coverage up to $|\eta|=2.7$.
 
ST muons are identified by requiring that an ID track extrapolated to the MS satisfies tight angular matching requirements to at least one reconstructed MS segment. A successfully-matched ID track is identified as a muon candidate, and the muon parameters are taken directly from the ID track fit.
 
Finally, CT muons are identified by extrapolating ID tracks through the calorimeters
to search for energy deposits consistent with a minimum-ionising particle. Such
deposits are used to tag the ID track as a muon, and the muon parameters are
again taken directly from the ID track fit. While the other muon reconstruction
algorithms make use of ID tracks with $\pt$ down to $2~\GeV$,
a $\pt$ threshold of $5~\GeV$ is applied for CT muon reconstruction due
to the large background contamination at low $\pt$.
 
The muon reconstruction described here features several improvements compared to
that described in Ref.~\cite{PERF-2015-10}:
\begin{itemize}
\item The use of a parabolic trajectory in the pattern
recognition provides better matching between the segments in the different stations
than the straight-line trajectory used previously.
\item The introduction of SiF muons allows
better use of the ID near the boundaries of its acceptance.
\item Alignment uncertainties are now accounted for in the track fits via constrained nuisance
parameters describing translational and rotational chamber displacements.
\item The calorimeter-tagging algorithm has been retuned for improved purity in the
region of limited MS coverage, $|\eta|<0.1$, and an additional, looser working
point has been introduced targeting high efficiency for use in tag-and-probe studies.
\end{itemize}
 
\section{Identification}
\label{sec:identification}
 
\subsection{Identification criteria}
\label{subsec:pid}
 
After reconstruction, high-quality muon candidates used for physics analyses are selected by a set of requirements on the number of hits in the different ID subdetectors and different MS stations, on the track fit properties, and on variables that test the compatibility of the individual measurements in the two detector systems. A given set of requirements for each of the muon
types defined in Section~\ref{sec:reconstruction} is referred to as a selection \emph{working point} (WP).
Several WPs are defined to suit the needs of the wide variety of physics analyses involving
final states containing muons.
Different analyses have different requirements in terms of efficiency of prompt-muon
identification, resolution of the momentum measurement, and rejection of background due to non-prompt muons.
Among non-prompt muons, an explicit distinction is made between muon candidates
originating from the semileptonic in-flight decay of light hadrons and those from hadrons containing heavy flavours.
The selection WPs target the rejection of light hadrons, which in general result in lower-quality muon tracks, due to the change in trajectory stemming from the in-flight decay within the detector.
Bottom and charm decays produce good-quality muon tracks and these can be distinguished from prompt muons, which
are more closely associated with the primary vertex and more isolated in the ID and/or in the calorimeters.
 
\subsubsection{Design rationale for selection working points}
 
The selection efficiency and purity in simulation are among the main metrics considered in the optimisation of the requirements defining each WP. In particular, the prompt muon efficiency of a selection WP represents the probability that a prompt muon traversing the detector is reconstructed as a muon and satisfies the WP. In a similar way, the purity of a selection WP is one minus the hadron misidentification rate, where the hadron misidentification rate is the fraction of light hadrons reconstructed as muons and satisfying the WP.
 
Three standard selection WPs are designed to cover the needs of the majority of physics analyses. In order
of increasing purity and decreasing efficiency, these are the \loose, \medium, and \tight\ WPs,
where the muons passing the \medium\ (\tight) WP requirements constitute a subset
of those passing \loose\ (\medium). The \medium\ WP provides an efficiency and purity suitable
for a wide range of analyses, while keeping the systematic uncertainties in the prompt-muon
efficiency and background rejection small. The \loose\ selection WP was optimised for the reconstruction
of Higgs boson decays in the four-muon final state, which, due to the high muon multiplicity and
large signal-to-background ratio, benefits from a higher efficiency at the cost of less purity
and larger systematic uncertainties. Finally, the \tight\ selection WP provides the highest purity,
offering a substantially improved background rejection at the cost of a few percent efficiency loss
for prompt muons compared to \medium. The \tight\ WP benefits analyses that are limited by background from non-prompt muons.
 
Two additional selection WPs are designed for analyses targeting extreme phase space regions.
The \highPt\ WP ensures an optimal momentum measurement for muons with $\pt$ above
$100~\GeV$. Optimised for $W^\prime$ and $Z^\prime$ searches, this WP provides the best
momentum resolution and an optimal rejection of poorly reconstructed tracks affected by large
uncertainties. The \lowPt\ WP targets the lowest-$\pt$ muons, which are
less likely to be reconstructed as full tracks in the MS, so that identification
based on MS segments is necessary. For these muons, the background from non-prompt muons can be large,
and the \lowPt\ WP exploits a set of variables providing a good separation between
prompt muons and light-hadron decays to obtain an optimal background rejection while maintaining high
efficiency. Two versions of the \lowPt\ WP have been developed: a \emph{cut-based} selection, which reduces the kinematic dependencies of the background efficiencies, simplifying the implementation of data-driven estimates, and a \emph{multivariate} (MVA) WP, maximising the overall performance. Typical analyses that benefit from the use of the \lowPt\ WP are measurements of Standard Model parameters in the quark-mixing sector~\cite{BPHY-2018-01}, and searches for supersymmetry with compressed mass spectra~\cite{Aad:2019qnd}.
 
In the following, the number of \emph{precision stations} of a muon is defined as the number of MS stations in which the muon has at least three hits in the MDT or CSC detectors.
A \emph{precision hole station} is defined as a station where the muon has less than three hits and is missing at least three hits
that are expected given its trajectory and the detector layout and operational status. The $q/p$ compatibility is defined for CB and IO muons with an MS track as:
\begin{equation}\nonumber
q/p~\mathrm{compatibility} = \frac{|q/p_\mathrm{ID}-q/p_\mathrm{MS}|}{\sqrt{\sigma^2(q/p_\mathrm{ID})+\sigma^2(q/p_\mathrm{MS})}},
\end{equation}
where $q/p_\mathrm{ID}$ and $q/p_\mathrm{MS}$ are the measurements in the ID and MS of the ratio of the charge $q$ to the momentum $p$ of the muon, expressed at the IP, while $\sigma(q/p_\mathrm{ID})$ and $\sigma(q/p_\mathrm{MS})$ are the corresponding uncertainties.
Finally, $\rho^\prime$ is defined for CB and IO muons with an MS track
as the absolute difference between the ID and MS $\pt$ measurements divided by the $\pt$
of the combined track:
\begin{equation}\nonumber
\rho^\prime = \frac{|p_{\mathrm{T,ID}}-p_{\mathrm{T,MS}}|}{p_{\mathrm{T,CB}}},
\end{equation}
where $p_{\mathrm{T,ID}}$ and $p_{\mathrm{T,MS}}$ are respectively the muon $\pt$ measured in the ID and in the MS, while $p_{\mathrm{T,CB}}$ is the value resulting from the combined track fit. No requirements on the $q/p$ compatibility and $\rho^\prime$ variables are considered for muons without an ID or MS track, for which these variables are not defined.
 
All CB, IO, ST, and CT muons are subject to a common set of requirements on the ID track for
all WPs. At least one hit in the pixel detector and at least five hits in the SCT detector
are required, and at most two missing hits are allowed in total in these detectors. A missing hit is counted
where the muon trajectory crosses an active sensor that does not register a hit.
An exception is made for SiF muons, for which at least one pixel hit but only
at least four hits in total in the pixel and SCT detectors are required at the reconstruction stage.

\subsubsection{The \loose, \medium, and \tight\ selection working points}
 
Within the ID acceptance $|\eta|<2.5$, the \medium\ WP accepts only CB and IO muons. These are
required to have at least two precision stations, except in
the region $|\eta|<0.1$, where muons with only one precision station are also included
provided they have at most one precision hole station. The $q/p$ compatibility is required
to be less than seven to ensure a loose agreement between the ID and MS measurements.
The acceptance is extended outside the ID coverage by including ME and SiF muons,
required to have at least three precision stations, in the range $2.5<|\eta|<2.7$.
Among prompt muons passing the \medium\ WP in \ttbar\ events, more than $98\%$ are
CB muons.
 
The \loose\ selection WP accepts all the muons passing the \medium\ WP. In addition, it includes CT and ST
muons in the range $|\eta|<0.1$, where the gap in the MS coverage leads to a loss of
efficiency for CB muon reconstruction. To increase the efficiency of the \loose\ criteria for low-$\pt$ muons, IO muons with $\pt$ below $7~\GeV$ and only one precision station are accepted
in the range $|\eta|<1.3$, provided they are independently reconstructed also as ST muons. Requiring that IO muons are independently confirmed by the ST reconstruction strategy significantly increases their purity.
Among prompt muons passing the \loose\ WP in \ttbar\ events, about $97\%$ are
CB or IO muons. Approximately $1.5\%$ are CT and ST muons in the region $|\eta|<0.1$, among
which the majority are CT muons. The efficiency increase of the \loose\ WP compared to
\medium\ is around $20\%$ for $3~\GeV<\pt<5~\GeV$ and approximately $1$--$2\%$ for higher $\pt$.
 
Among the muons passing the \medium\ selection WP, only CB and IO muons with at least two precision stations
are accepted for the \tight\ WP. The normalised $\chi^2$ of the combined track fit is required
to be less than 8 to reject pathological tracks due to hadron decays in flight. Further requirements are placed on the $q/p$
compatibility and $\rho^\prime$ depending on the $\pt$ and $|\eta|$ of the muon. These are
optimised to provide better background rejection for lower-$\pt$ muons, because of the higher
expected non-prompt background at low $\pt$. In the optimisation, the rejection of non-prompt
muons is maximised for a given target prompt-muon efficiency that rises from approximately $91\%$ at $\pt=4~\GeV$
to $95\%$ at $\pt=9~\GeV$ and approaches $96\%$ as the $\pt$ approaches $20~\GeV$.
For the region $6~\GeV<\pt<20~\GeV$, the \tight\ WP achieves a background reduction of more than
$50\%$ compared to \medium, with a corresponding efficiency loss for prompt muons of approximately $6\%$.
 
\begin{figure}[!h]
\begin{center}
\includegraphics[width=.49\linewidth]{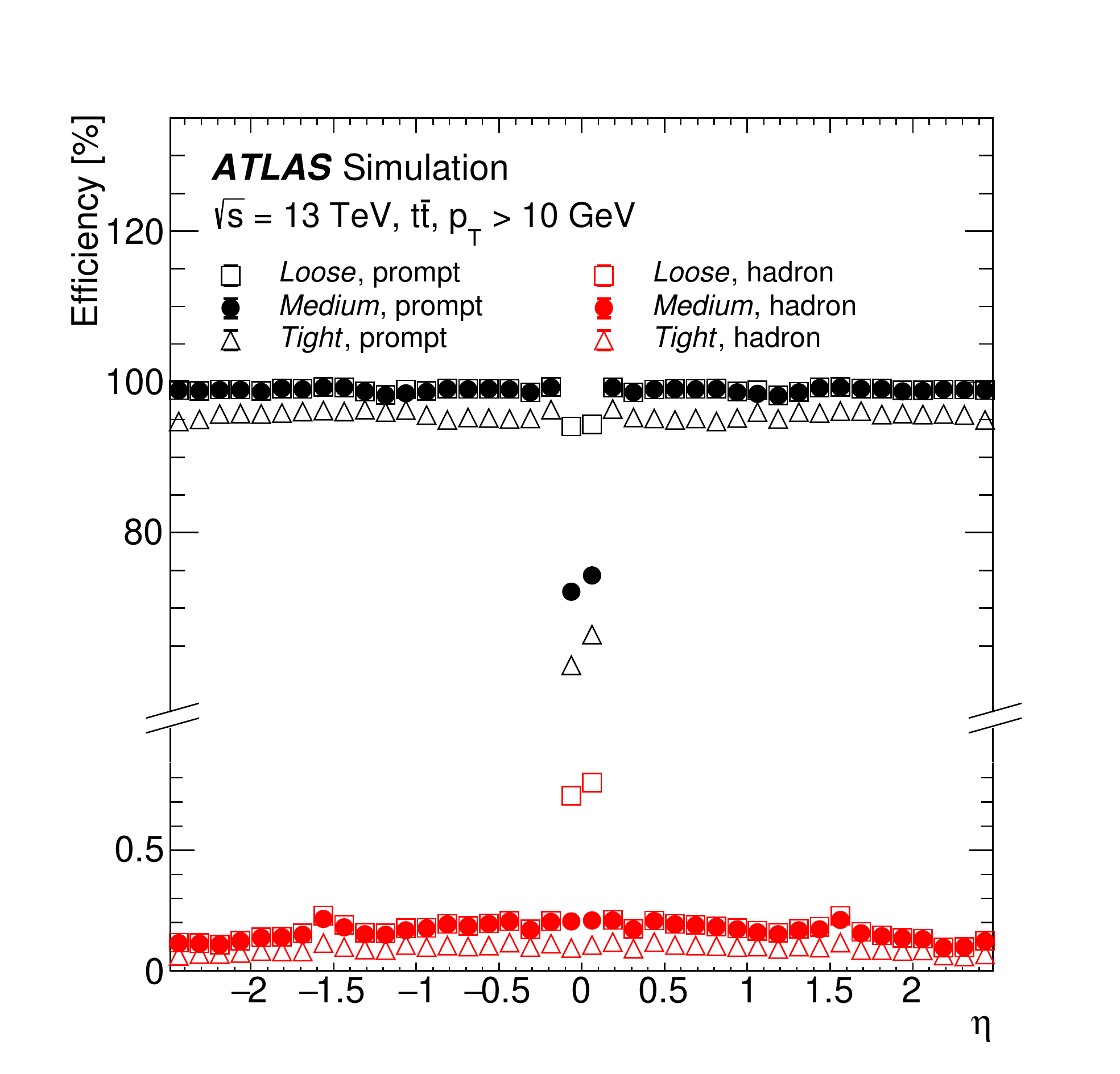}
\includegraphics[width=.49\linewidth]{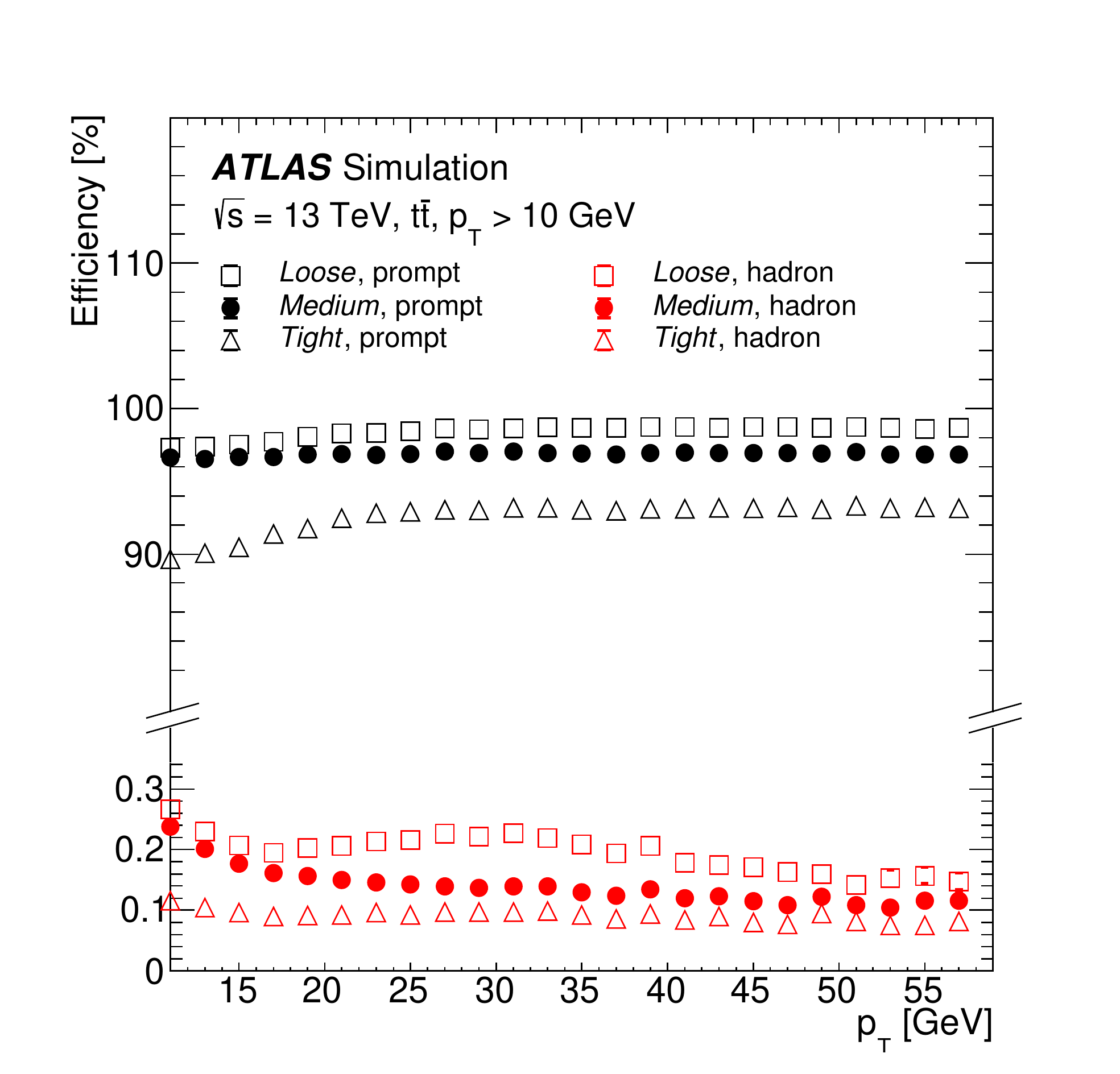}
\end{center}
\caption{Efficiency as a function of $\eta$ (left) and $\pt$ (right) of the ID track for the \loose, \medium\ and \tight\ WP
requirements in simulated $\ttbar$ events, shown separately for prompt muons and muons from light hadron decays.
The efficiency is calculated as the fraction of ID tracks that are associated with a reconstructed muon passing the given WP requirements.
The ID tracks are matched, respectively, to generator-level prompt muons or light hadrons.}
\label{fig:LMTrates}
\end{figure}

The performance of the \loose, \medium\ and \tight\
selection WPs for tracks with $\pt>10~\GeV$ in simulation is illustrated in Figure~\ref{fig:LMTrates}.
 
\subsubsection{The \highPt\ selection working point}
 
In the reconstruction of very high $\pt$ muons with almost straight trajectories, the limiting
factors are the intrinsic detector resolution of the individual measurements along the track and
the knowledge of the relative alignment between the corresponding detector elements. The
design resolution for stand-alone momentum measurements in the MS can only be achieved
for muons with hits in at least three precision stations. For muons with only two precision stations,
the resolution of the stand-alone measurement deteriorates significantly, but some of the loss in momentum resolution can be recovered
through the combined track fit, which uses the hits in the ID as well.
 
Only CB and IO muons passing the \medium\ WP requirements are accepted for the \highPt\ WP.
At least three precision stations are required, with the following exceptions:
\begin{itemize}
\item For muons traversing
the $B$-field inversion zones instrumented with additional chambers, at least four precision stations are
required due to the particular trajectory of muons in this region.
\item Muons with only two precision stations are accepted provided the missing hits are in the inner station, as this category of tracks shows a better momentum resolution than other tracks with less than three precision stations. They are, however, restricted to the $|\eta|<1.3$ region, where the effects of relative misalignments between the ID and MS on muons with two precision stations are less pronounced.
\end{itemize}
Muons are rejected if their $\eta$ and $\phi$ coordinates correspond to regions of the MS
where the relative alignment between the traversed chambers is not known with sufficient precision.
For this reason, all muons in the barrel--endcap overlap region $1.0 < |\eta| < 1.1$ are rejected, while partial acceptance losses also occur in $1.1 < |\eta| < 1.3$, and in the $|\eta|<1.0$ region corresponding to the detector support structures, around $\phi=-1.2$ and $\phi=-2.0$.
 
\begin{figure}
\centering
\includegraphics[width=0.49\textwidth]{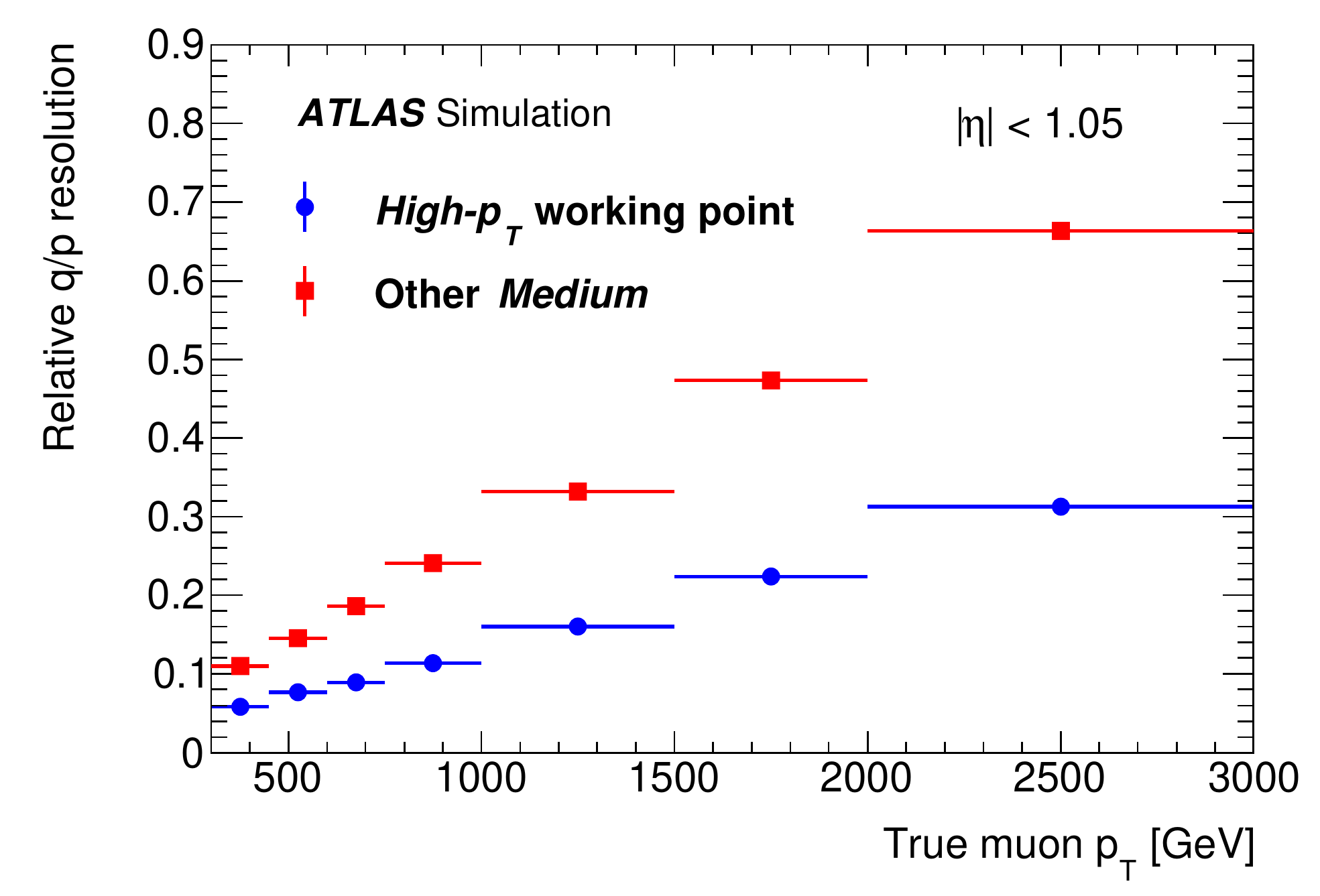}
\includegraphics[width=0.49\textwidth]{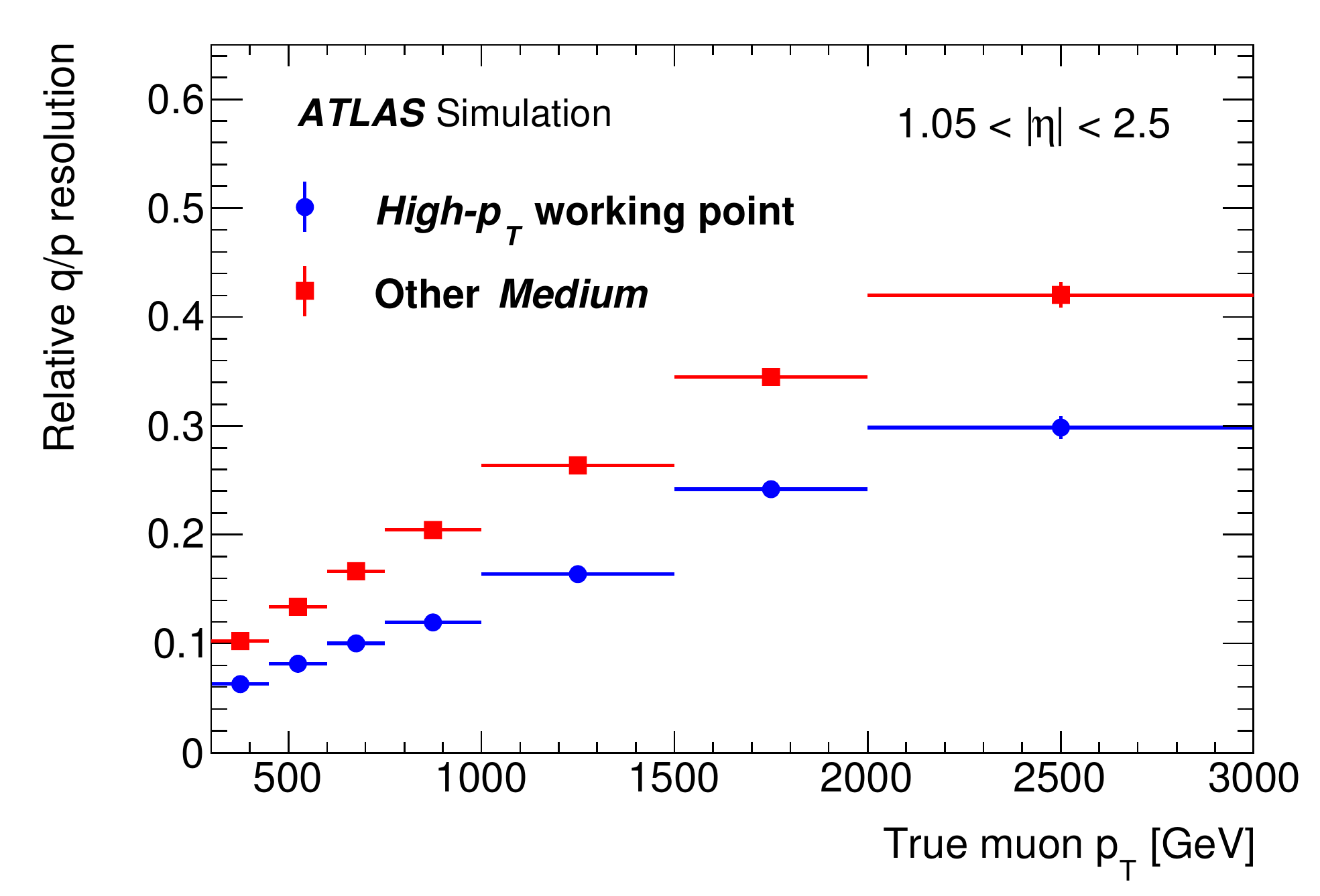}
\caption{The $q/p$ resolution as a function of the transverse momentum of muons in the barrel (left)
and endcaps (right), shown separately for muons passing the \highPt\ WP
requirements and for the rest of the muons passing the \medium\ WP requirements. The resolutions are evaluated in MC samples that include realistic
misalignments corresponding to the data-taking conditions in 2015.}
\label{fig:highPtResolution}
\end{figure}

The resolution of high-$\pt$ muons is evaluated in MC samples that include a realistic
simulation of relative misalignments between the MS chambers and between the ID and MS. The
resolution is extracted from a Gaussian fit to the core of the distribution of relative
residuals $[(q/p)_\mathrm{reco} - (q/p)_\mathrm{truth}]/(q/p)_\mathrm{truth}$, with $(q/p)_\mathrm{reco}$
and $(q/p)_\mathrm{truth}$ being the reconstructed and generated $q/p$ values, respectively.
Figure~\ref{fig:highPtResolution} shows the resolution as a function of $\pt$ for muons passing
the \highPt\ WP requirements, and for comparison, the resolution for muons failing the
\highPt\ requirements but passing the \medium\ WP ones.
As expected, superior resolution is obtained for the muons
passing the \highPt\ WP requirements, while the resolution for the rest of the muons passing
\medium\ is worse by up to roughly a factor of two depending on the detector region
and $\pt$.
 
An additional selection is placed on the estimated momentum uncertainty from the combined track
fit to reject suboptimal momentum measurements. Specifically, the relative $q/p$ uncertainty $\sigma_\mathrm{rel}(q/p)=\sigma(q/p)/|q/p|$ is required to be below a given threshold, defined as a $\pt$-dependent
coefficient multiplied by the expected momentum resolution. The $\pt$-dependent coefficient is optimised separately for muons with different numbers of precision stations, and follows a decreasing trend for $\pt$ greater than $1~\TeV$ due to the presence of larger resolution tails at very high $\pt$. The expected resolution is parameterised
as a function of $\pt$ in five $|\eta|$ regions, separately for the muons with two precision
stations and those with at least three. The resulting criterion, referred to as the $\sigma_\mathrm{rel}(q/p)$ selection hereafter, is more
than $99\%$ efficient at $\pt=1~\TeV$ for muons with at least three precision stations. The efficiency decreases slightly for higher $\pt$, reaching approximately $96\%$ at $\pt=2~\TeV$ and $89\%$ at $\pt=2.5~\TeV$. For muons with only two precision stations, the $\sigma_\mathrm{rel}(q/p)$ selection is more stringent, with an efficiency of around $50\%$ for $\pt$ between $1$ and $2~\TeV$.

A $\pt$-dependent uncertainty in the efficiency of the $\sigma_\mathrm{rel}(q/p)$ selection is assigned. As the impact of the selection becomes sizeable only at very high $\pt$, where the available number of muons from $Z\to\mu\mu$ decays is limited, the uncertainty is evaluated from an inclusive sample of muons with high $\pt$. All muons in the sample are required to satisfy the \highPt\ WP criteria, not including the
$\sigma_\mathrm{rel}(q/p)$ selection. The fraction of muons that also pass the $\sigma_\mathrm{rel}(q/p)$ selection
is compared between data and Drell--Yan dimuon MC samples covering the invariant mass range up to
several $\TeV$. The difference is assigned as the uncertainty in the selection efficiency, which
becomes a dominant source of uncertainty at very high $\pt$, approaching for example $55\%$
for $\pt$ above $3~\TeV$ in the region $|\eta|<1.3$.
 
The overall reconstruction and selection efficiency of the \highPt\ WP criteria for muons in $Z/\gamma^{*}$ events, simulated with realistic detector misalignments corresponding to the data taking conditions in 2015, is about $80\%$ at $\pt=100~\GeV$, and decreases approximately to $76\%$ at $\pt=500~\GeV$, $72\%$ at $\pt=1~\TeV$, and $68\%$ at $\pt=2~\TeV$.
 
\subsubsection{The \lowPt\ selection working point}
 
Only CB and IO muons are used in the \lowPt\ selection WP. Muons, on average, lose roughly $3~\GeV$ of their energy while traversing the calorimeters. At very low $\pt$, a muon may not reach the middle
station of the MS, or even the MS itself, leading to a loss of efficiency for stand-alone MS track reconstruction.
For this reason, a significant fraction of muons in this $\pt$ region are reconstructed
only by the IO algorithm, and these are required to be independently
reconstructed also as ST muons for increased purity. At least one precision station is required,
except in the region $|\eta|>1.3$, where muons with $\pt$ greater than $3$~\GeV{} generally have enough energy to reach the second station and thus the requirement is at least two. For $\pt$ above $10~\GeV{}$,
the efficiency improvement relative to \medium\ becomes marginal, and the \lowPt\ WP
is defined to be identical to \medium\ above $\pt=18~\GeV$.
 
Further selection requirements are imposed to reject light-hadron decays. CB and IO muon tracks resulting
from hadron decays in flight are characterised by a distinctive \emph{kink} along the trajectory in the ID
due to the momentum carried away by the undetected neutrino. Several variables offering good
discrimination between prompt and non-prompt muons are exploited in the \lowPt\ WP. For the cut-based WP, selection requirements are imposed independently on the individual discriminating variables, while the multivariate WP further exploits correlations by combining several discriminating variables in a
boosted decision tree (BDT).
 
Three variables quantifying the presence of a kink on the muon track are used to define
the cut-based WP: the momentum balance significance (MBS), the scattering neighbour
significance (SNS), and the scattering curvature significance (SCS). The MBS is defined as:
\begin{equation}\nonumber
\mathrm{MBS} = \frac{\left|p_\mathrm{ID}-\hat{p}_\mathrm{MS}-E_\mathrm{loss}\right|}{\sigma(E_\mathrm{loss})},
\end{equation}
where $p_{\mathrm{ID}}$ and $\hat{p}_{\mathrm{MS}}$ are respectively the momentum measured in the ID and in the MS, with the latter expressed at the entrance of the MS, $E_\mathrm{loss}$ the energy loss in the calorimeter system, and $\sigma(E_\mathrm{loss})$ its uncertainty. For muons with no momentum measured in the MS, MBS is set to 0. The SNS and SCS are variables estimating the significance of a change in trajectory (kink) along the track under the hypothesis of a decay vertex between adjacent hits, as expected in the presence of a hadron decaying to a muon. The SNS is defined as the largest value of scattering angle significance over the entire track. Scattering angle significance is computed considering pairs of adjacent hits along the track, and evaluated as the angular distance in the bending plane between the two half tracks ending/starting at each of the hits, divided by the corresponding uncertainty. The SCS looks for the most pronounced discontinuity along the track by evaluating the integral of the scattering angle significances before/after the hypothesized decay vertex. It considers all possible pairs of partial tracks starting/ending at each of the hits, and is computed as the maximum, in absolute value and among all pairs, of the difference between the two sums of significances along each partial track. It is then normalized to the square root of the total number of pairs.
For the cut-based \lowPt\ WP, each of the three significance variables is
required to be below three. Furthermore, in the region 
$|\eta|>1.55$, the \medium\ WP requirements have a high efficiency for low-$\pt$ muons,
and are applied in addition for further reduction of the background from non-prompt muons
in this region.
 
The multivariate \lowPt\ selection WP is based on a gradient BDT which is trained on
separate samples containing prompt muons from $W$ boson decays and non-prompt muons from light-hadron
decays, respectively, in both cases from simulated $\ttbar$ events.
The training is performed separately for muons reconstructed by the CB and IO algorithms, using in both cases the same set of discriminating variables. A total of eight variables are deployed, which provide good discriminating power between prompt and non-prompt muons, and are well modelled in the MC simulation.
The variables used include SCS, SNS, and MBS, as well as additional ones that take advantage of different information from the detector: the energy loss in the calorimeters, the number of MS segments associated with the muon and their direction relative to the track in the ID, and the number of missing precision hits in the middle MS station.
 
The modelling of all variables in simulation is verified by a comparison with data in dedicated control regions
with a high purity of low-$\pt$ prompt muons and muons from hadron decays.
The modelling for prompt muons is evaluated using a selection targeting the
$J/\psi$ resonance. 
The modelling for muons from hadron decays is evaluated using a selection targeting the
decay $B_s^0\to J/\psi\, \phi$ with subsequent decays $J/\psi\to \mu\mu$ and $\phi\to K^+ K^-$.
The two muons are required to satisfy the \medium\ WP requirements and have an invariant mass
close to the $J/\psi$ mass. A $B_s^0$ candidate is reconstructed by matching the muons to a common
vertex with two ID tracks that have an invariant mass close to the $\phi$ mass.
A high purity of $B_s^0$ events is attained by selecting candidates where the four-particle
invariant mass is close to the $B_s^0$ mass, and the corresponding sideband regions are used
to estimate the background. The modelling is checked for muon candidates matched to the ID
tracks forming the $\phi$ candidate.
 
\begin{figure}
\centering
\includegraphics[width=.45\textwidth]{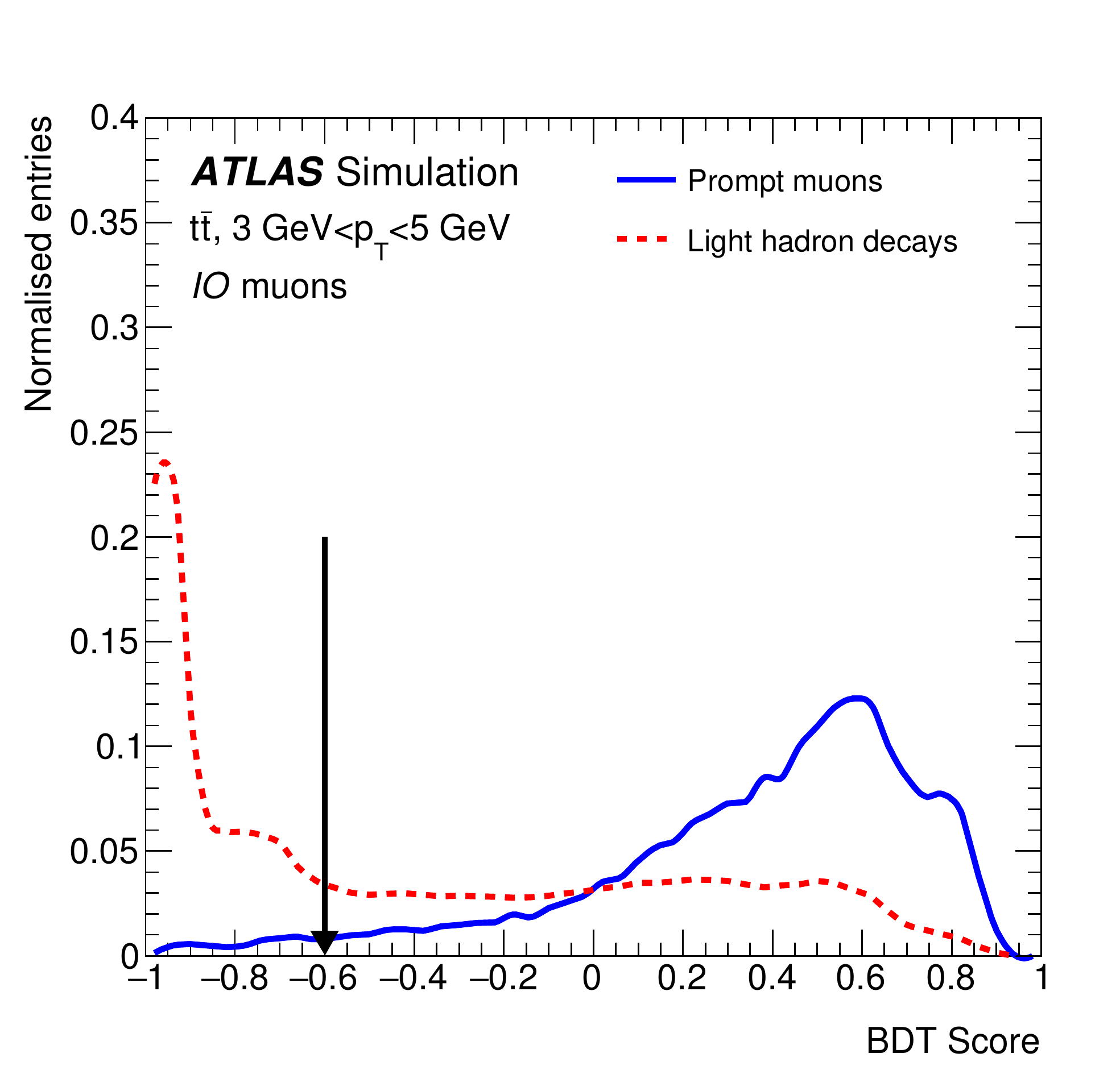}
\includegraphics[width=.45\textwidth]{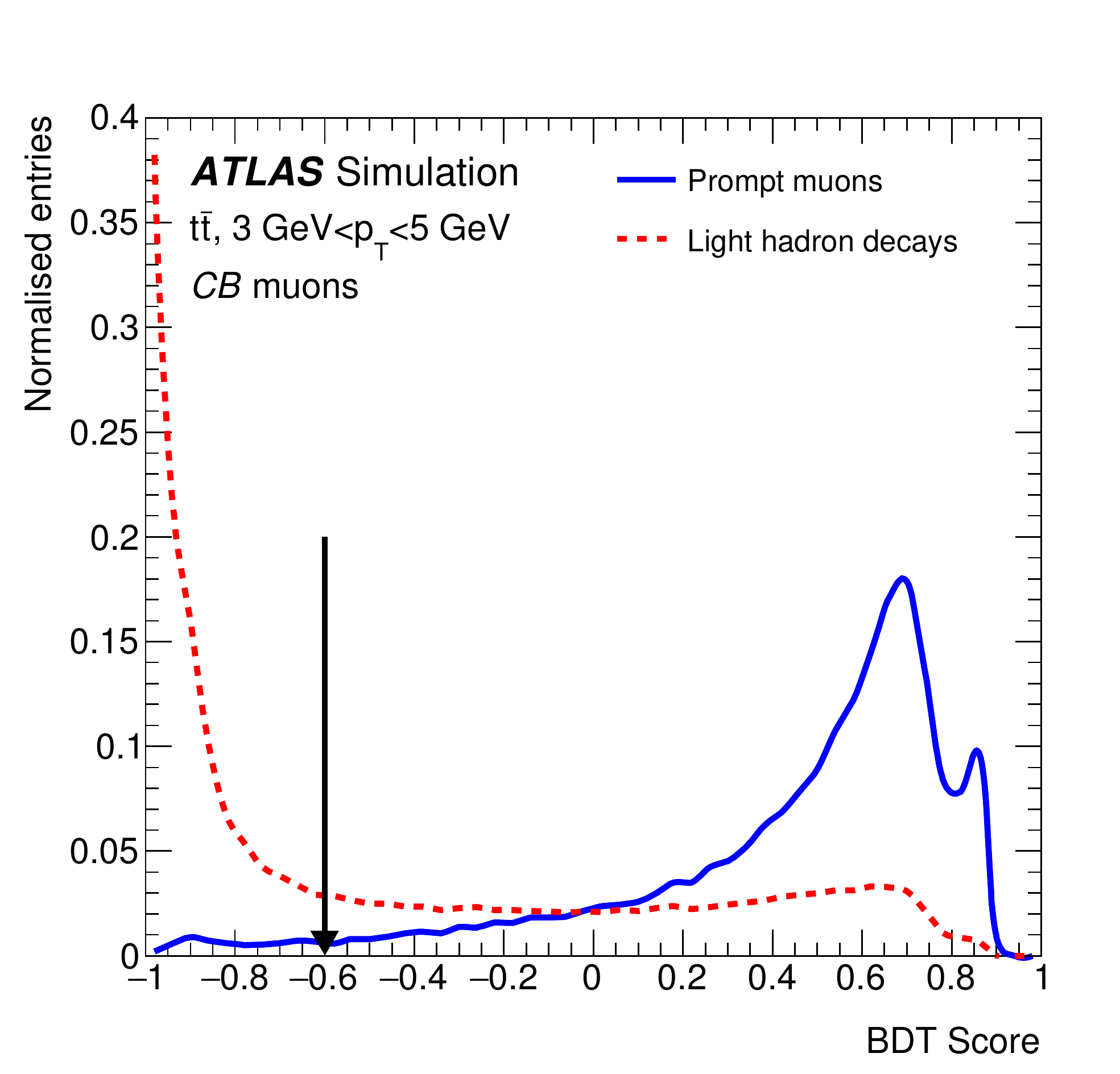}
\caption{Distributions of the gradient BDT score for muons reconstructed with the IO algorithm (left) and CB algorithm (right) in simulated $t\bar{t}$ events. The distributions are shown for prompt muons (full line, blue), and for light hadron decays (dashed line, red). The black arrows indicate the values of the requirements that define the multivariate \lowPt\ selection.}
\label{fig:BDTGscore}
\end{figure}

The distributions of the gradient BDT score for prompt and non-prompt
muons are shown in Figure~\ref{fig:BDTGscore}, where good separation between the two categories
is observed. Good agreement is observed when comparing the distributions obtained from the event sample
used for the BDT training to those extracted from a statistically independent sample, indicating that there is
no overtraining of the BDT.
 
\begin{figure}
\centering
\includegraphics[width=.48\textwidth]{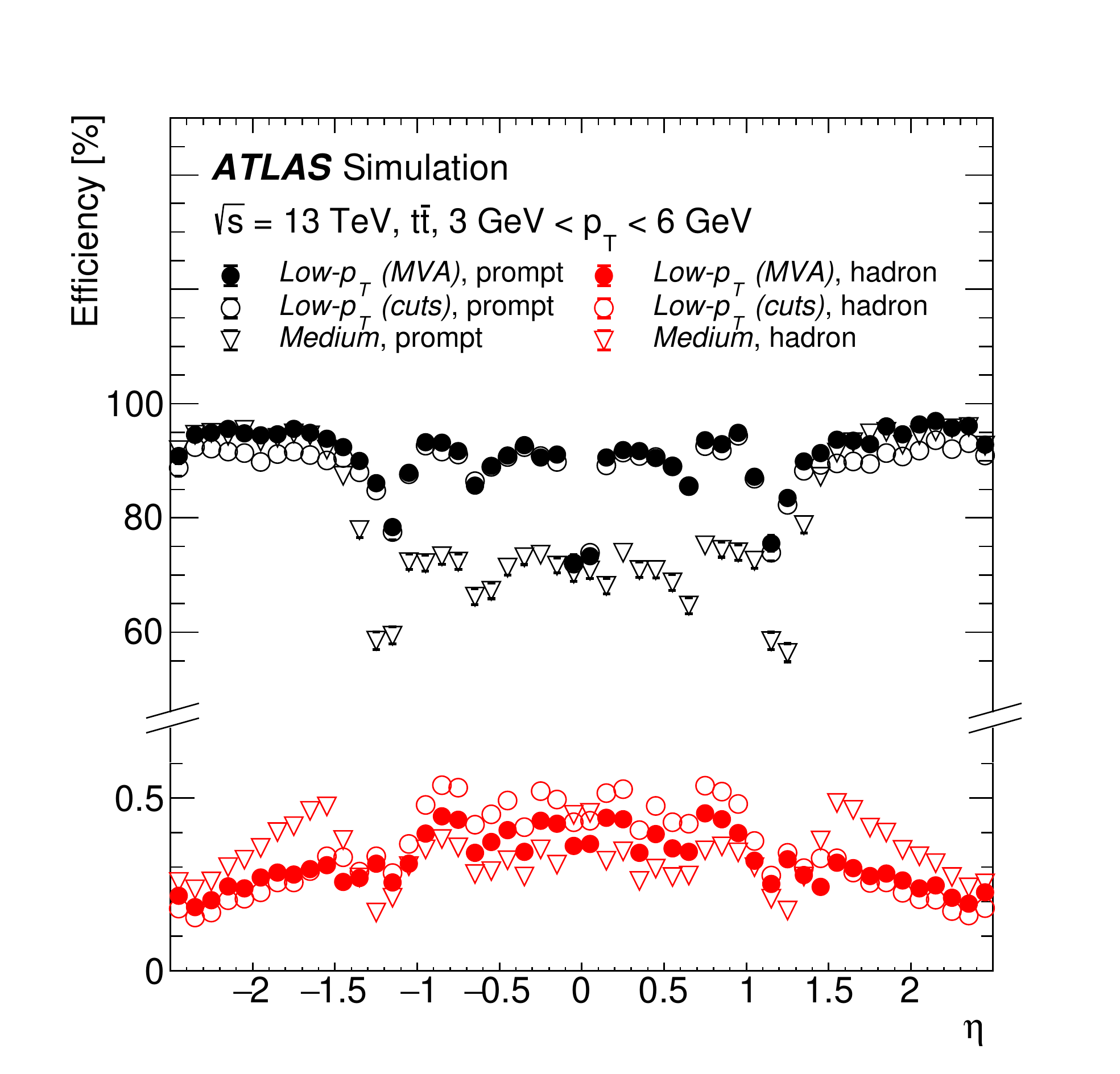}
\includegraphics[width=.48\textwidth]{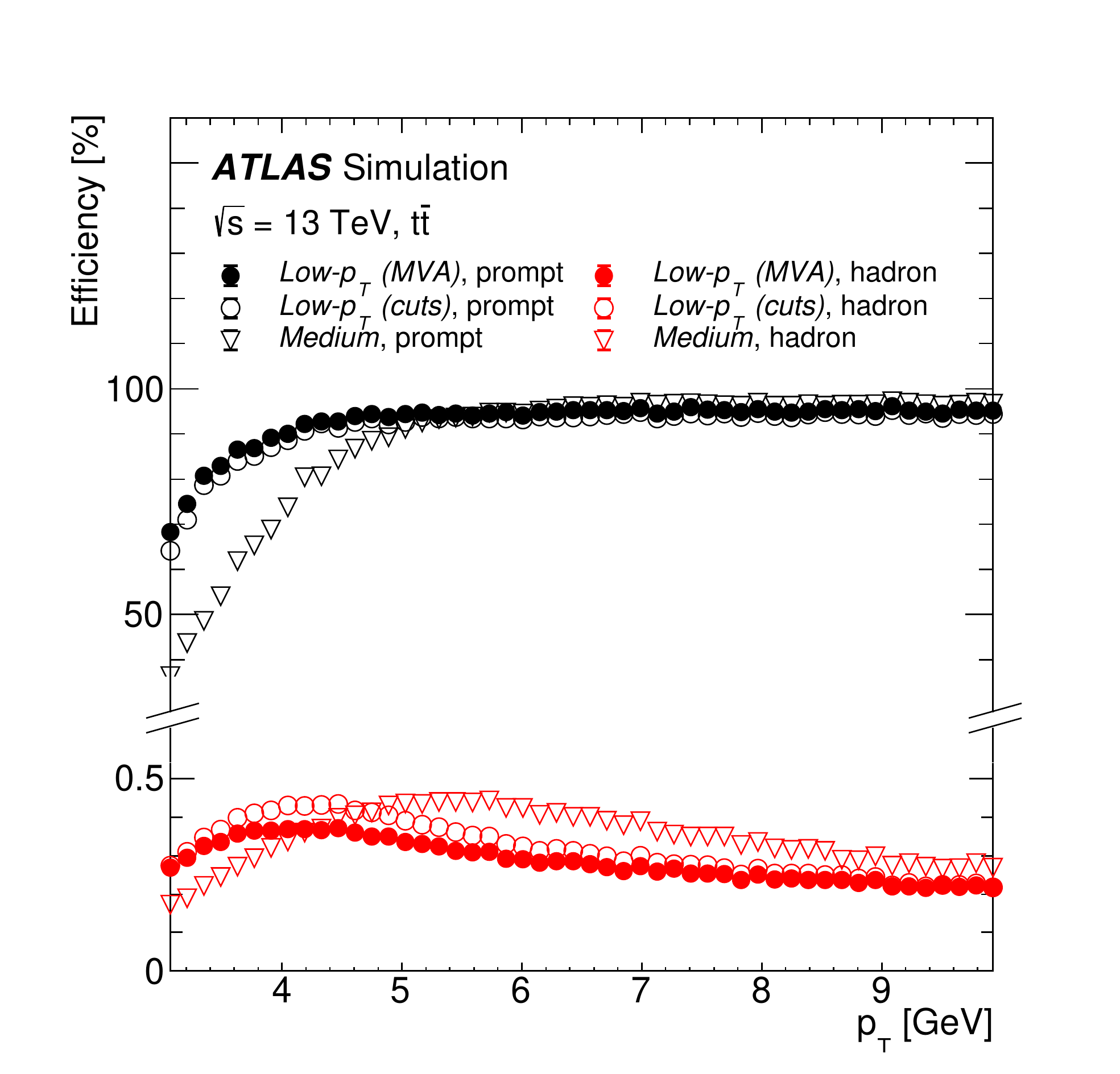}
\caption{Efficiency as a function of $\eta$ (left) and $\pt$ (right) of the ID track for the \lowPt\ and \medium\ WP
requirements in simulated $\ttbar$ events, shown separately for prompt muons and muons from light hadron decays.
The efficiency is calculated as the fraction of ID tracks that are associated with a reconstructed muon passing the given WP requirements.
The ID tracks are matched, respectively, to generator-level prompt muons or light hadrons. Both the cut-based and multivariate \lowPt\ WPs are shown. }
\label{fig:lowPtMediumComparison}
\end{figure}

The performance of the cut-based and multivariate \lowPt\
selection WPs in simulation is compared with that of the \medium\ selection WP in Figure~\ref{fig:lowPtMediumComparison}.
Relative to \medium, the cut-based \lowPt\ WP achieves a substantial increase in the prompt-muon
efficiency in the barrel region while retaining good rejection of non-prompt muons. In the
endcap regions, improved rejection of light-hadron decays is achieved at the cost of
a small prompt-muon efficiency loss. Relative to the cut-based \lowPt\ WP, the multivariate WP achieves better
rejection of non-prompt muons in the barrel region and a higher prompt-muon efficiency in the
endcap regions. Overall, compared to the \medium\ selection WP,
the cut-based (multivariate) \lowPt\ WP accepts an additional 16\% (18\%) of the prompt muons with $3~\GeV<\pt<5$~\GeV{}, while the corresponding increase for light hadrons is approximately 0.2\% (0.1\%).

\subsubsection{Efficiencies and misidentification rates}
 
The prompt muon efficiencies and light-hadron misidentification rates for muons in the region $|\eta|<2.5$ are shown in Table~\ref{tab:MCrates}. In this case, the efficiency is calculated for each selection WP as the fraction of ID tracks associated with a reconstructed muon passing the given WP requirements. It is evaluated in a $\ttbar$
MC sample, for ID tracks matched to generator-level prompt muons from $W$ boson decays. Similarly, the misidentification rate is calculated using ID tracks matched to generator-level hadrons.
\begin{table}
\centering
\caption{Prompt-muon efficiencies $\epsilon_\mu$ and light-hadron misidentification rates $\epsilon_\mathrm{had}$
for the different selection working points, evaluated in a $\ttbar$ MC sample
in different $\pt$ regions for $|\eta|<2.5$. It should be noted that the \tight\ WP by construction does not
select any muons with $\pt<4~\GeV$, which is reflected in the corresponding efficiency in the first $\pt$ region. The statistical uncertainties are at least one order of magnitude smaller than the last digit reported.}
\begin{small}
\begin{tabular}{|c||c|c||c|c||c|c||c|c|}
\hline
\hline
& \multicolumn{2}{c||}{$3<\pt~[\GeV]<5$} & \multicolumn{2}{c||}{$5<\pt~[\GeV]<20$} & \multicolumn{2}{c||}{$20<\pt~[\GeV]<100$} & \multicolumn{2}{c|}{$\pt>100~\GeV$} \\ \hline
Selection WP     & $\epsilon_\mu$ [\%] & $\epsilon_\mathrm{had}$ [\%] & $\epsilon_\mu$ [\%] & $\epsilon_\mathrm{had}$ [\%]
& $\epsilon_\mu$ [\%] & $\epsilon_\mathrm{had}$ [\%]  & $\epsilon_\mu$ [\%] & $\epsilon_\mathrm{had}$ [\%] \\
\hline
\loose  &  90  &  1.17  &  98  &  1.06  &  99  &  0.25  &  98  &  0.12 \\ \hline
\medium  &  70  &  0.63  &  97  &  0.85  &  97  &  0.17  &  97  &  0.07 \\ \hline
\tight  &  36  &  0.15  &  90  &  0.38  &  93  &  0.12  &  93  &  0.04 \\ \hline
\lowPt\ (cut-based)  &  86  &  0.82  &  95  &  0.71  &  97  &  0.17  &  97  &  0.07 \\ \hline
\lowPt\ (multivariate)  &  88  &  0.73  &  96  &  0.66  &  97  &  0.17  &  97  &  0.07 \\ \hline
\highPt  &  45  &  0.34  &  79  &  0.60  &  80  &  0.13  &  80  &  0.05 \\ \hline
\hline
\end{tabular}
\label{tab:MCrates}
\end{small}
\end{table}
 
As expected, the highest prompt-muon efficiency is achieved for the \loose\ selection WP, while the \tight\ WP
achieves the lowest misidentification rate. In the region $3~\GeV<\pt<5~\GeV$, the \lowPt\ WP offers an
efficiency close to that of \loose, with a significantly lower misidentification rate. The efficiency of
the \highPt\ WP is significantly lower than that of all the other WPs due to the strict requirements
necessary to achieve optimal momentum resolution. The misidentification rates are further reduced by approximately one order of magnitude, or more, after the application of vertex association and isolation requirements, discussed in the following sections.
 
\subsection{Vertex association criteria}
\label{subsec:ttva}
 
Selection requirements are imposed on the impact parameters of the muon track to reject muons originating from
hadron decays in flight as well as muons not originating from the hard-scattering proton--proton interaction, for
example those due to pile-up interactions or cosmic rays. The transverse impact parameter $|d_0|$ is the distance
from the beamline to the point of closest approach of the muon track in the transverse plane. It is measured
relative to the actual beam position rather than the reconstructed primary vertex,\footnote{
Collision vertices are reconstructed from ID tracks that satisfy $\pt > 0.5$~\GeV,
and the primary vertex is chosen as the vertex with the largest $\sum \pt^2$ for the tracks associated with this vertex.}
as the beam width is smaller
than the typical uncertainty in the reconstructed primary vertex position in the transverse plane.
The longitudinal impact parameter $z_0$ is the coordinate along the beam axis of the point of closest approach of the
muon track to the beamline, measured relative to the reconstructed primary vertex position. Consequently, the shortest
distance from the muon track to the primary vertex in a longitudinal projection is $|z_0|\sin\theta$, where
$\theta$ is the polar angle of the muon track. For tracks with $\pt > 10$~\GeV, the impact parameter resolution approaches asymptotically a value of about $10~\mu\mathrm{m}$ in the transverse plane and $50~\mu\mathrm{m}$ in the longitudinal direction, while it degrades progressively at lower transverse momenta as a consequence of multiple scattering in the detector material.
 
The transverse impact parameter selection requirement is defined in terms of the $d_0$ \emph{significance},
$|d_0|/\sigma(d_0)$, which is required to be less than three. Due to the excellent tracking resolution for
muons with intermediate to high $\pt$, the beam width is not negligible compared to the
estimated uncertainty in $d_0$ from the track fit, and is hence accounted for in the total uncertainty $\sigma(d_0)$.
Finally, the muon track is ensured to be compatible
with originating from the reconstructed primary vertex by the requirement $|z_0|\sin\theta<0.5~\mathrm{mm}$.

\subsection{Isolation requirements}
\label{subsec:iso}
 
Muons from prompt decays of SM bosons or hypothetical BSM particles can be discriminated from muons from hadronic sources
by measuring the amount of hadronic activity in their vicinity.
The transverse energy (or momentum if considering only tracks) reconstructed in a cone around a muon and divided by the muon \pT defines the muon isolation.
Depending on the topology, most non-prompt muons that can be rejected using isolation criteria originate from heavy-flavour hadron decays. Conversely, the contributions from light-hadron decays or hadrons misidentified as muons are generally efficiently suppressed by the
selection requirements described in Sections~\ref{subsec:pid} and~\ref{subsec:ttva}.
Isolation can be measured independently using either the ID (with ID tracks~\cite{ATL-PHYS-PUB-2015-018}) or the calorimeters (with topological cell clusters~\cite{PERF-2014-07}), or through a combination of the two (with particle flow~\cite{PERF-2015-09}). Several isolation WPs are defined,
balancing prompt-muon acceptance, rejection of non-prompt muons, and performance in close proximity to other objects.
 
Track-based isolation is defined as the scalar sum of the transverse momenta of the ID tracks associated with the primary vertex in an $\eta$--$\phi$ cone of a given size $\Delta R$ around the muon, excluding the muon track itself.
Depending on the isolation selection criterion, $\Delta R$ is either $0.2$, labelled as \ptcone[20], or min(10 \GeV/$\pT^{\mu}$, 0.3), labelled as \ptvarcone[30],
where the latter is optimised for topologies where jets or other leptons are expected in close proximity to an energetic muon~\cite{Rehermann:2010vq}.
In order to increase the rejection of hadronic activity, some isolation selection criteria use \ptvarcone[30] for $\pT^{\mu} < 50$~\GeV\ and \ptcone[20] for $\pT^{\mu} > 50$~\GeV.
The minimum transverse momentum of tracks used in the calculation varies for each isolation criterion and can be either 500~\MeV\ or 1~\GeV.
Track-based isolation variables are largely independent of pile-up, due to the rejection of tracks originating from pile-up vertices or with large transverse impact parameters relative to the primary vertex.
All muon isolation WPs include a selection on one track-based isolation variable, with or without an additional criterion for calorimeter-based or particle-flow-based isolation.
 
Calorimeter-based isolation, labelled as $\topoetcone[20]$, is defined as the sum of the transverse energy of
topological cell clusters in a cone of size $\Delta R$~=~0.2 around the position of the muon, extrapolated to the calorimeters,
after subtracting the contribution from the energy deposit of the muon itself and correcting for pile-up effects.
Contributions from pile-up and the underlying event are estimated using an ambient energy-density technique and are corrected on an event-by-event basis, similarly to the pile-up correction performed in the ATLAS jet calibration~\cite{PERF-2014-03}.
This technique, although it corrects calorimeter-based isolation for the effects of pile-up on average, results in poor energy resolution due to the large size of the pile-up correction relative to the average calorimeter isolation values.
As a result, the performance of calorimeter-based isolation tends to have more pile-up dependence than track-based isolation, and
all criteria that include a requirement on calorimeter-based isolation also apply a more stringent selection on track-based isolation.
 
Combining selections on track-based and calorimeter-based isolation generally results in better performance than employing one but not the other,
as the two isolation variables provide complementary information.
Track-based isolation has better resolution and lower pile-up dependence than calorimeter isolation,
and the ID provides a better transverse momentum scale and resolution than the calorimeters for individual soft hadrons.
On the other hand, calorimeter-based isolation includes neutral particles and particles below the ID track \pT threshold,
which are otherwise ignored by track isolation.
However, track and calorimeter isolation measure hadronic activity in a redundant manner,
as charged particles are measured by both the calorimeters and the ID. The particle-flow algorithm allows removal of overlapping contributions from the track-based and calorimeter-based isolation,
decreasing the correlation between the two variables.
 
The particle-flow-based isolation variable is defined as the sum of track-based isolation, chosen in the configuration with \ptvarcone[30] for $\pT^{\mu} < 50$~\GeV\ and \ptcone[20] for $\pT^{\mu} > 50$~\GeV, and the transverse energy of neutral particle-flow objects in a cone of size $\Delta R$ = 0.2 around the muon, labelled as \pflowiso[20]. The latter is corrected for the contribution
from the energy deposit of the muon itself and for pile-up effects, and is assigned a weighting factor $w=0.4$, optimised to maximise the rejection of heavy-flavour hadron decays in the desired range of prompt-muon efficiencies. Contributions from pile-up and the underlying event are estimated using the ambient energy-density technique and are corrected for on an event-by-event basis.
As with the calorimeter isolation, this pile-up correction can result in poor resolution.
However, due to the removal of the contribution from charged particles, the average contribution from pile-up to neutral particle-flow isolation is much smaller than the contribution to calorimeter isolation, and as a result, the pile-up dependency of the efficiency for isolation selections based on neutral particle flow is decreased.
 
\begin{figure}[h!]
\centering
\includegraphics[width=0.59\textwidth]{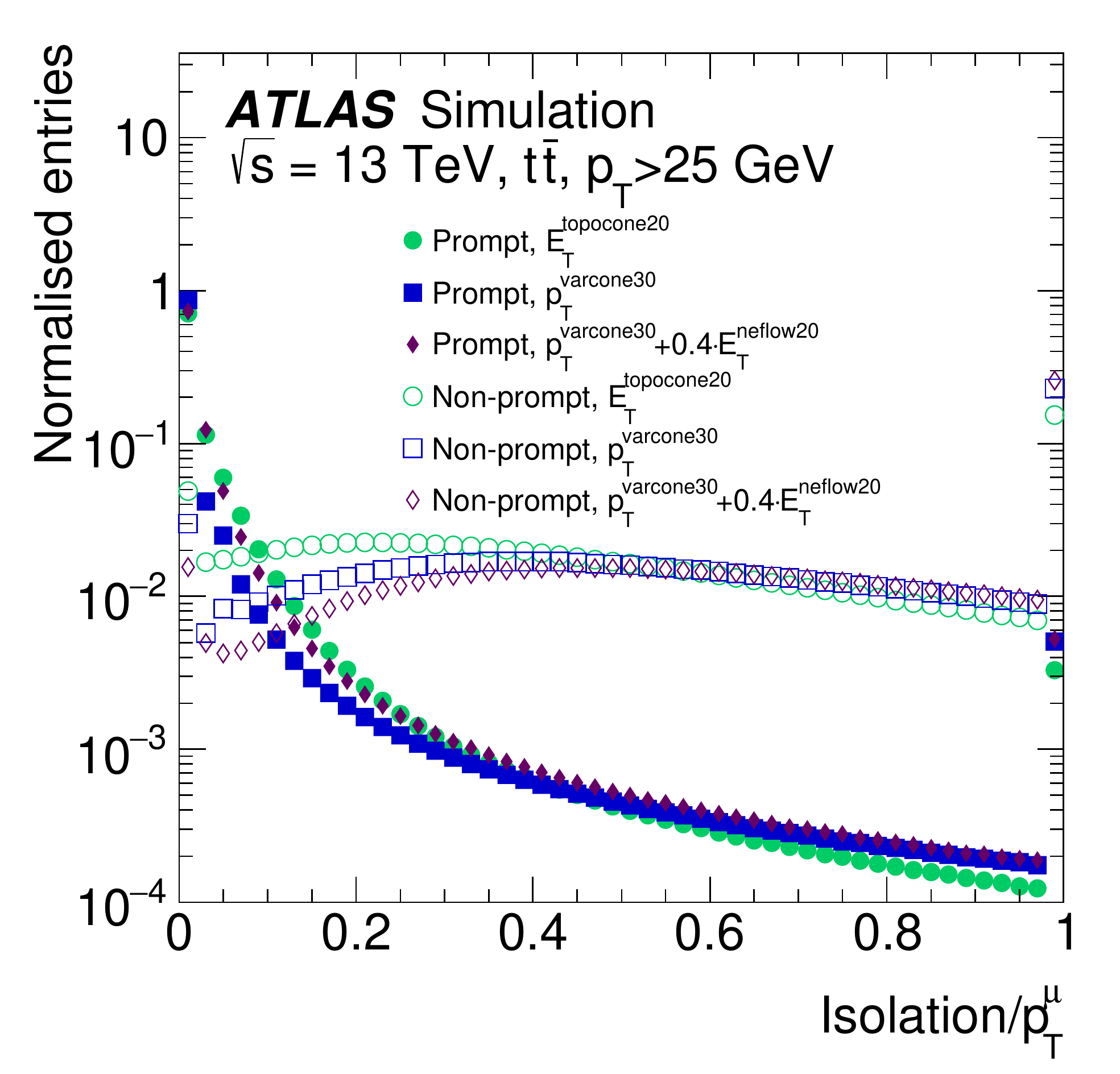}
\caption{Distributions of the isolation variables defined in Section~\ref{subsec:iso}, after dividing their value by the \pT of the muon. Prompt and non-prompt muons are extracted from simulated $t\bar{t}$ events. The distributions are normalised to unit area. The rightmost bin contains all the events exceeding the range of the horizontal axis.}
\label{fig:isoVars}
\end{figure}

Figure~\ref{fig:isoVars} shows the behaviour of the previously defined isolation variables, after dividing their values by the \pT of the muon, for prompt and non-prompt muons extracted from simulated $t\bar{t}$ events.
As expected, the distributions for prompt muons show a sharp peak near zero, while those for non-prompt muons are relatively flat.
 
A multivariate discriminant, \emph{prompt lepton BDT}~\cite{HIGG-2017-02}, is developed for physics analyses that need the highest rejection
of non-prompt muons such as $t\bar{t}H$ searches~\cite{HIGG-2017-02,HIGG-2018-50} and $WWW$ measurements~\cite{STDM-2017-22}.
This discriminant is based on a BDT exploiting eight input variables to maximise the rejection power for non-prompt muons from heavy-flavour hadrons: calorimeter and track isolation,
information about tracks within a cone of size $\Delta R$ = 0.4 around the muon
including the track multiplicity,
and the likelihood of originating within a jet stemming from a $b$-hadron decay, calculated using the DL1mu or RNNIP algorithms~\cite{ATL-PHYS-PUB-2017-013}.
The training is performed using the $t\bar{t}$ MC sample with two separate ranges of muon transverse momentum, 3~\GeV{}~$<\pt<$~10~\GeV{} and $\pt>10$~\GeV{}, to account for the drastic change in the distributions of the input variables.
 
The various isolation WPs are summarised in Table~\ref{tab:iso_wp_def}.
A track-only isolation WP is the most robust with respect to pile-up and suffers the lowest drop in efficiency from nearby objects.
Two loose isolation WPs are defined using track isolation and either calorimeter or neutral particle-flow isolation and
are optimised for cases where high prompt-muon efficiency is prioritised over rejection of non-prompt muons.
Two tight isolation WPs are defined using track isolation and either calorimeter or neutral particle-flow isolation and
are optimised for cases suffering from large backgrounds from non-prompt muons.
Moreover, two isolation WPs are defined using the \emph{prompt lepton BDT}: \textit{PLBDTLoose} and \textit{PLBDTTight}.
In addition to a loose cut on the track isolation, a \pT-dependent BDT threshold selection is applied in each of these to achieve the same
prompt-muon efficiency as the \textit{TightTrackOnly} and \textit{Tight} isolation WPs, respectively.
 
\begin{table}[htbp]
\centering
\caption{Definitions of the muon isolation WPs. The criteria used are listed in the second column, while the requirement on the minimum track \pT is shown in the third column. WPs marked with * exist in two variants: one with the cone $\Delta R$ parameter decreasing with $\pT^{\mu}$ as min(10~\GeV/$\pT^{\mu}$, 0.3), the other remaining constant at $\Delta R = 0.2$ for $\pT^{\mu} > 50$~\GeV.}
\begin{tabular}{|c|c|c|}
\hline
Isolation WP & Definition & Track \pT requirement \\
\hline
\emph{PflowLoose}* & $(\ptvarcone[30] +0.4\cdot \pflowiso[20]) < 0.16 \cdot \pT^{\mu}$  & \multirow{2}{*}{$\pT>500$ \MeV} \\
\emph{PflowTight}* & $(\ptvarcone[30] +0.4\cdot \pflowiso[20]) < 0.045 \cdot \pT^{\mu}$ & \\
\hline
\emph{Loose}* & $\ptvarcone[30] < 0.15 \cdot \pT^{\mu}$, $\topoetcone[20] < 0.3 \cdot \pT^{\mu}$  & \multirow{2}{*}{$\pT>1$ \GeV} \\
\emph{Tight}* & $\ptvarcone[30] < 0.04 \cdot \pT^{\mu}$, $\topoetcone[20] < 0.15 \cdot \pT^{\mu}$ & \\
\hline
\emph{HighPtTrackOnly} & $\ptcone[20] < 1.25$ \GeV & \multirow{2}{*}{$\pT>1$ \GeV} \\
\emph{TightTrackOnly}* & $\ptvarcone[30] < 0.06 \cdot \pT^{\mu}$ &  \\
\hline
\multirow{2}{*}{\emph{PLBDTLoose (PLBDTTight)}} & $\ptvarcone[30] < \max(1.8~\GeV,0.15 \cdot \pT^{\mu})$               & \multirow{2}{*}{$\pT>1$ \GeV} \\
& BDT cut to mimic \textit{TightTrackOnly} (\textit{Tight}) efficiency & \\
\hline
\end{tabular}
\label{tab:iso_wp_def}
\end{table}

The efficiencies for prompt muons and muons from bottom and charm semileptonic decays in $\ttbar$ MC events are summarised in Table~\ref{tab:MCrates_iso}, for tracks satisfying the \medium\ identification and the vertex association selections. For these tracks, the suppression factor, defined as the inverse of the efficiency, for muons from bottom and charm semileptonic decays in $\ttbar$ simulation ranges from 8 (12) at very low $\pt$ to 20 (100) for $\pt>25$~\GeV{} for the \emph{PflowLoose} (\emph{PflowTight}) criteria. The highest suppression factor achieved is 250, and is obtained with the \emph{PLBDTTight} criteria around $\pt=30$~\GeV{}.
 
\begin{table}[h]
\centering
\caption{Isolation efficiencies for prompt muons, $\epsilon_\mu$, and muons from bottom and charm semileptonic decays, $\epsilon_\mathrm{HF}$,
for the different isolation working points, evaluated in a $\ttbar$ MC sample
in different $\pt$ regions for tracks satisfying the \medium\ identification and the vertex association criteria. The isolation working points considered correspond to the variants with the cone size remaining constant at $\Delta R$ = 0.2 for $\pT^{\mu} > 50$~\GeV. The statistical uncertainties are at least one order of magnitude smaller than the last digit reported.}
\begin{tabular}{|c||c|c||c|c||c|c||c|c|}
\hline
\hline
& \multicolumn{2}{c||}{$3<\pt~[\GeV]<5$} & \multicolumn{2}{c||}{$5<\pt~[\GeV]<20$} & \multicolumn{2}{c||}{$20<\pt~[\GeV]<100$} & \multicolumn{2}{c|}{$\pt>100~\GeV$} \\ \hline
Isolation WP & $\epsilon_\mu$ [\%] & $\epsilon_{\text{HF}}$ [\%] & $\epsilon_\mu$ [\%] & $\epsilon_{\text{HF}}$ [\%] & $\epsilon_\mu$ [\%] & $\epsilon_{\text{HF}}$ [\%] & $\epsilon_\mu$ [\%] & $\epsilon_{\text{HF}}$ [\%] \\ \hline
\emph{Loose} & 63 & 14.3 & 86 & 7.2 & 97 & 6.1  & 99 & 12.7~~ \\ \hline
\emph{Tight} & 53 & 11.9 & 70 & 4.2 & 89 & 1.0  & 98 & 1.6 \\ \hline
\emph{PflowLoose} & 62 & 12.9 & 86 & 6.8 & 97 & 5.0  & 99 & 9.1 \\ \hline
\emph{PflowTight} & 45 & ~~8.5 & 63 & 3.1 & 87 & 0.9  & 97 & 0.8 \\ \hline
\emph{HighPtTrackOnly} & 92 & 35.9 & 92 & 17.2~~ & 92 & 4.5  & 92 & 0.6 \\ \hline
\emph{TightTrackOnly} & 80 & 19.9 & 81 & 7.0 & 94 & 3.2  & 99 & 3.3 \\ \hline
\emph{PLBDTLoose} & 81 & 17.4 & 83 & 5.1 & 93 & 1.3  & 98 & 1.7 \\ \hline
\emph{PLBDTTight} & 57 & ~~9.6 & 69 & 2.7 & 87 & 0.5  & 98 & 1.7 \\ \hline
\hline
\end{tabular}
\label{tab:MCrates_iso}
\end{table}
 
\section{Methodology for efficiency measurements}
\label{sec:efficiency}
 
Two different methods are used to measure the reconstruction, identification, isolation and vertex association efficiencies with high precision.
 
In the $|\eta|<2.5$ region, corresponding to the ID acceptance, two independent detectors are available and the tag-and-probe method detailed in Section~\ref{subsec:tagandprobe} is used. Section~\ref{subsec:ztagandprobe} describes the measurements of the reconstruction and identification efficiencies for muons with $\pt$ greater than $15$~\GeV{}, and of the isolation and vertex association efficiencies down to $\pt=3$~\GeV{}, with an almost pure sample of $Z\to\mu\mu$ events. $J/\psi\to\mu\mu$ events, selected as detailed in Section~\ref{subsec:jpsitagandprobe}, are deployed to further extend the reconstruction and identification efficiency measurements down to $\pt=3$~\GeV{}.
 
In the $2.5<|\eta|<2.7$ region, muons are reconstructed only as ME of SiF muons. The level of agreement between collision data and detector simulation is measured via the method summarised in Section~\ref{subsec:forward}.
 
\subsection{The tag-and-probe method in the $|\eta|<2.5$ region}
\label{subsec:tagandprobe}
 
The tag-and-probe method is based on the selection of a sample containing dimuon pairs, for example from $Z\to\mu\mu$ decays, via a set of requirements on the event topology used to reduce the background contamination. One leg of the decay, the \textit{tag}, is required to satisfy stringent identification criteria and to have triggered the online event selection. The second muon candidate in the pair, the \textit{probe}, is used to test the efficiency of a certain reconstruction algorithm or of certain selection criteria. Probes are usually required to be reconstructed with a detector subsystem independent of the one under study.
 
Several types of probes are used to measure the various efficiencies:
\begin{itemize}
\item ID probes are ID tracks used to measure the reconstruction efficiency in the MS, or of specific identification algorithms.
\item MS probes are ME tracks used to test the efficiency of the ID reconstruction.
\item CT probes are ID tracks also satisfying the calo-tagging reconstruction algorithm described in Section~\ref{sec:reconstruction}. In the same way as the ID probes, they are used to measure the reconstruction efficiency in the MS, or of specific muon identification algorithms.
\item ST probes are ID tracks also satisfying the segment-tagging reconstruction algorithm described in Section~\ref{sec:reconstruction}. In the same way as the ID probes, they are used to measure the reconstruction efficiency in the MS, or of specific muon identification algorithms.
\item \emph{Two-track} probes are  MS tracks required to be within $\Delta R=0.05$ of an ID track. They are used to measure the combined reconstruction efficiency of a muon candidate with ID and MS tracks, or the efficiency of specific identification criteria.
\item \emph{Loose} probes are muon candidates satisfying the \emph{Loose} identification requirements. They are used to measure the isolation and vertex association efficiencies.
\end{itemize}
The efficiency of a certain algorithm is measured using a matching requirement of $\Delta R < 0.05$ between the given probe and any muon candidate reconstructed and identified with the algorithm of interest. The efficiency is then computed as the number of probes $P$ that are successfully matched to a muon reconstructed and identified according to the $X$ criterion, $N_{P}^{X}$, divided by the total number of selected probes $N_{P}^\mathrm{All}$:
\begin{equation}\nonumber
\epsilon\left(X|P\right) = \frac{N_{P}^{X}} {N_{P}^\mathrm{All}}.
\end{equation}
Probes are counted in data events after the subtraction of the backgrounds, using different techniques in the $J/\psi\to\mu\mu$ and in the $Z\to\mu\mu$ data sets. In simulation, to eliminate any background contamination, both the tag and the probe muons are required to be a prompt muon at generator level. 
 
The level of agreement between the efficiency measured in data for a given algorithm $X$, $\epsilon^{\mathrm{Data}}\left(X\right)$, and the corresponding efficiency in simulation, $\epsilon^{\mathrm{MC}}\left(X\right)$, is assessed via the ratio of these two numbers, called the efficiency scale factor (SF):
\begin{equation}\nonumber
\mathrm{SF} = \frac{\epsilon^{\mathrm{Data}}\left(X\right)}{\epsilon^{\mathrm{MC}}\left(X\right)}.
\end{equation}
In the ratio, possible biases introduced by the measurement method which appear both in data and MC simulation cancel out. The SF quantifies the deviation of the simulation from the real detector behaviour, and is therefore used in physics analysis to correct the simulation.
 
\subsection{Efficiency measurements with $Z\to\mu\mu$ decays}
\label{subsec:ztagandprobe}
 
To select $Z\to\mu\mu$ decays, the invariant mass $m_{\mathrm{tag-probe}}$ of the tag-and-probe pair is required to be between 61 and 121~\GeV. The tag muon is required to satisfy the \textit{Medium} identification criteria, $\pt>27$~\GeV, $|\eta|<2.5$, and the single-muon trigger requirements described in Section~\ref{sec:datamc}. In order to suppress the contamination from misidentified candidates from jet activity, the tag must fulfil the \textit{Tight} isolation criteria. Furthermore, the vertex association criteria ensure a maximal purity of tags originating from the hard-scattering proton--proton collision. The probes have to satisfy $\pt>3$~\GeV\ and $|\eta|<2.5$. Additional requirements that are specific to the different measurements described in Sections~\ref{subsubsec:ztprecoeff},~\ref{subsubsec:ztpttvaeff}, and~\ref{subsubsec:ztpisoeff} are also employed.
 
\subsubsection{Reconstruction and identification efficiencies}
\label{subsubsec:ztprecoeff}
 
To minimise the systematic uncertainties, a maximal purity in the selection of $Z\to\mu\mu$ decays is mandatory, and additional criteria are applied to the probe: $\pt>10$~\GeV\ is required, and the impact parameters of the probe track must satisfy $\left|d_{0}/\sigma\left(d_{0}\right)\right|<3$ and $\left|z_{0}\right|<10$~mm. Moreover, the probe must carry opposite charge relative to the tag, and fulfil an isolation selection which consists of stricter calorimeter-based and looser track-based isolation criteria than the \emph{Tight} isolation WP.
 
After tightening the tag-and-probe selection, the purity of $Z\rightarrow\mu\mu$ decays in the probe sample is about 99.9\%. Diboson production involving $Z\rightarrow\mu\mu$ decays and contributions from \mbox{$Z\rightarrow\tau\tau$} and $\ttbar$ are considered as signal in the efficiency calculation, and account for about 0.06\% of the selected sample. The contributions arising from processes such as \mbox{$W(\rightarrow \mu\nu)+$\,jets} and multi-jet events, where the probes stem from pion, kaon or heavy-flavour decays, account for the remaining fraction and amount to less than 0.05\%.
 
The method described in Refs.~\cite{PERF-2014-05} and~\cite{PERF-2015-10} expressed the efficiency of the reconstruction and identification algorithm $X$ ($X=$ \emph{Loose}, \emph{Medium}, \emph{Tight}, \emph{Low-$\pt$}, \emph{Low-$\pt$-MVA}, \emph{High-$\pt$}) in terms of conditional probabilities:
\begin{equation}\label{eq:reco-eff-IDMS_OLD}
\epsilon\left(X\right) = \epsilon\left(\mathrm{ID}\right) \times \epsilon\left(X|\mathrm{ID}\right) \simeq \epsilon\left(X|\mathrm{CT}\right) \times \epsilon\left(\mathrm{ID}| \mathrm{MS}\right).
\end{equation}
In Eq.~\eqref{eq:reco-eff-IDMS_OLD}, $\epsilon\left(X\right)$ is factorised as the measured tracking efficiency in the ID, $\epsilon\left(\mathrm{ID}\right)$, multiplied by the conditional probability $\epsilon\left(X|\mathrm{ID}\right)$ that a muon track reconstructed in the ID is also reconstructed and identified with the $X$ algorithm. The validity of this procedure is guaranteed by the fact that the track reconstruction in the ID is independent of that in the MS, and of the other details of the muon identification algorithms.
However, $\epsilon\left(\mathrm{ID}\right)$ cannot be measured directly. It is therefore replaced, in Eq.~\eqref{eq:reco-eff-IDMS_OLD}, by $\epsilon\left(\mathrm{ID}| \mathrm{MS}\right)$, the conditional efficiency for a muon reconstructed by the MS to be also reconstructed in the ID. The quantity $\epsilon\left(\mathrm{ID}| \mathrm{MS}\right)$ is computed as the fraction of MS probes matched to an ID track.
To further reduce the background contamination, the $\epsilon\left(X|\mathrm{ID}\right)  \simeq \epsilon\left(X|\mathrm{CT}\right)$ approximation was used, replacing ID probes with the more pure CT probes, and a systematic uncertainty was assigned to cover for the small bias introduced.
 
In order to improve the precision of the measurement, the approach above was revised, and four different contributions to $ \epsilon\left(X\right)$ are now considered explicitly:
\begin{equation}\label{eq:reco-effi-IDMS}
\begin{split}
\epsilon\left(X\right) &\simeq \epsilon\left(X | \mathrm{ID}\wedge\mathrm{MS}\right)\times  \epsilon\left(\mathrm{MS} |\mathrm{ID}\right)  \times \epsilon\left(\mathrm{ID}| \mathrm{MS}\right) + \epsilon\left(X \wedge \neg \mathrm{MS} | \mathrm{ID}\right) \times \epsilon\left(\mathrm{ID}| \mathrm{MS}\right) \\
&\simeq \left[ \epsilon\left(X | \mathrm{ID}\wedge\mathrm{MS}\right)\times  \epsilon\left(\mathrm{MS} |\mathrm{CT}\right) + \epsilon\left(X \wedge \neg \mathrm{MS} | \mathrm{CT}\right)\right] \times \epsilon\left(\mathrm{ID}| \mathrm{MS}\right).
\end{split}
\end{equation}
Each of the terms appearing on the right-hand side of Eq.~\eqref{eq:reco-effi-IDMS} is measured separately. The first term, $\epsilon\left(X | \mathrm{ID}\wedge\mathrm{MS}\right)$, is the component of the $X$ reconstruction and identification efficiency conditional on the combined reconstruction in the ID and MS, and it is thus measured via \emph{two-track} probes. Conversely, $\epsilon\left(X \wedge \neg \mathrm{MS} | \mathrm{CT}\right)$ describes the contribution to the $X$ efficiency from muons reconstructed without a full track in the MS, as measured with CT probes. The MS reconstruction efficiency $\epsilon\left(\mathrm{MS} |\mathrm{CT}\right)$ is also measured with CT probes, while the ID reconstruction efficiency $\epsilon\left(\mathrm{ID} |\mathrm{MS}\right)$ is measured with MS probes. The advantage of this method is that the bias in Eq.~\eqref{eq:reco-eff-IDMS_OLD} stemming from $\epsilon\left(X|\mathrm{CT}\right)$, due to the neglected correlation between the probability of reconstructing a CT muon and that of fulfilling $X$ criterion, is substantially mitigated. The residual bias in Eq.~\eqref{eq:reco-effi-IDMS} was tested using generator-level information in detector simulation, and found to be about a factor five smaller than that achieved in Ref.~\cite{PERF-2015-10}, which in the $|\eta|>0.1$ region ranged between 0.1\% and 0.7\%.

\begin{figure}
\centering
\includegraphics[width=0.49\textwidth]{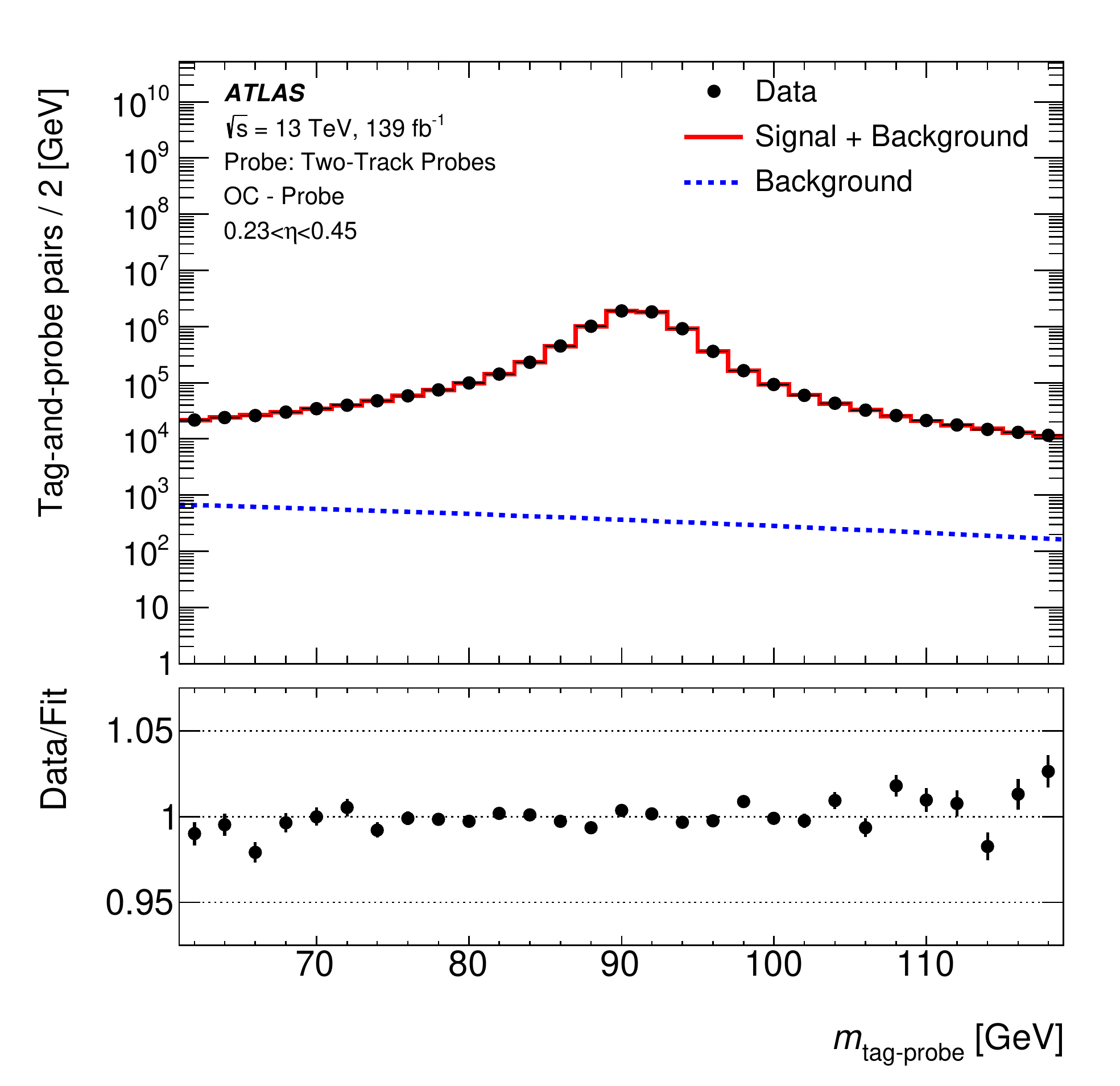}
\includegraphics[width=0.49\textwidth]{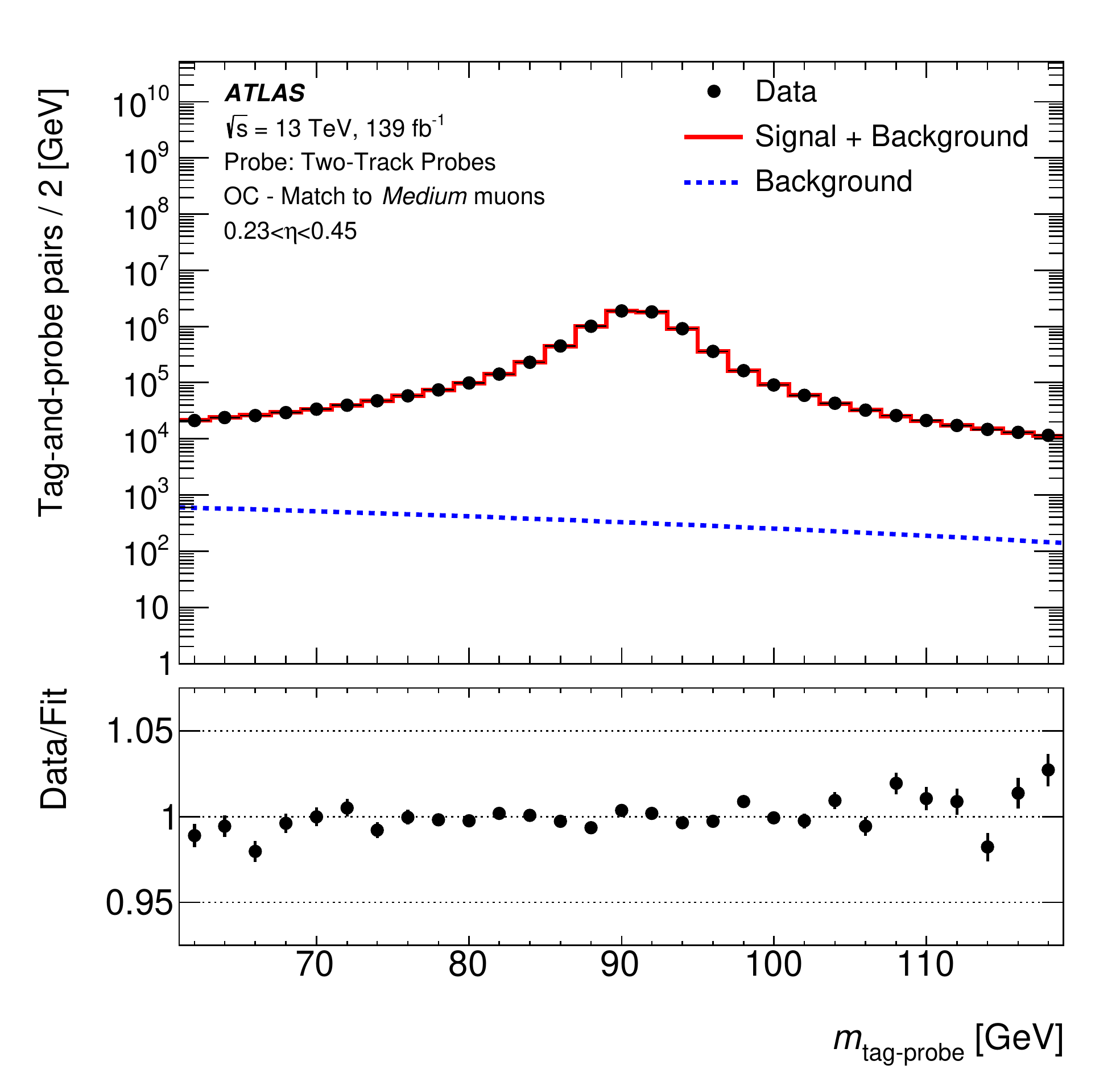}
\caption{Fit to the $m_{\mathrm{tag-probe}}$ distribution for the selected probes (left), and the probes matched to the \medium\ selection (right), for \emph{two-track} probes with $0.23<\eta<0.45$ and opposite-charge (OC) tag-and-probe pairs. The blue line indicates the fitted non-prompt muon background component, while the red line represents the total fitted signal plus the background contribution. The panel beneath shows the data to fit-function ratio for each bin. The error bars indicate the statistical uncertainty.}
\label{fig:exampleFitZ}
\end{figure}

The signal and background contributions in data are extracted via a fit to the $m_{\mathrm{tag-probe}}$ spectrum in the 61--121~\GeV{} range, separately for the samples of all selected probes, and for the samples of matched probes. An example is shown in Figure~\ref{fig:exampleFitZ}. All SM processes producing an opposite-charge pair of prompt muons are treated as signal, and modelled with a $m_{\mathrm{tag-probe}}$ template obtained using MC simulation. The background contribution stemming from all processes involving non-prompt muons is modelled using the following functional form:
\begin{equation}\label{eq:mcp-reco-fit-function}
f(m_{\mathrm{tag-probe}})=\left( 1-\frac{m_{\mathrm{tag-probe}}}{\Lambda} \right)^{p_1} \cdot \left(\frac{m_{\mathrm{tag-probe}}}{\Lambda}\right)^{p_2},
\end{equation}
where the $\Lambda$ parameter, approximating the energy necessary to produce the dimuon pair~\cite{Aaltonen:2008dn}, is set as 2.5 times the upper boundary of the considered $m_{\mathrm{tag-probe}}$ spectrum. The $p_1$ and $p_2$ parameters are instead obtained via a separate fit using a sample of same-charge tag-and-probe pairs, satisfying all the selection criteria except the isolation requirements.
 
\subsubsection{Vertex association efficiencies}
\label{subsubsec:ztpttvaeff}
Run conditions and ID distortions and misalignments impact the efficiency of the vertex association criteria detailed in Section~\ref{subsec:ttva}. For its measurement, the same strategy used for the reconstruction and identification efficiencies is deployed, with some relevant adjustments.
$J/\psi\rightarrow\mu\mu$ events are not suited for measuring the muon vertex association efficiencies, due to the sizeable displacement of the $J/\psi$ mesons originating for example from the decay of $b$-hadrons. Therefore, $Z\rightarrow\mu\mu$ decays are exploited down to low muon transverse momentum by selecting probes with $\pt>3$~\GeV{} which satisfy the \loose\ identification criteria. Further, no requirements on the longitudinal and transverse impact parameters of the probe can be applied. To reduce the consequent increase in background contamination, the tag-and-probe invariant mass region considered is restricted to 86--96~\GeV{}. The tag muon selection is instead identical to the one used for measuring the reconstruction and identification efficiencies, as well as for the template fit method and the efficiency and SF computation.
 
\subsubsection{Isolation efficiencies}
\label{subsubsec:ztpisoeff}
 
Similarly to the measurement of the vertex association efficiencies, $Z\rightarrow\mu\mu$ decays with $\pt>3$~\GeV{} probes are used to measure the muon isolation efficiencies in the full transverse momentum range of interest. To improve the background rejection at low $\pt$, \emph{Loose} identification criteria and the standard vertex association requirements are additionally applied to the probe. Moreover, the two muons originating from the $Z$ boson decay are required to be separated by $\Delta R_{\mu\mu}>0.3$, to reject events where the tag muon lies inside the isolation cone of the probe.

The fitting procedure used to extract the background contribution in data is identical to that used in the measurement of the reconstruction and identification efficiencies, but an $m_{\mathrm{tag-probe}}$ interval of 81--101~\GeV{} is instead considered. The non-prompt muon background is obtained by fitting same-charge data with the functional form of Eq.~\eqref{eq:mcp-reco-fit-function}. In the efficiency computation, the contribution of $Z\rightarrow\mu\mu$ events is separated from that of all other processes producing pairs of oppositely charged prompt muons, as modelled with MC simulation. Restricting the measurement of the isolation efficiencies to a clean and well-defined process such as $Z\rightarrow\mu\mu$ does not affect the generality of the results obtained, provided that the most relevant kinematic dependencies of the efficiencies are accounted for in the derived SFs, but simplifies the evaluation of the associated uncertainties.
 
To ensure applicability to a wide range of physics processes, the measured efficiencies and SFs are studied as a function of the muon $\pt$, $\eta$, and angular distance \deltarmuj{} from the closest reconstructed jet. Jets are reconstructed from calorimeter topological energy clusters~\cite{PERF-2014-07} in the region $|\eta| < 4.5$ using the anti-$k_t$ algorithm~\cite{Cacciari:2008gp,Fastjet} with radius parameter $R = 0.4$. The jets are required to have $\pt>20$~\GeV{} after being calibrated~\cite{PERF-2016-04} and after subtracion of the expected energy contribution from pile-up according to the jet area~\cite{PERF-2014-03}. In order to suppress jets due to pile-up, jets with $\pt < 120~\GeV$ and $|\eta| < 2.5$ are required to satisfy the \emph{Medium} working point of the jet vertex tagger~\cite{PERF-2014-03}, which uses information from the tracks associated with the jet.
 
A dedicated procedure is used to resolve reconstruction ambiguities between probes and jets. When a jet overlaps with a selected probe within $\deltarmuj < 0.4$, if the ratio of the probe's $\pt$ to the jet's $\pt$ (\textit{probe-to-jet} $\pt$ ratio) is below $0.5$, or if the ratio of the probe's $\pt$ to the magnitude of the summed vector $\pt$ of all tracks associated with the jet (\textit{probe-to-jet-tracks} $\pt$ ratio) is below $0.7$, the probe is rejected to suppress bottom and charm hadron decays.

\subsubsection{Systematic uncertainties}
\label{subsubsec:ztp_syst}
 
The main contributions to the systematic uncertainty in the measurement of the reconstruction and identification efficiency SFs using $Z\rightarrow\mu\mu$ events are discussed below:
 
\begin{itemize}

\item \emph{T\&P method}: possible biases in the tag-and-probe method, such as biases due to different kinematic distributions between reconstructed probes and generated muons or correlations between ID and MS efficiencies, are estimated in simulation by comparing the measured efficiency with the fraction of generator-level muons that are successfully reconstructed. This type of bias is expected to affect data and simulation in a similar way, and therefore to approximately cancel out in the SF computation. Half of the observed difference is nevertheless assigned as the SF uncertainty, in order to conservatively account for possible imperfections of the simulation. The use of \emph{two-track} probes reduces this uncertainty to below approximately $0.1\%$, with a progressive decrease as $\pt$ rises.
 
\item \emph{Probe matching}: the default $\Delta R$-based matching procedure is varied in order to assess an uncertainty in how much a given probe type contributed to a certain type of reconstructed muon candidate.
This is done by comparing the nominal fraction of matched probes with the fraction of probe tracks for which muon candidate reconstruction is successful.
 
\item \emph{Template shape}: the uncertainty in the shape of the template modelling the non-prompt muon background is evaluated by simultaneously varying the $p_{1}$ and $p_{2}$ parameters in Eq.~\eqref{eq:mcp-reco-fit-function} by their fit uncertainties. The consequent deviation of the SFs from their nominal value is taken as the systematic uncertainty. 
 
\item \emph{$\Lambda$-SC}: the numerical value of the $\Lambda$ parameter in Eq.~\eqref{eq:mcp-reco-fit-function} guarantees a well-behaved, smooth function across $m_{\mathrm{tag-probe}}$. Possible effects on the SFs are estimated by varying its value by $\pm 20\%$.
 
\item \emph{Background fit}: to cover effects associated with the fitting procedure used to extract the contribution of the non-prompt muon background, the change in the SFs obtained when varying the fitted non-prompt muon background by its corresponding fit uncertainty is assigned as a systematic uncertainty.
 
\item \emph{Cross-section}: the uncertainty in the cross-sections of the simulated processes impacts the shape of the signal template by altering its composition, especially in the mass sidebands. Therefore, the normalisation of each MC sample is varied by the measured cross-section's uncertainty~\cite{TOPQ-2016-01,STDM-2015-19,STDM-2016-01}.
 
\item \emph{Luminosity}: the invariant-mass template used to model the non-prompt muon background is corrected for the contamination from same-charge prompt muons, estimated from simulation normalised to the integrated luminosity of the data. The procedure is therefore sensitive to uncertainties in the cross-section of the simulated samples, and in the integrated luminosity of the same-charge data sample. The uncertainty in the combined 2015--2018 integrated luminosity is 1.7\%~\cite{ATLAS-CONF-2019-021}, obtained using the LUCID-2 detector~\cite{LUCID2} for the primary luminosity measurements. To evaluate the impact of this systematic effect, the normalisation factor for the whole simulation is varied accordingly. 
 
\end{itemize}
 
\begin{figure}
\centering
\includegraphics[width=0.49\textwidth]{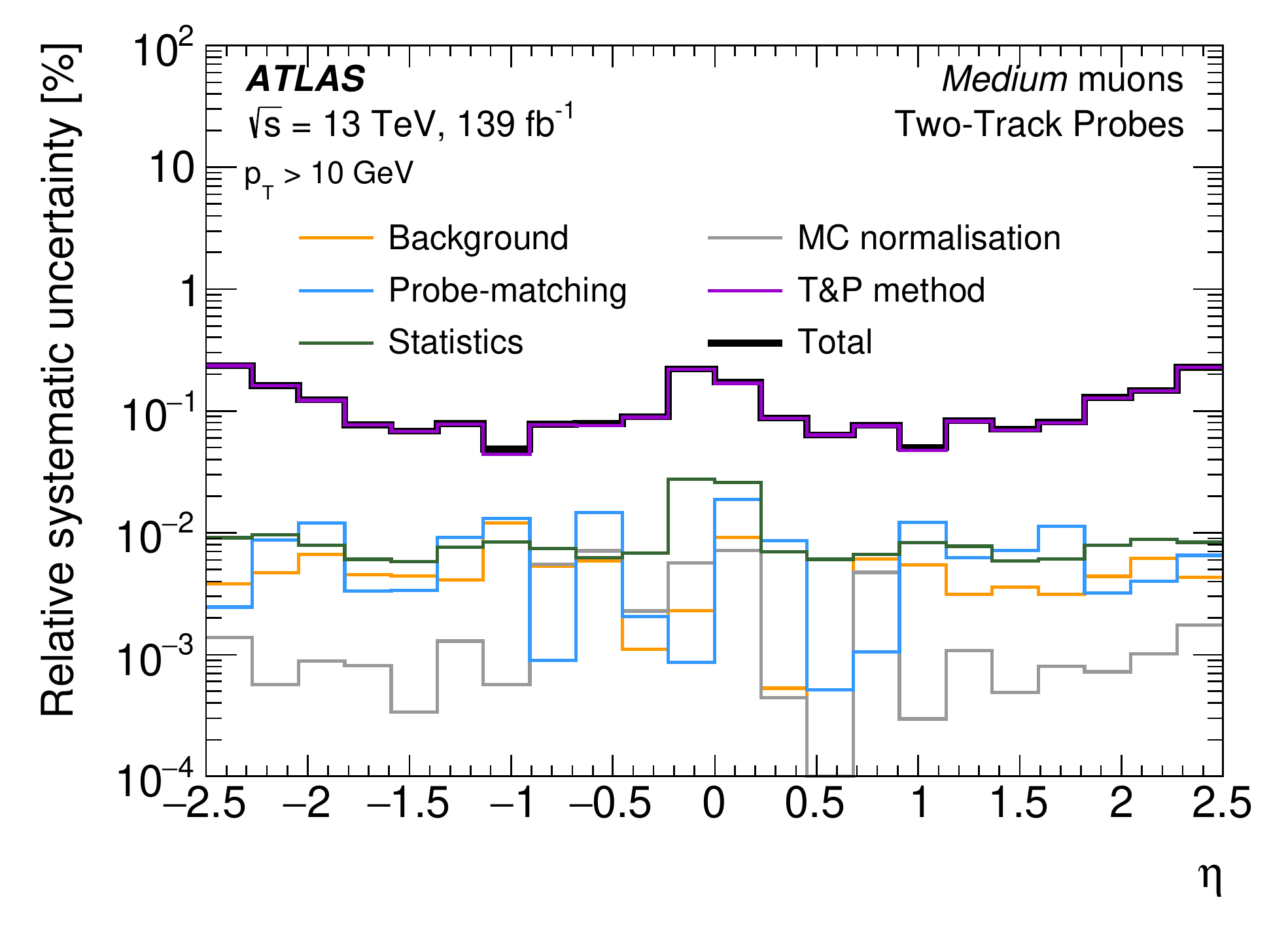}
\includegraphics[width=0.49\textwidth]{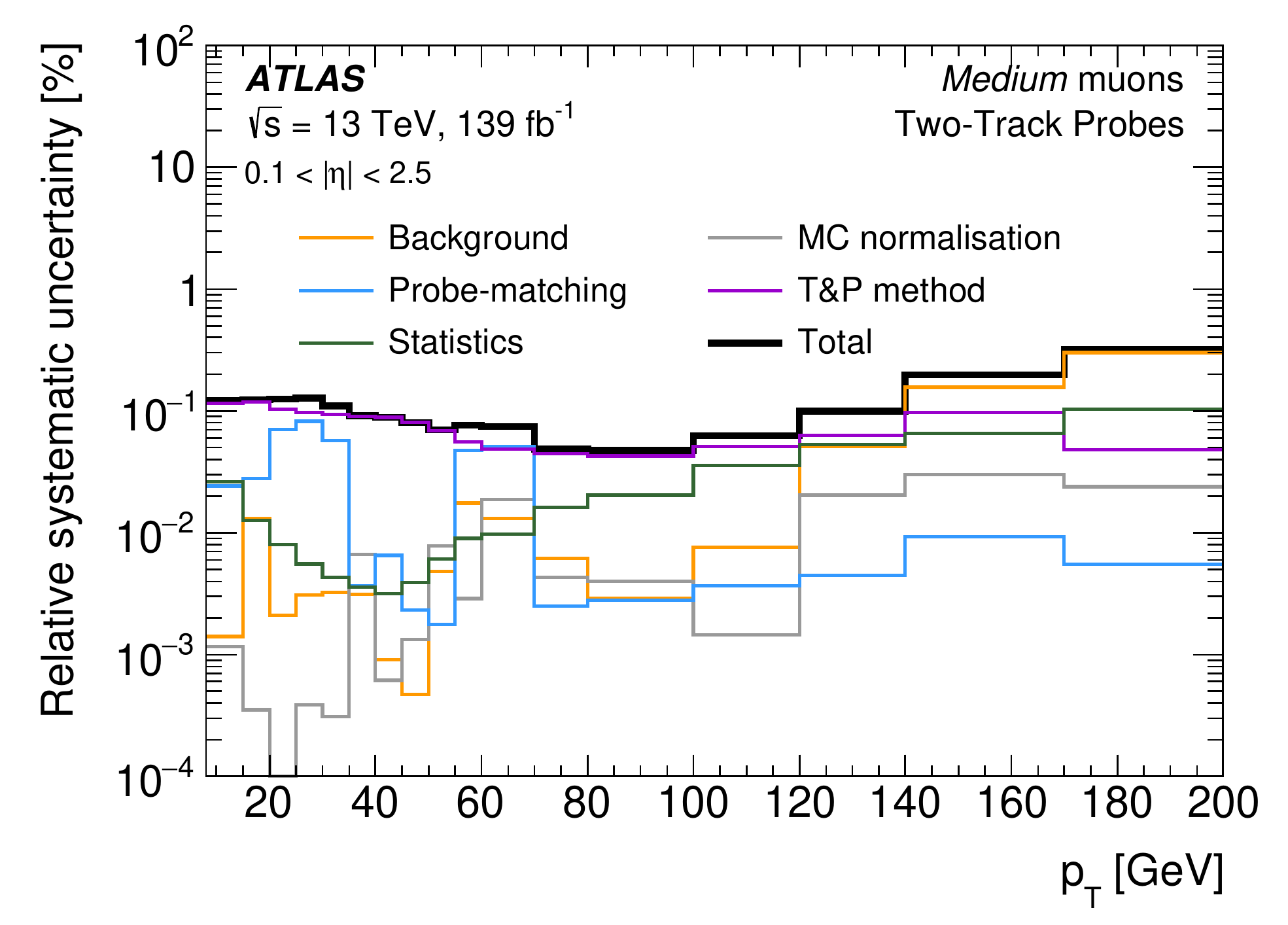}
\caption{Relative contributions to the systematic uncertainty in the efficiency SFs for \medium\ muons measured with $Z\rightarrow\mu\mu$ decays, as a function of $\eta$ (left) and $\pt$ (right) for muons with $\pt>10$~\GeV{}, and integrated over the other kinematic observables. The uncertainty depicted as \emph{Background} is the sum in quadrature of the \emph{Template shape}, \emph{$\Lambda$-SC}, and \emph{Background fit} uncertainties, whereas the \emph{MC normalisation} comprises the \emph{Cross-section} and \emph{Luminosity} uncertainties. The total uncertainty is the sum in quadrature of the individual contributions.}
\label{fig:MediumSyst}
\end{figure}

Figure~\ref{fig:MediumSyst} shows the relative contributions to the systematic uncertainty in the reconstruction and identification efficiency SFs of \medium\ muons, as a function of $\eta$ and $\pt$. For muons with $\pt<100$~\GeV{}, the largest relative contribution stems from the \emph{T\&P method} uncertainty, while for muons with $\pt>150$~\GeV{} the uncertainties in the non-prompt muon background estimates become dominant.
 
The impact of the muon momentum resolution and scale uncertainties in simulation~\cite{PERF-2015-10}, which can lead to small alterations of the shape of the signal template, was found to be negligible. In the momentum range considered, the measured SFs do not show a significant dependence on $\pt$. For muon tracks with $\pt>500$~\GeV{}, the muon reconstruction efficiency is expected to progressively decrease as $\pt$ increases, due to a higher likelihood of large energy losses in the calorimeters that can impair the combined muon reconstruction. To account for possible imperfections in the simulation of such extreme energy losses, the full decrease of reconstruction efficiency as a function of $\pt$ in simulation is conservatively assigned as a systematic uncertainty.
 
All the uncertainties detailed above apply also to the measurement of the vertex association and isolation efficiency SFs, with the exception of the \emph{T\&P method} and \emph{Probe matching} uncertainties. Conversely, other sources of systematic uncertainty are considered in the measurement of the isolation efficiencies and SFs:
 
\begin{itemize}
 
\item \emph{Probe PID}: the choice of probe identification working point influences the background contamination level and the signal yield, and therefore is a source of systematic uncertainty. This uncertainty is computed as the difference between the scale factors obtained using \emph{Loose} versus \emph{LowPt} probes, for $\pT<15$~\GeV, or \emph{Loose} versus \emph{Tight} probes, for $\pT>15$~\GeV.
 
\item \emph{Mass window}: a variation of the $m_{\mathrm{tag-probe}}$ range considered in the template fit could lead to different SF values. An uncertainty is assigned as the largest difference observed after changing the nominal $m_{\mathrm{tag-probe}}$ interval of 81--101~\GeV{} for the $\eZ$ computation to 86--96~\GeV{} and 71--111~\GeV{}.
 
\item \emph{Jet modelling}: to account for possible differences in the modelling of isolation criteria between different MC generators, a dedicated uncertainty is computed as the efficiency difference in $Z\rightarrow\mu\mu$ events simulated with \textsc{Sherpa 2.2.1} and \textsc{Powheg+Pythia8}.
 
\item \emph{\deltarmuj{}}: as the isolation efficiencies and scale factors are found to depend on the angular distance between the probe and the closest jet, the procedure used to resolve muon--jet reconstruction ambiguities is a source of systematic uncertainty. To account for it, the criteria for the \emph{probe-to-jet} and \emph{probe-to-jet-tracks} $\pt$ ratios are independently dropped, and the largest change in the SFs is taken as the uncertainty.
 
\end{itemize}
 
\begin{figure}
\centering
\includegraphics[width=0.49\textwidth]{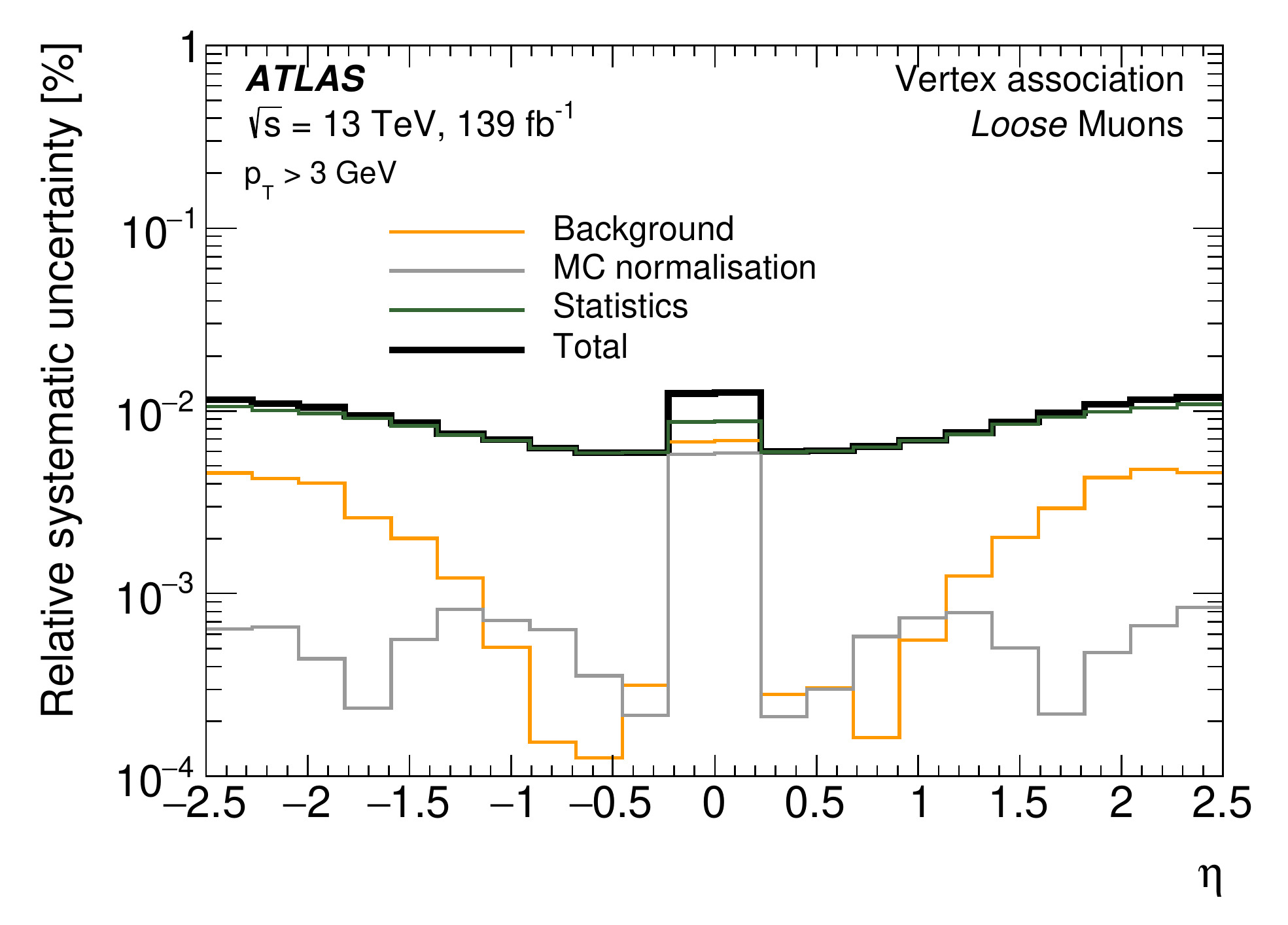}
\includegraphics[width=0.49\textwidth]{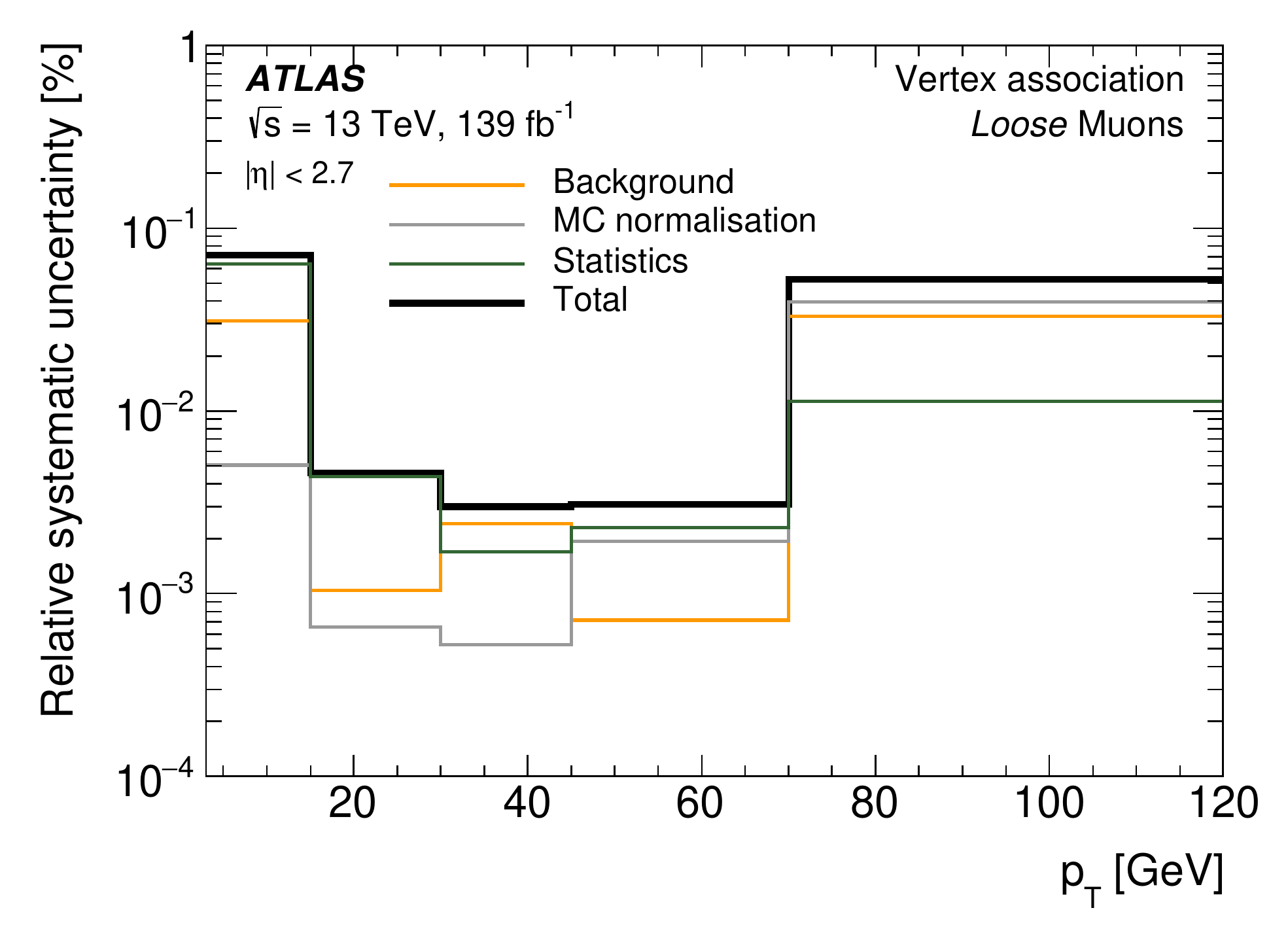}
\caption{Relative contributions to the systematic uncertainty in the efficiency SFs for the vertex association criteria measured with $Z\rightarrow\mu\mu$ decays, as a function of $\eta$ (left) and $\pt$ (right) for muons with $\pt>3$~\GeV{}, and integrated over the other kinematic observables. The uncertainty depicted as \emph{Background} is the sum in quadrature of the \emph{Template shape}, \emph{$\Lambda$-SC}, and \emph{Background fit} uncertainties, whereas the \emph{MC normalisation} comprises the \emph{Cross-section} and \emph{Luminosity} uncertainties. The total uncertainty is the sum in quadrature of the individual contributions.}
\label{fig:TTVASyst}
\end{figure}

\begin{figure}
\centering
\includegraphics[width=0.49\textwidth]{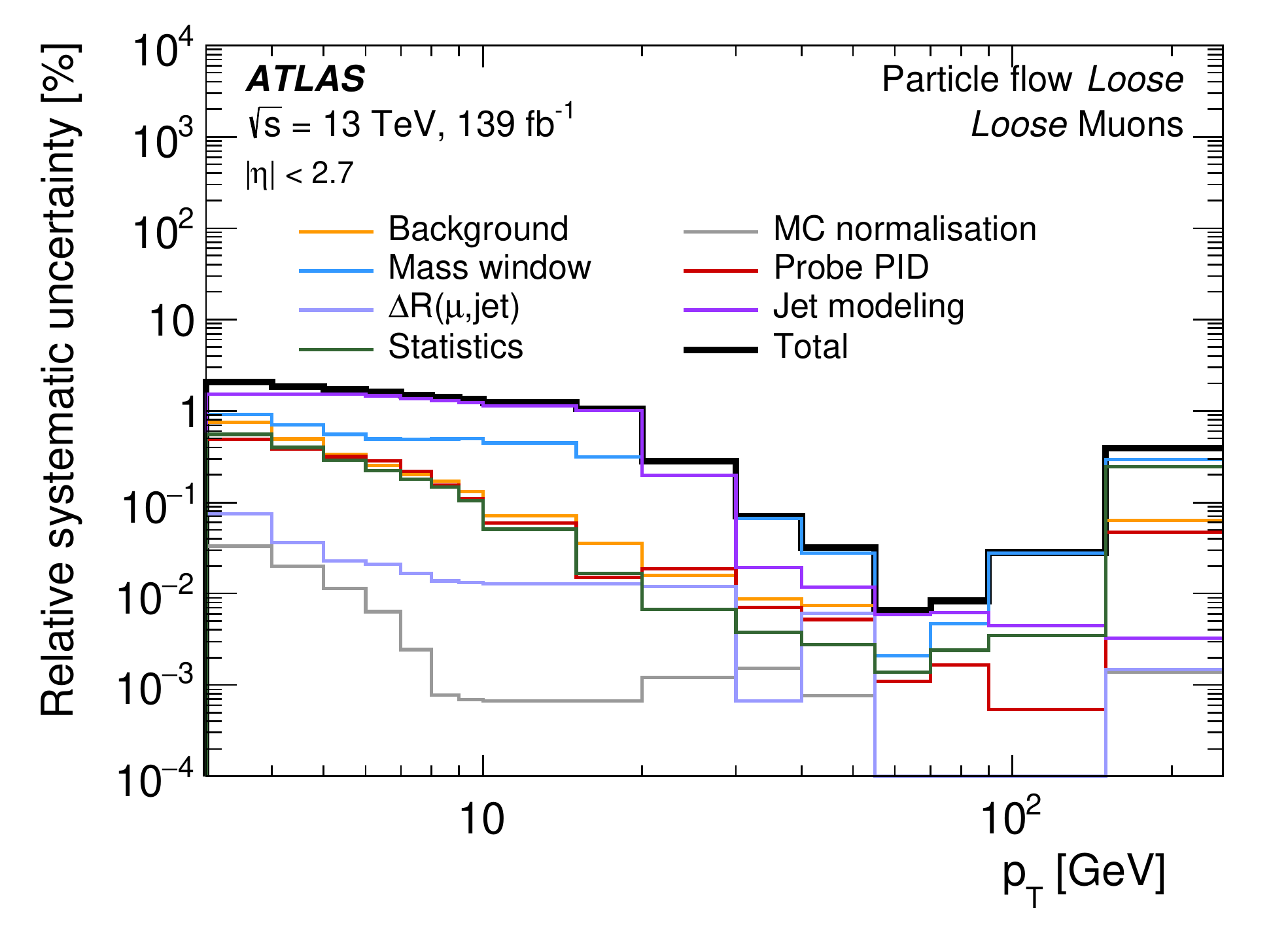}
\includegraphics[width=0.49\textwidth]{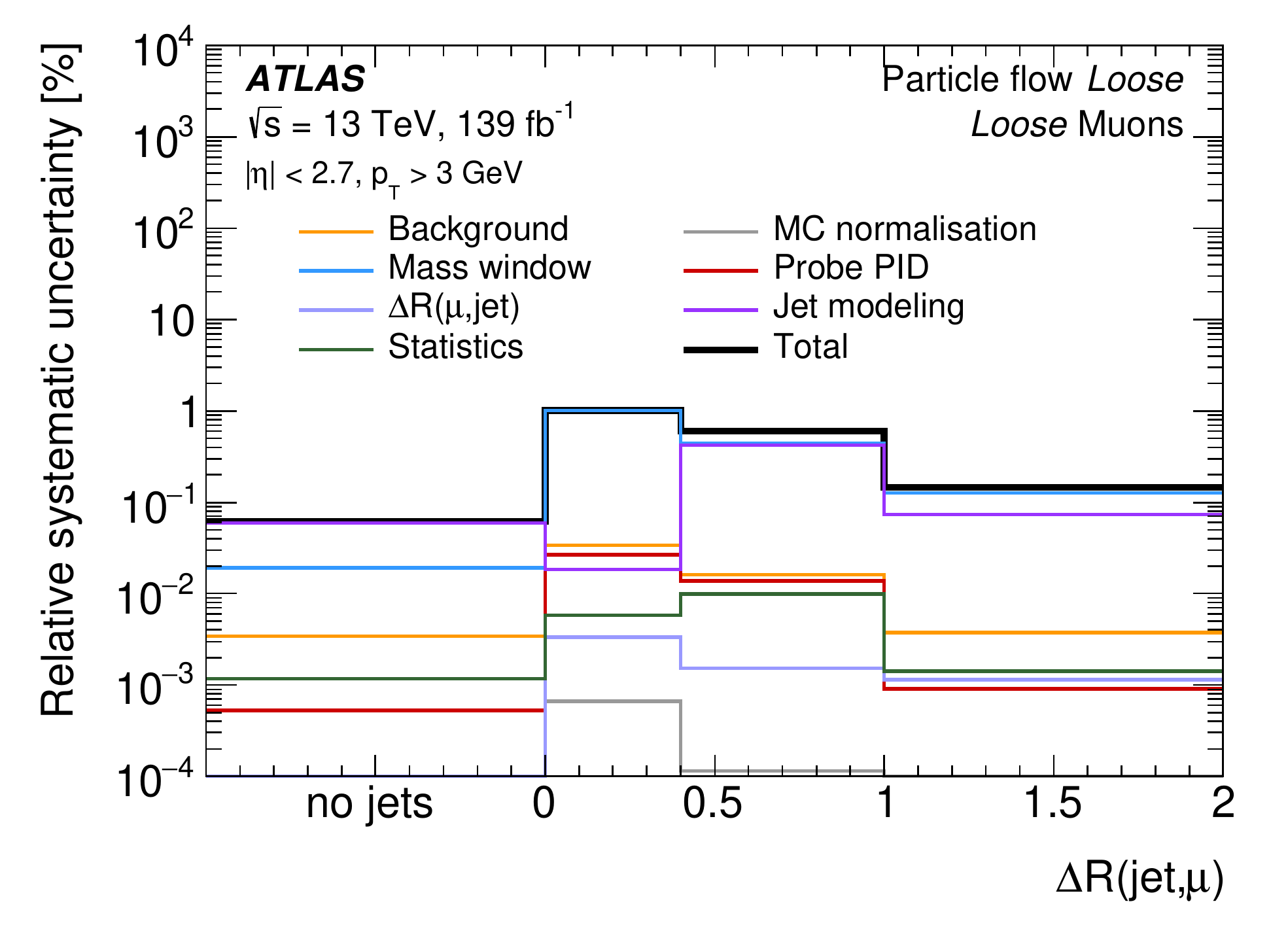}
\caption{Relative contributions to the systematic uncertainty in the efficiency SFs for the \emph{PflowLoose} isolation criteria, measured with $Z\rightarrow\mu\mu$ decays, as a function of $\pt$ (left) and \deltarmuj{} (right) for muons with $\pt>3$~\GeV{}, and integrated over the other kinematic observables. The uncertainty depicted as \emph{Background} is the sum in quadrature of the \emph{Template shape}, \emph{$\Lambda$-SC}, and \emph{Background fit} uncertainties, whereas the \emph{MC normalisation} comprises the \emph{Cross-section} and \emph{Luminosity} uncertainties. The total uncertainty is the sum in quadrature of the individual contributions.}
\label{fig:PflowSyst}
\end{figure}

Figures~\ref{fig:TTVASyst} and~\ref{fig:PflowSyst} summarise the uncertainties in the measurement of the vertex association and \emph{Pflow Loose} isolation efficiency SFs, respectively. In the former case, the precision of the measurement is limited by the available statistics. The \emph{Jet modelling} and \emph{Mass window} uncertainties instead show the largest impact on the \emph{Pflow Loose} isolation efficiency SFs.
 
\subsection{Efficiency measurements with $J/\psi\to\mu\mu$ decays}
\label{subsec:jpsitagandprobe}
 
The $J/\psi\to\mu\mu$ events offer a large sample suited to measuring the muon reconstruction and identification efficiencies in the 3--20~\GeV{} transverse momentum range with small statistical uncertainties. To cope with the larger background contamination at very low muon $\pt$, ST probes are deployed. Contrary to CT probes, ST probes are available also for ID tracks with $\pT < 5$~\GeV{}.
 
Tag-and-probe pairs are selected within the invariant mass window of 2.7--3.5~\GeV{}. The tag muon is required to have $\pt>6$~\GeV{}, to satisfy the \emph{Tight} muon identification selection, and to have triggered the read-out of the event. In order to avoid low-momentum curved tracks sharing the same trigger region, tag and probe tracks are extrapolated to the MS trigger plane furthest from the interaction point and the extrapolated positions are required to be $\Delta R >$ 0.2 apart. Finally, events are selected with $|z_0^{T} - z_0^{P}| < 5$~mm, where $z_0^{T}$ ($z_0^{P}$) is the longitudinal impact parameter of the tag (probe) track, to suppress backgrounds. A probe is required to have $\pt>3$~\GeV{}, and is considered successfully reconstructed if a selected muon is found within a cone of size $\Delta R = 0.05$ around the probe track.
 
\subsubsection{Reconstruction and identification efficiencies}
 
With the introduction of ST probes, Eq.~\eqref{eq:reco-eff-IDMS_OLD} needs to be modified as follows:
\begin{equation}\nonumber
\begin{split}
\epsilon\left(X\right) &\simeq \epsilon\left(X|\text{ID}\right) \times \epsilon\left(\text{ID}|\text{MS}\right) = \frac{ \epsilon\left(X|\text{ST}\right) }{ \epsilon\left(\text{ST}|X\right)  } \times \epsilon\left(\text{ST} |\text{ID} \right) \times \epsilon\left(\text{ID}|\text{MS}\right)\\
&\simeq  \epsilon\left(X|\text{ST}\right) \times \epsilon\left(\text{ST} |\text{ID}\lor \text{CT if\ } \pt>5\,\GeV  \right) \times \epsilon\left(\text{ID}|\text{MS}\right).
\end{split}
\end{equation}
This approach allows the measurement of the efficiency of the $X$ requirements given an ST probe, $\epsilon\left(  X | \textrm{ST}  \right)$, to be separated from that of terms common to all selection criteria, such as the ID efficiency $\epsilon\left( \textrm{ID} | \textrm{MS} \right)$ and the ST efficiency $\epsilon\left(\text{ST} |\text{ID}\lor \text{CT if\ } \pt>5\,\GeV  \right)$ given an ID probe ($\pt<5$~\GeV{}) or a CT probe ($\pt>5$~\GeV{}). The $\epsilon\left(\text{ST}|X\right)$ term accounts for the conditional probability of a successful ST reconstruction given a muon fulfilling the $X$ criterion. In the kinematic regime of interest, it is measured to be compatible with one within its uncertainty.
 
\begin{figure}
\centering
\includegraphics[width=0.49\textwidth]{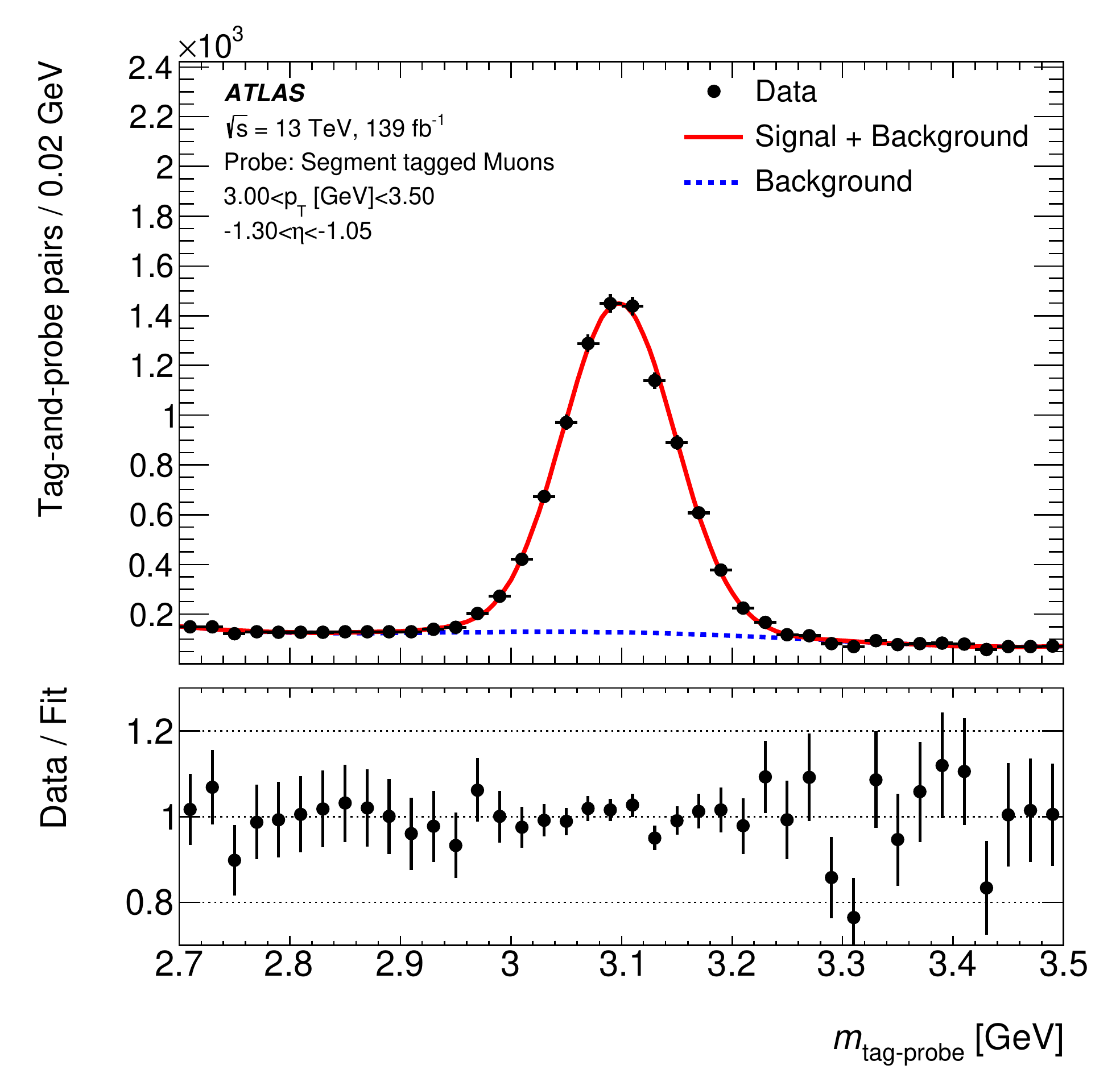}
\includegraphics[width=0.49\textwidth]{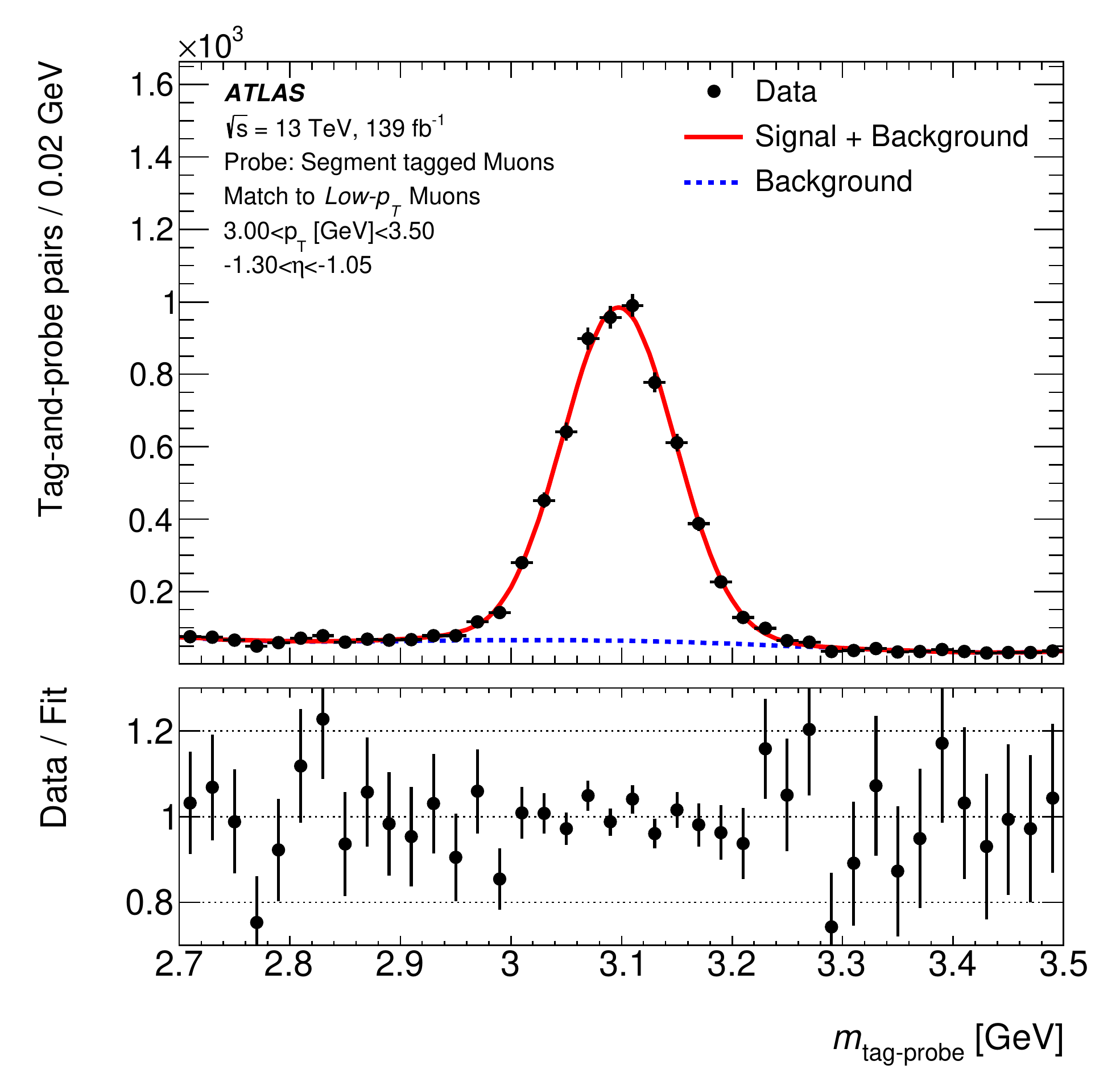}
\caption{Fit to the $m_{\mathrm{tag-probe}}$ distribution for the selected probes (left), and the probes matched to the \lowPt\ selection (right), for ST probes with $-1.30<\eta<-1.05$ and $3.0<\pt<3.5$~\GeV{}. The blue line indicates the fitted non-prompt muon background component, while the red line represents the total fitted signal plus the background contribution. The panel beneath shows the data to fit-function ratio for each bin.}
\label{fig:exampleFitJPsi}
\end{figure}

The muon reconstruction and identification efficiency and the background contamination are measured with a simultaneous maximum-likelihood fit of the tag-and-probe invariant mass in the all-probes and matched-probes samples. An example is shown in Figure~\ref{fig:exampleFitJPsi}. A Crystal Ball~\cite{Oreglia:1980cs} function is used to model the signal. For the background model, an iterative procedure is deployed, where second-, third- and fourth-order polynomial functions are tested and the one resulting in the fit with the smallest $\chi^{2}/N_{\text{d.o.f.}}$ is retained.
 
\subsubsection{Systematic uncertainties}
 
\begin{figure}
\centering
\includegraphics[width=0.49\textwidth]{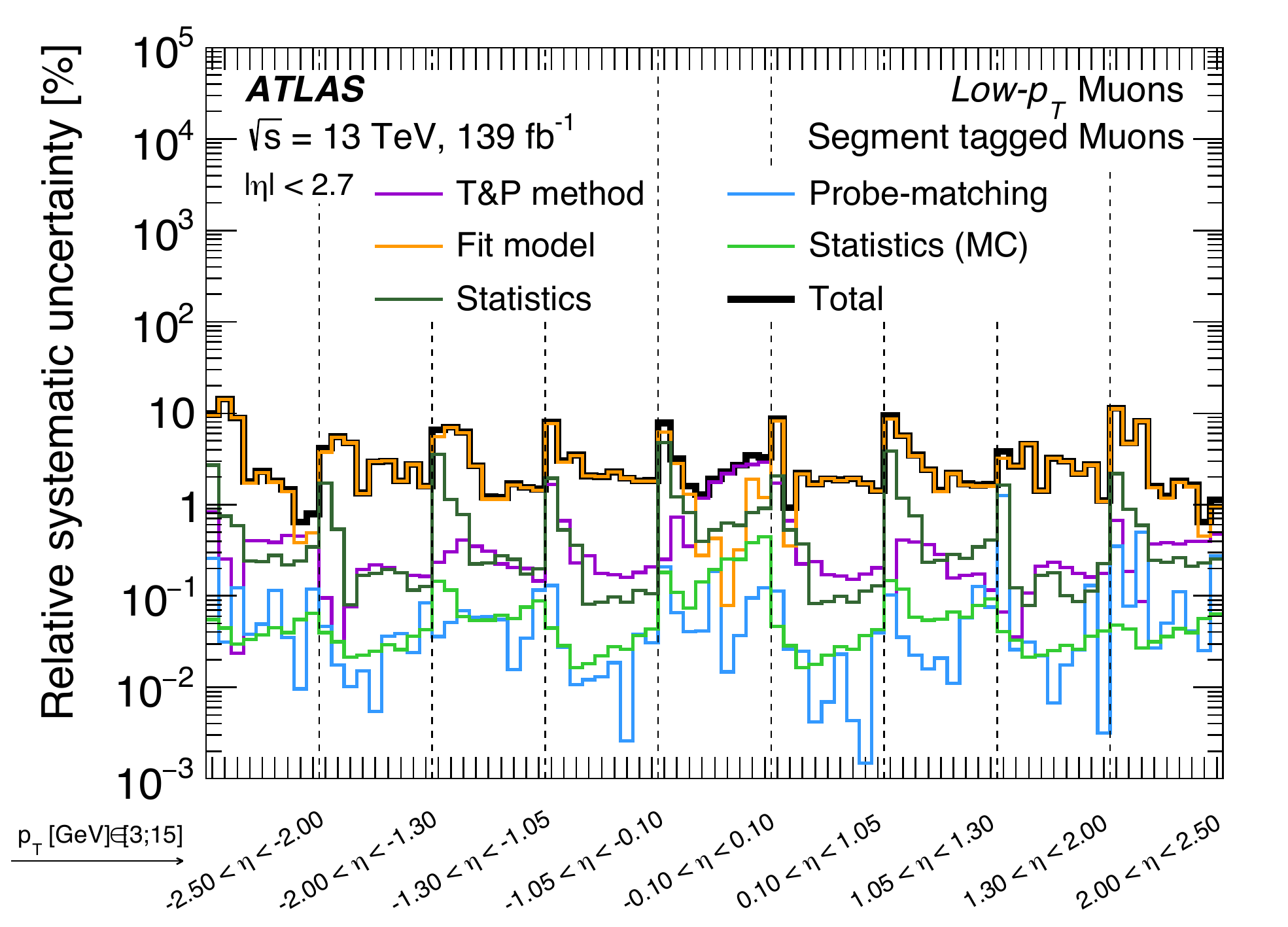}
\includegraphics[width=0.49\textwidth]{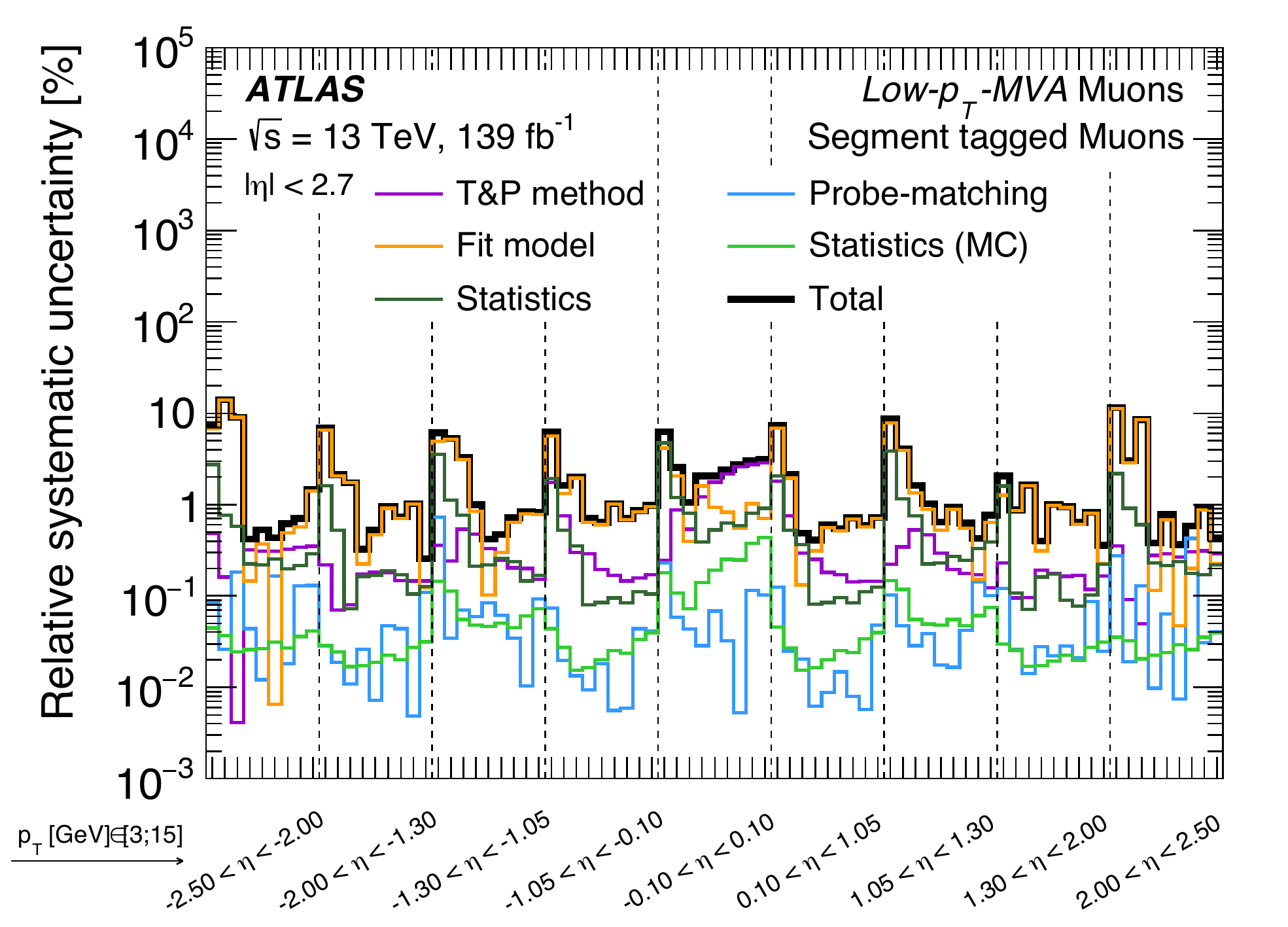}
\caption{Relative contributions to the systematic uncertainty in the efficiency SFs for muons with $3<\pt<15$~\GeV{} fulfilling the cut-based (left) and multivariate (right) \lowPt\ selection criteria, obtained from $J/\psi\to\mu\mu$ data, as a function of $\eta$ and $\pt$, and integrated over the other kinematic observables. The resulting values are plotted as distinct measurements in each $\eta$ bin with $\pt$ increasing from 3 to 15~\GeV{} going from left to right. The total uncertainty is the sum in quadrature of all the individual contributions.}
\label{fig:LowPtSyst}
\end{figure}

The main contributions to the systematic uncertainty in the measurement of the reconstruction and identification efficiency SFs using $J/\psi\to\mu\mu$ events are shown in Figure~\ref{fig:LowPtSyst}.
 
In addition to the \emph{T\&P method} and \emph{Probe matching} uncertainties discussed in Section~\ref{subsubsec:ztp_syst}, a \emph{Fit model} systematic uncertainty is assigned to cover the possible biases introduced by the fitting procedure. This uncertainty is estimated by generating pseudo-data mimicking the level of background observed in each invariant mass bin in the data, and injecting it into the corresponding $m_{\mathrm{tag-probe}}$ distribution from $J/\psi$ MC simulation. The latter is then fit using the same procedure as deployed for data, and the difference between the fitted efficiencies and those obtained by counting the fraction of probes in $J/\psi$ MC simulation that satisfy the selection of interest is assigned as the uncertainty. This is found to be the dominant systematic uncertainty in the lowest $\pt$ bins, where the background contamination is larger.
 
\subsection{The double-ratio method for the $2.5<|\eta|<2.7$ region}
\label{subsec:forward}
 
The \emph{Loose} and \emph{Medium} selections accept muons within the full MS acceptance, allowing physics analyses to benefit from ME and SiF muons in the $2.5<|\eta|<2.7$ region. As the ID coverage is limited to $|\eta|<2.5$, a tag-and-probe method involving the two independent detectors is not a viable option in this region. A direct measurement of the muon efficiency SF is instead performed, using the same technique as described in Ref.~\cite{PERF-2014-05}, and detailed below.
 
The SF is calculated from the double ratio:
\begin{equation}\label{eq:high_eta_SF}
SF = {\left[\frac{N_X^{\mathrm{Data}}}{N_X^{\mathrm{MC}}}\right]}^{Z\rightarrow \mu \mu} _{|\eta|>2.5}\left/{\left[\frac{N_X^{\mathrm{Data}}}{N_X^{\mathrm{MC}}}\right]}^{Z\rightarrow \mu \mu}_{2.2<|\eta |<2.5}\textrm{.}\right.
\end{equation}
The numerator of Eq.~\eqref{eq:high_eta_SF} is the ratio of the number of $Z\rightarrow \mu \mu $ candidates in data to the number in MC simulation, where one of the muons (called a \emph{forward} muon hereafter) is identified according to the $X$ criterion in the $2.5<|\eta |<2.7$ region, while the other leg of the $Z$ decay (called a \emph{central} muon) is required to have $|\eta|<2.4$.
The denominator is instead the data-to-MC ratio of $Z\rightarrow \mu \mu $ candidates with the \emph{forward} muon lying in the $2.2<|\eta|<2.5$ region, and the \emph{central} muon in the $|\eta|<2.4$ region.
 
The main assumption behind this method is that the ratio of the number of $Z\rightarrow \mu \mu$ events with one \emph{forward} muon to the number of $Z\rightarrow \mu \mu$ events with two \emph{central} muons, prior to detector effects, is well modelled in the simulation. In the double ratio, theoretical and experimental uncertainties common to the numerator and the denominator cancel out to first order.
 
The \emph{central} muon is required to have triggered the event read-out, to fulfil the \emph{Medium} selection, the standard vertex association and the \emph{Tight} isolation criteria, and to have $\pt$ greater than $25$~\GeV{}. The \emph{forward} muon is selected with $\pt$ above $10$~\GeV{} and opposite charge to the \emph{central} muon. Their invariant mass must be in the 81--101~\GeV{} interval. The simulation of muons with $|\eta|<2.5$ is corrected using the standard SF described in the previous section.
 
To evaluate the systematic uncertainties affecting the measured SF, the pseudorapidity requirement on the \emph{forward} muon used in the denominator is changed to $2.0<|\eta|<2.2$, $2.0<|\eta|<2.5$, and $|\eta|<2.2$, and the largest deviation from the nominal SF is taken as the uncertainty. Furthermore, the isolation requirements on the \emph{central} muon are changed to \emph{Loose}, and independently its minimum $\pt$ is raised to 35~\GeV{}, computing in either case a symmetric uncertainty as half the deviation from the nominal SF. The uncertainty in the background processes is estimated by subtracting from data the doubled and halved background contributions predicted with MC simulation: half of the largest deviation from the nominal SF is assigned as the uncertainty.The impact of the limited number of MC events is also accounted for. Finally, theoretical uncertainties such as the uncertainty in the knowledge of the true parton distribution functions are evaluated as the variation in the double ratio observed after reweighting the CT10 PDF set used in the $Z\rightarrow \mu \mu$ MC simulation to MSTW2008~NLO~\cite{Martin:2009iq}, and after considering the uncertainties associated to the MSTW2008~NLO PDF set.
 
 
\section{Results}
\label{sec:results}
 
\FloatBarrier
The muon reconstruction, identification, vertex association and isolation efficiencies were measured for tracks with $\pt>3$~\GeV{} and $|\eta|<2.7$. In the following paragraphs, the observed performance of the various selection algorithms and the level of agreement between the predicted efficiencies and the corresponding measurements in collision data are discussed.
 
\subsection{Reconstruction and identification efficiencies}
 
\begin{figure}
\centering
\includegraphics[width=0.49\textwidth]{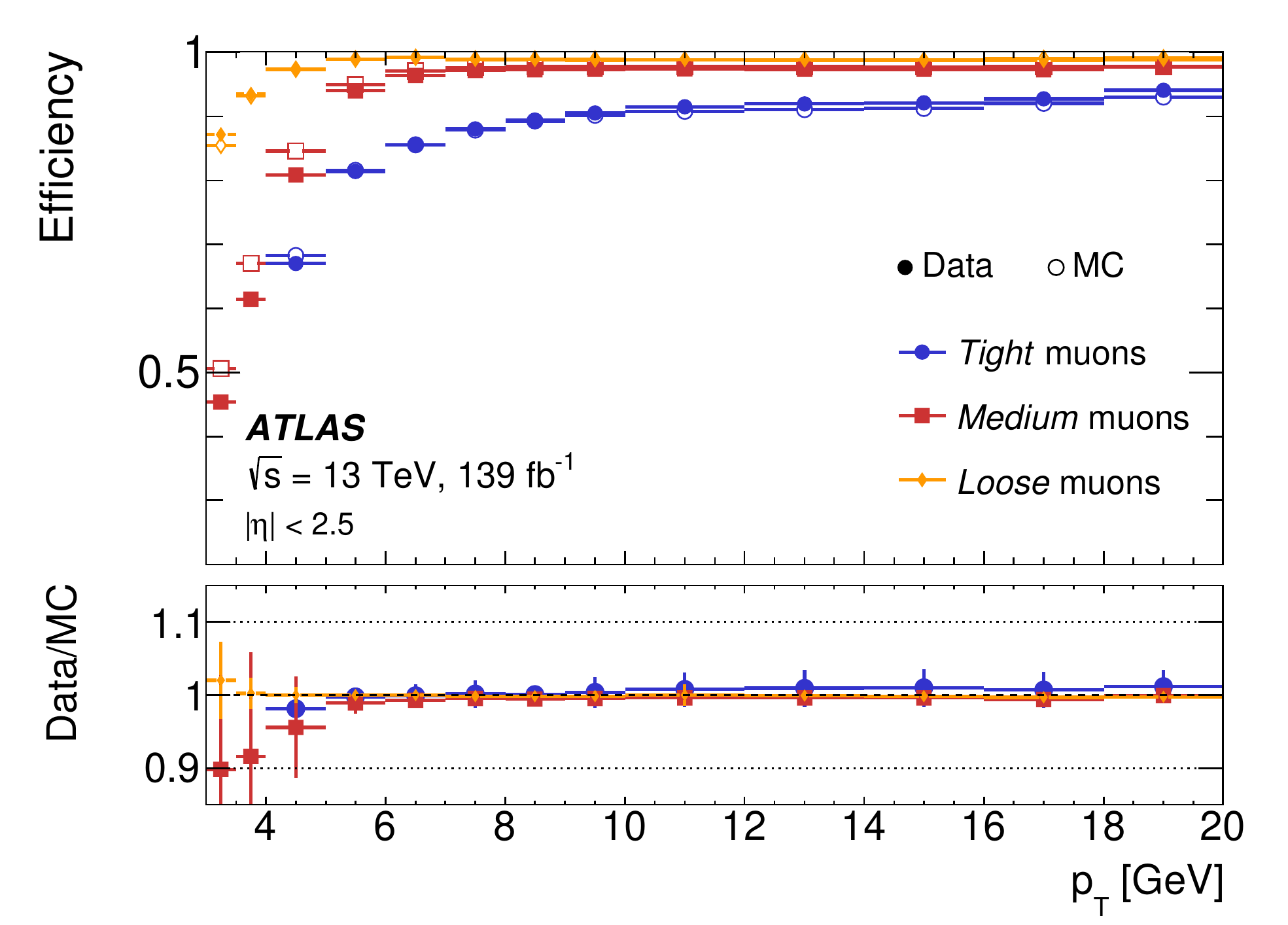}
\includegraphics[width=0.49\textwidth]{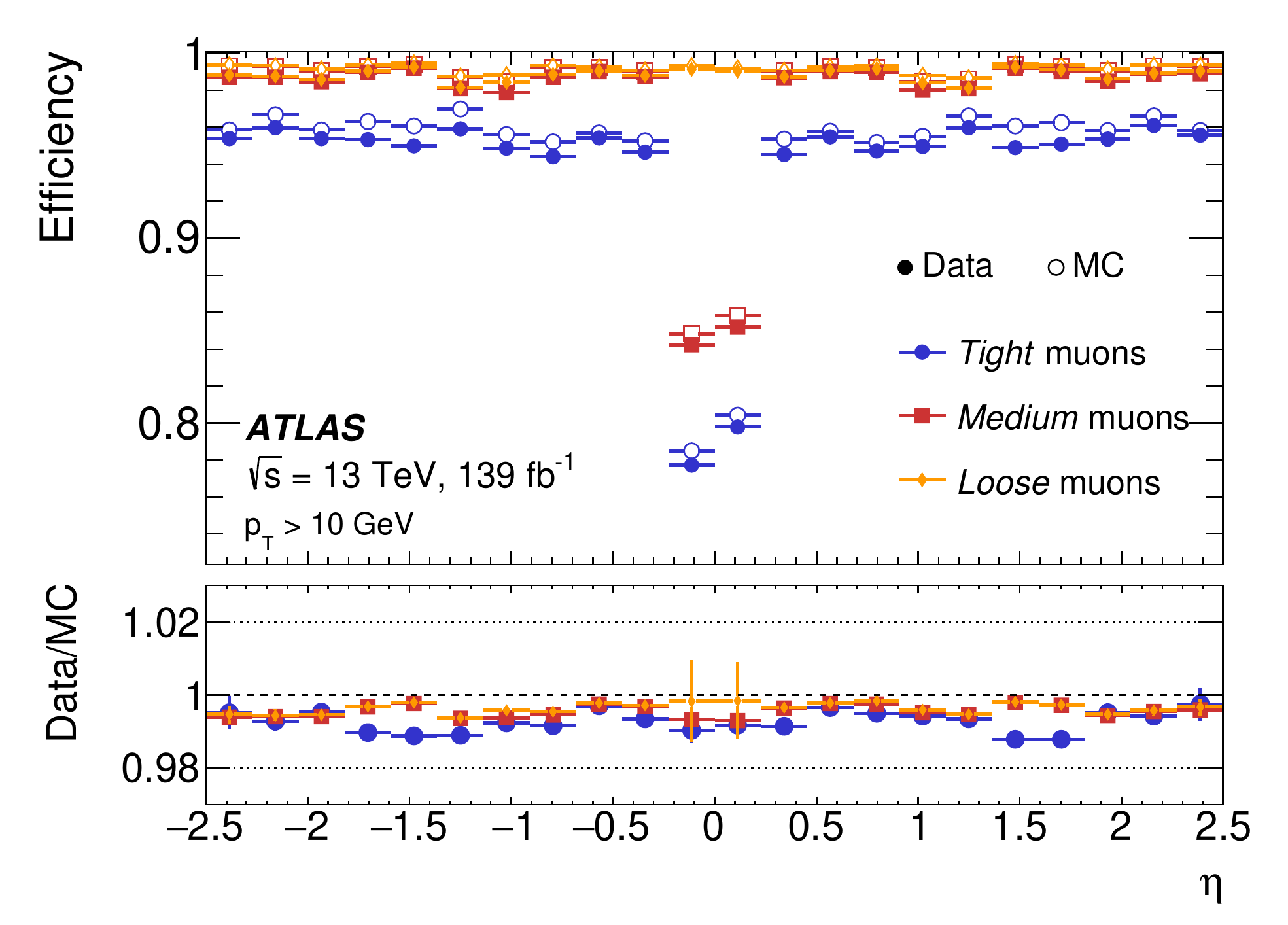}
\caption{Muon reconstruction and identification efficiencies for the \loose, \medium, and \tight\ criteria. The left plot shows the efficiencies measured in $J/\psi\to\mu\mu$ events as function of $\pt$. The right plot displays the efficiencies measured in $Z\rightarrow\mu\mu$ events as a function of $\eta$, for muons with $\pt>10$~\GeV{}. The predicted efficiencies are depicted as open markers, while filled markers illustrate the result of the measurement in collision data. The statistical uncertainty in the efficiency measurement is smaller than the size of the markers, and thus not displayed. The panel at the bottom shows the ratio of the measured to predicted efficiencies, with statistical and systematic uncertainties.}
\label{fig:Zeff_LMT}
\end{figure}

\begin{figure}
\centering
\includegraphics[width=0.59\textwidth]{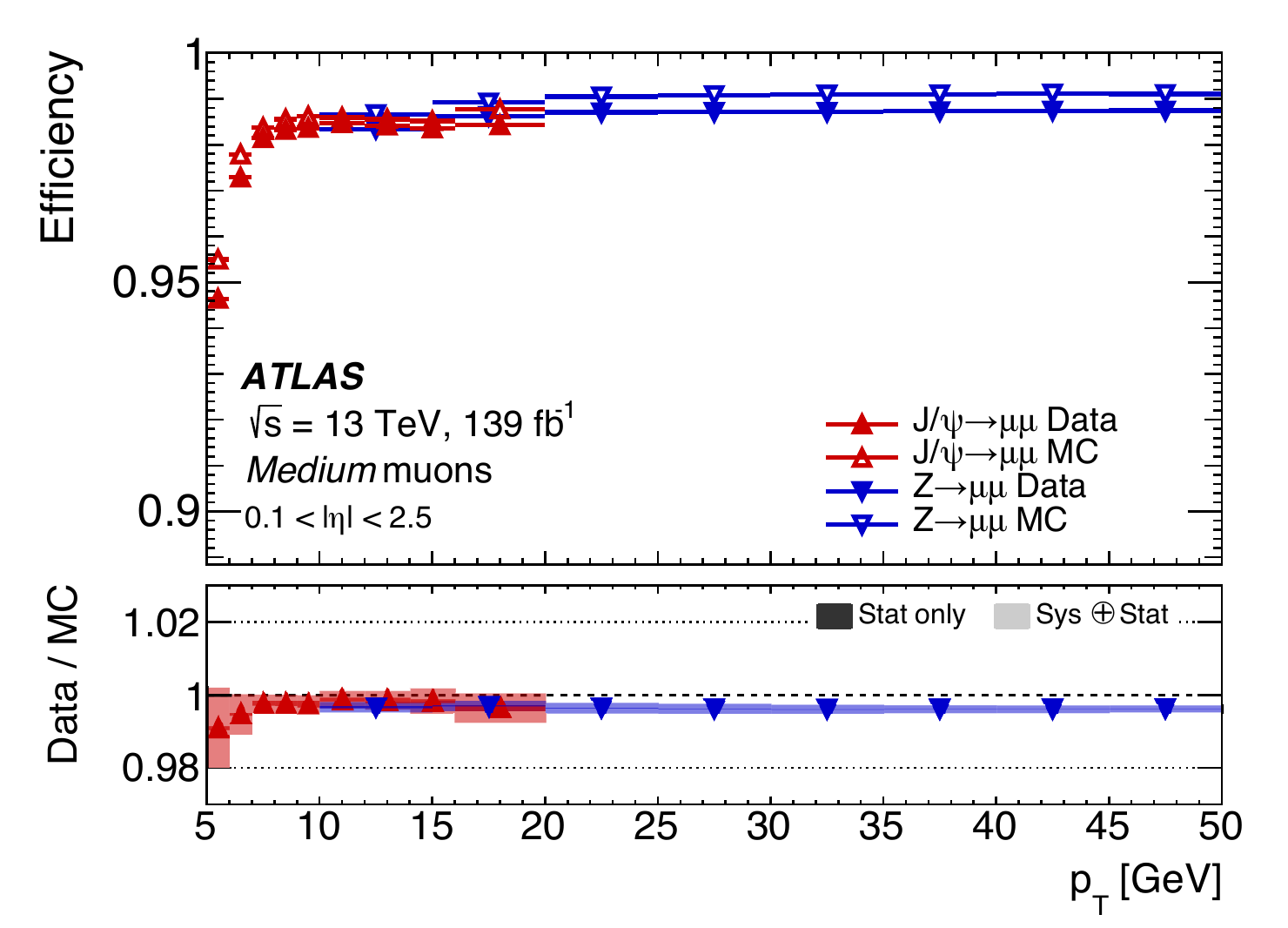}
\caption{Muon reconstruction and identification efficiencies for the \medium\ criteria measured in $J/\psi\to\mu\mu$ and $Z\rightarrow\mu\mu$ events as a function of $\pt$ for muons with $0.1<|\eta|<2.5$.  When not negligible, the statistical uncertainty in the efficiency measurement is indicated by the error bars. The panel at the bottom shows the ratio of the measured to predicted efficiencies, with statistical and systematic uncertainties.}
\label{fig:ZJPsieff}
\end{figure}

Figure~\ref{fig:Zeff_LMT} shows the muon reconstruction and identification efficiency for \loose, \medium, and \tight\ muons as measured in $J/\psi\to\mu\mu$ and $Z\rightarrow\mu\mu$ events. For muon tracks with $\pt$ greater than $10$~\GeV{}, the efficiencies and the data/MC agreement are stable for all selection levels. Conversely, the efficiencies drop significantly in the $\pt$ region below 5~\GeV{}, as soft muons crossing the calorimeters often do not have enough residual energy to reach the second station of precision MS chambers. The measurements from $J/\psi\to\mu\mu$ and $Z\rightarrow\mu\mu$ events agree within uncertainties in the overlap region between 10 and 20~\GeV, as visible in Figure~\ref{fig:ZJPsieff}. For all these reasons, the reconstruction and identification SFs and the corresponding uncertainties are computed in the ($\eta$,$\phi$) plane for muons with $\pt$ above 15~\GeV{}, using $Z\rightarrow\mu\mu$ events. In particular, sixteen $\phi$ bins are used, following the layout of the MS precision chambers. In the SF for the \highPt\ selection, the number of $\eta$ bins and the $\phi$ bin boundaries are adjusted to reflect the unique characteristics of the WP, which is the most sensitive to the presence or absence of hits in each of the MS stations. In the 3--15~\GeV{} transverse momentum range, the SFs are measured in $J/\psi\to\mu\mu$ events as a function of $\pt$ and $\eta$.
 
The \loose\ and \medium\ selections are characterised by very similar efficiency throughout the detector with the exception of the region $|\eta|<0.1$, where the \loose\ selection accepts CT and ST muons to fill the gap in the MS coverage. The efficiency of the \loose\ and \medium\ criteria exceeds 98\% for tracks with $0.1<|\eta|<2.5$. Excellent agreement between detector simulation and the collision data is observed, with differences on average at the level of 0.5\%. The efficiency of the \tight\ selection is measured to exceed 95\% for tracks with $0.1<|\eta|<2.5$, with differences between data and simulation at the 1\% level or below.
 
\begin{figure}[htb]
\centering
\includegraphics[width=0.49\textwidth]{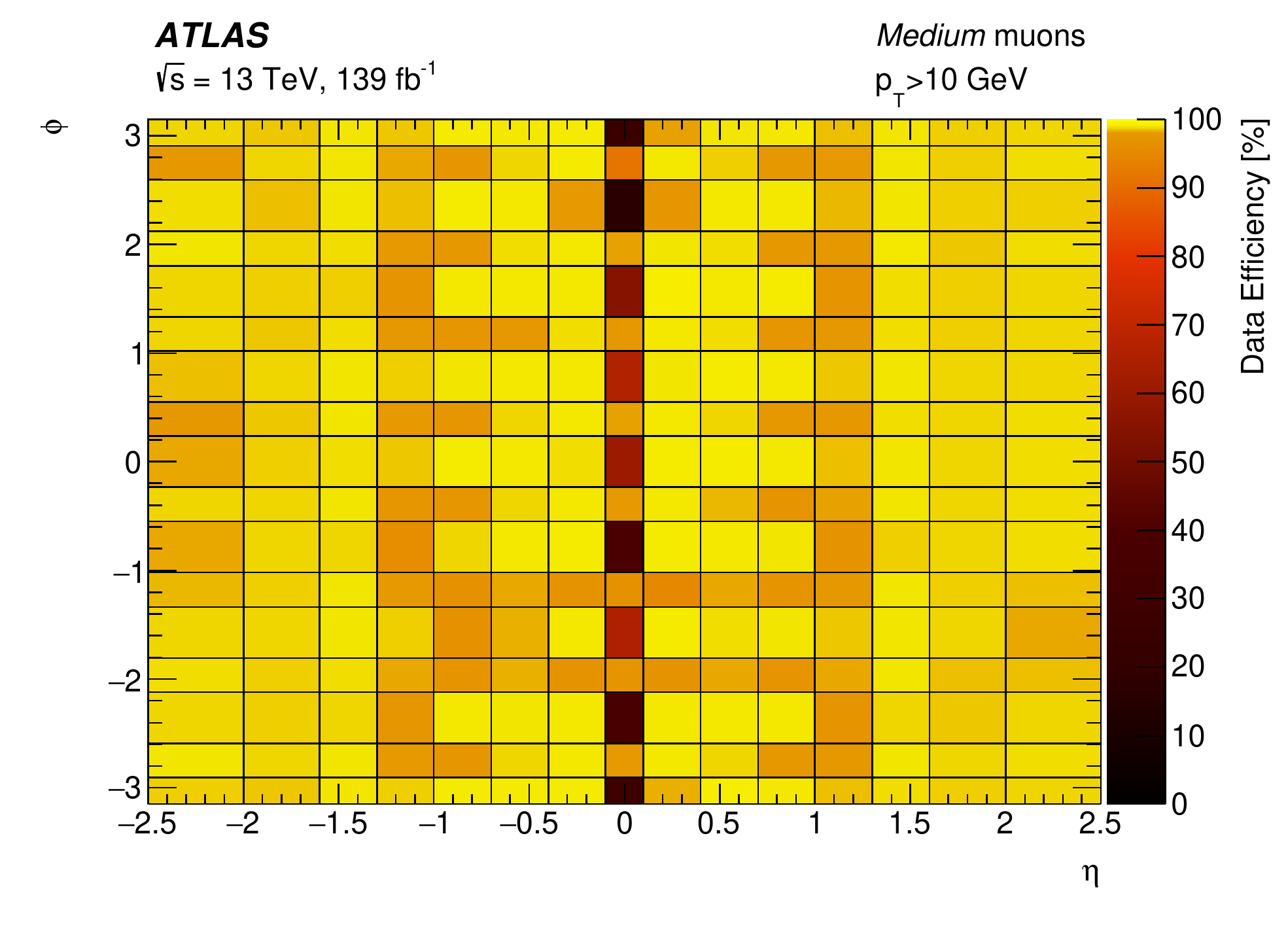}
\includegraphics[width=0.49\textwidth]{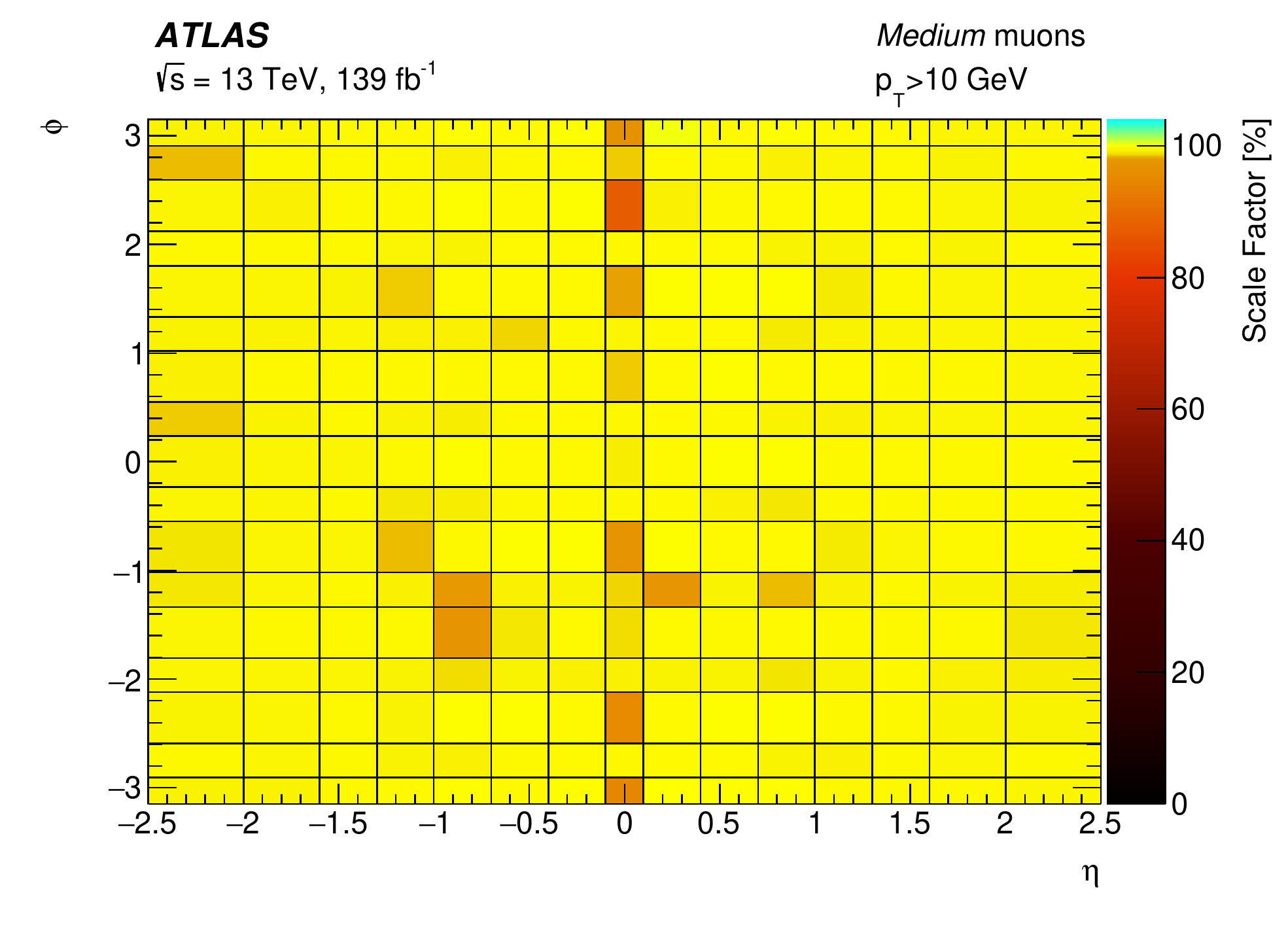}
\caption{Reconstruction and identification efficiency measured in collision data (left), and the data/MC efficiency scale factor (right) for \medium\ muons as a function of $\eta$ and $\phi$ for muons with $\pt>10$~\GeV{} in $Z\rightarrow\mu\mu$ events.}
\label{fig:ZSF_Medium}
\end{figure}

\begin{figure}[htb]
\centering
\includegraphics[width=0.49\textwidth]{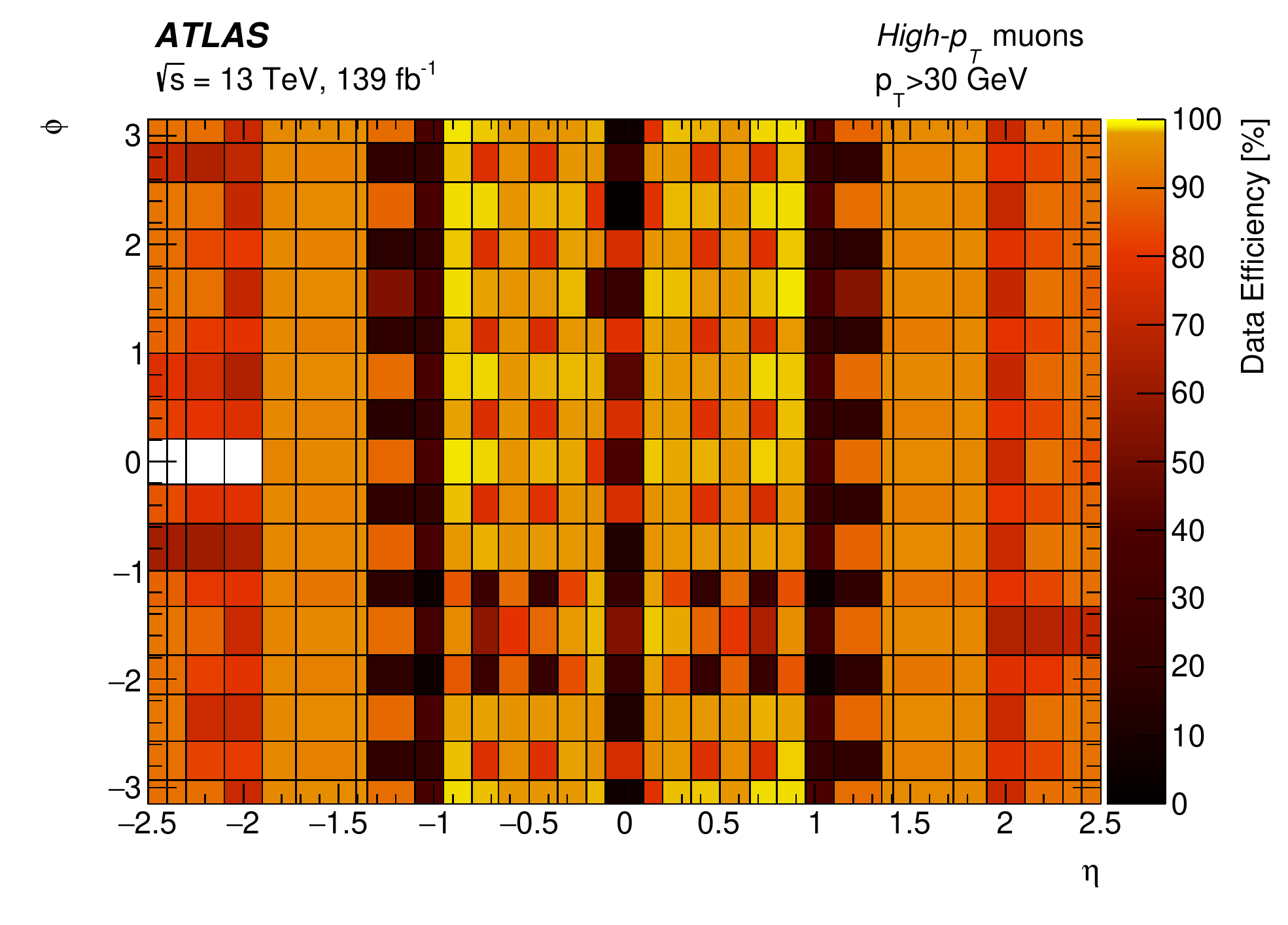}
\includegraphics[width=0.49\textwidth]{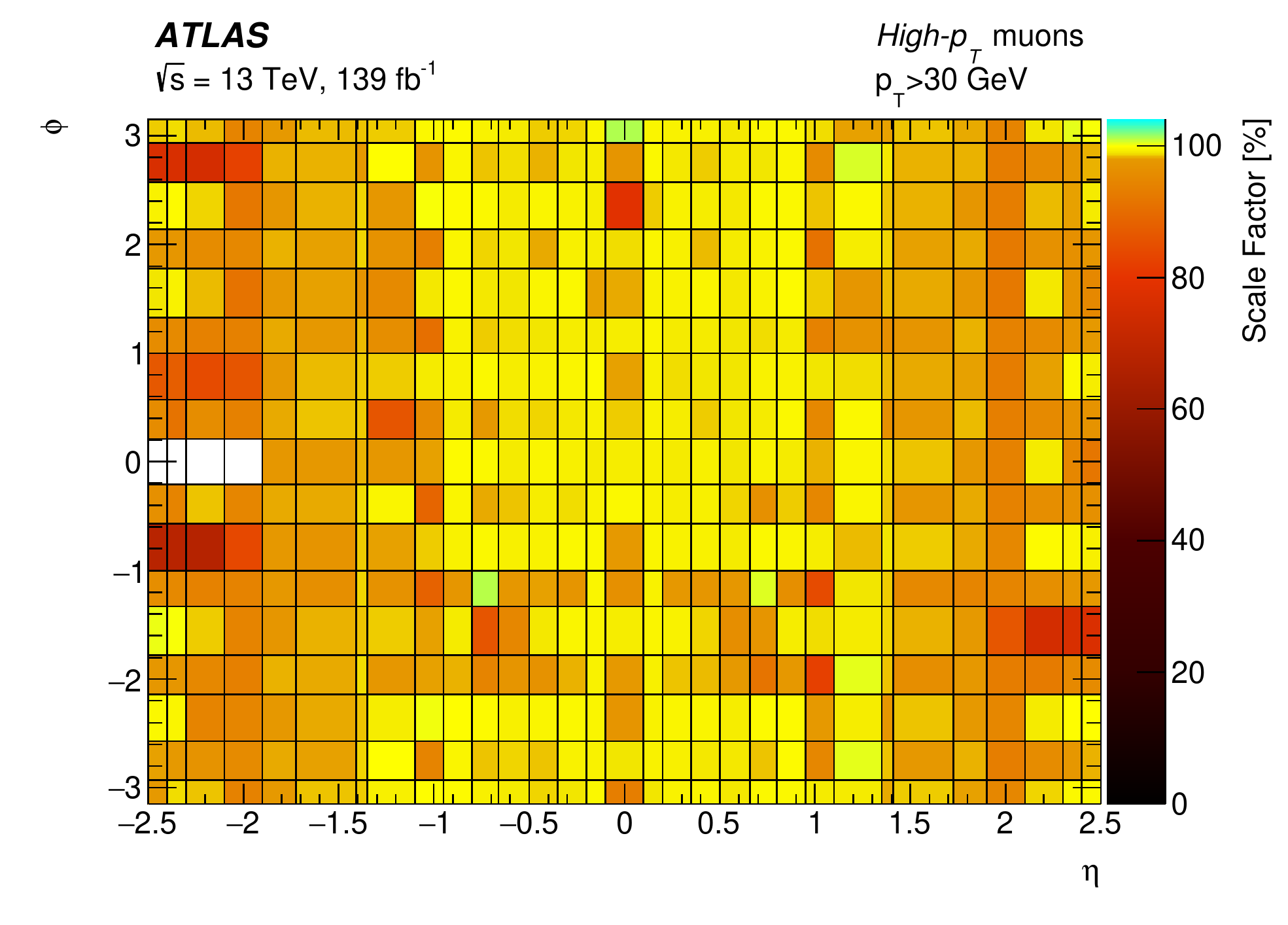}
\caption{Reconstruction and identification efficiency measured in collision data (left), and the data/MC efficiency scale factor (right) for \highPt\ muons as a function of $\eta$ and $\phi$ for muons with $\pt>30$~\GeV{} in $Z\rightarrow\mu\mu$ events. The white area in the $\eta<-2.0$ and $\phi\simeq0$ region corresponds to a vetoed problematic CSC.}
\label{fig:ZSF_HighPt}
\end{figure}

The efficiencies as measured in $Z\rightarrow\mu\mu$ data events and the corresponding SFs for the \medium\ and \highPt\ selections are shown as a function of $\eta$ and $\phi$ in Figures~\ref{fig:ZSF_Medium}~and~\ref{fig:ZSF_HighPt}, respectively. The efficiency of the \highPt\ selection is significantly lower, as a consequence of the strict requirements on momentum resolution.
 
For the \medium\ WP, overall agreement between data and simulation across the ($\eta$,$\phi$) plane is at the level of 0.5\%, with about 9\% of the measured SFs showing a deviation from unity greater than 1\%, and with only two bins with a deviation greater than 5\%, both in the $|\eta|<0.1$ region. The most pronounced local inefficiencies are also observed in the $|\eta|<1.0$ region, around $\phi=-1.2$ and $\phi=-2.0$, corresponding to the detector support structures that prevent complete coverage by the MS. Further inefficiencies for \medium\ muons are visible in the region $1.0<|\eta|<1.3$, characterised by poorly aligned MDT chambers in half of the $\phi$ sectors of the innermost station. Tracks reconstructed in these chambers are rejected by the \highPt\ selection, as visible in Figure~\ref{fig:ZSF_HighPt}. The efficiency drops localised around $(\eta,\phi)=(-2.5,0)$ and $(\eta,\phi)=(-2.5,-1)$ were traced to temporary faults in the corresponding CSCs during the 2017 data taking. Similarly, the efficiency drops around $(\eta,\phi)=(-2.5,+2.7)$ and $(\eta,\phi)=(+2.5,-1.5)$ are linked to malfunctioning CSCs during 2018. Finally, the visible efficiency loss in the \highPt\ selection for tracks with $|\eta|>2.0$ is associated with the stringent quality requirements imposed on the hits collected in the CSCs, which are particularly important for a robust muon momentum measurement due to the structure of the ATLAS magnetic field. The acceptance gap around $(\eta,\phi)=(-2.0,0)$ is instead linked to a problematic CSC chamber throughout most of Run~2.

\begin{figure}[t]
\centering
\includegraphics[width=0.49\textwidth]{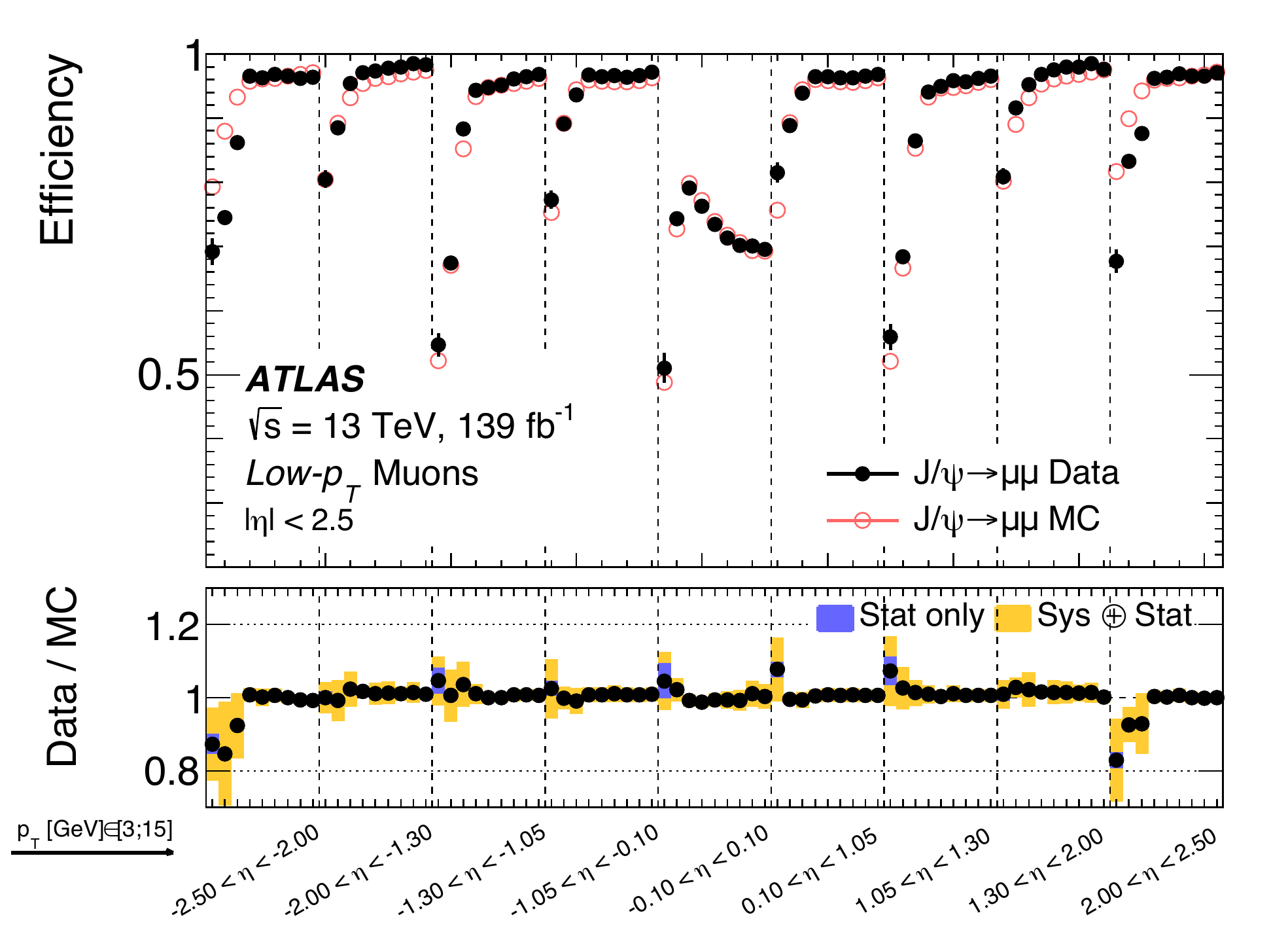}
\includegraphics[width=0.49\textwidth]{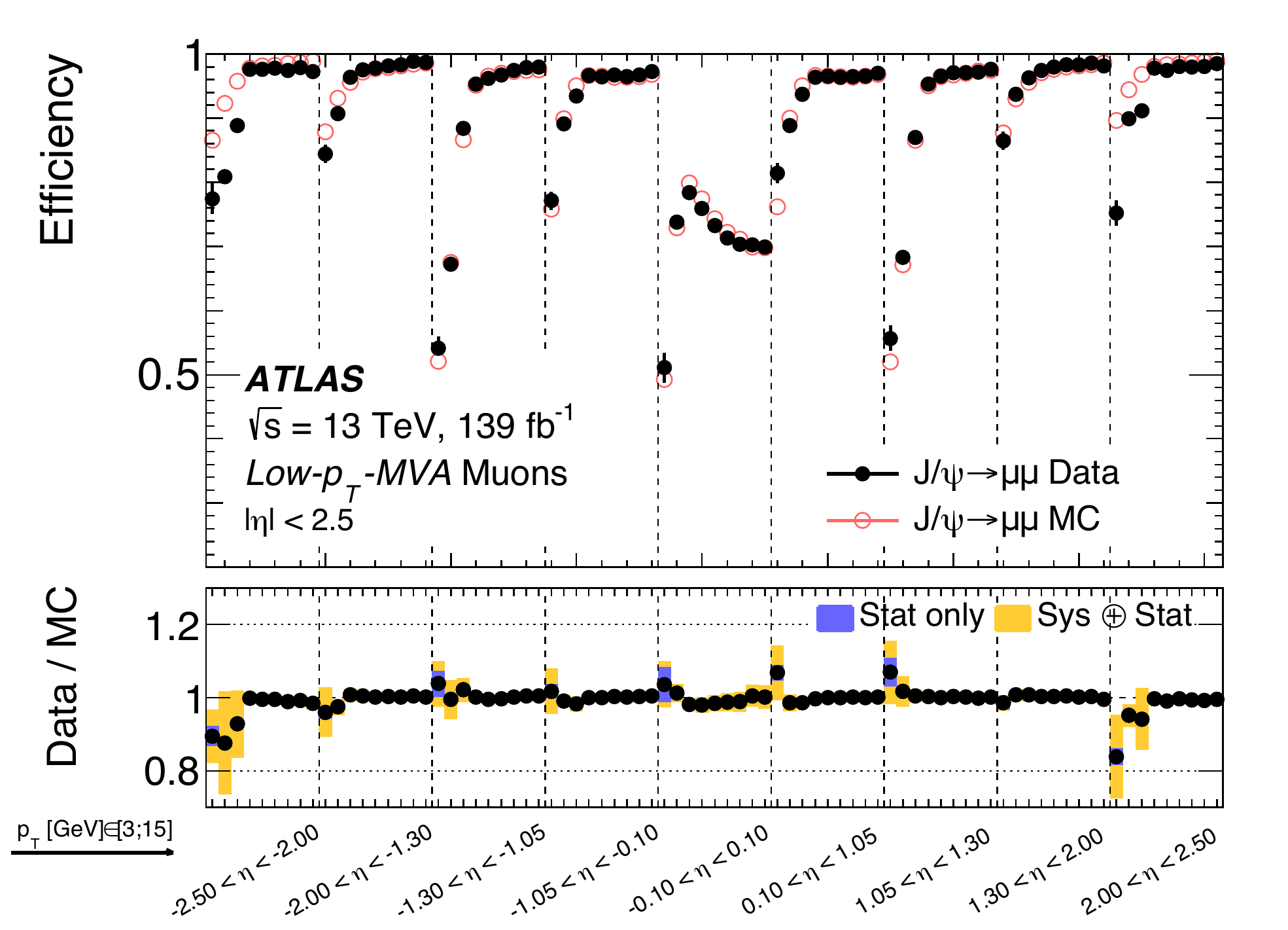}
\caption{Muon reconstruction and identification efficiency measured in $J/\psi\to\mu\mu$ events for the cut-based (left) and multivariate (right) \lowPt\ criteria. In the plots, within each $\eta$ region, the efficiency is measured in nine $\pt$ bins (3--3.5, 3.5--4, 4--5, 5--6, 6--7, 7--8, 8--10, 10--12, 12--15~\GeV{}). The resulting values are plotted as distinct measurements in each $\eta$ bin with $\pt$ increasing from 3 to 15~\GeV{} going from left to right.
When not negligible, the statistical uncertainty in the efficiency measurement is indicated by the error bars. The panel at the bottom shows the ratio of the measured to predicted efficiencies, with statistical and systematic uncertainties.}
\label{fig:LowPtEff}
\end{figure}

The efficiencies for the cut-based and multivariate \lowPt\ selections are shown in Figure~\ref{fig:LowPtEff}. For muons with $\pt$ greater than $10$~\GeV, the efficiencies of the two selections are very similar, as expected. Below 10~\GeV{} in muon $\pt$ the differences are more marked, with the multivariate selection having a larger efficiency especially in the forward $\eta$ regions. The multivariate criteria also show, in general, smaller uncertainties in the SFs, due to having more power to reject the non-prompt muon backgrounds contaminating the event sample used for the measurement. Good agreement is found between predicted and observed efficiencies, except in the $|\eta|>2.0$ region for tracks with $\pt$ below $4$~\GeV{}, where the differences are larger than 10\%. The efficiency drop in collision data is partly associated with the faulty CSCs discussed previously, which are not modelled by the detector simulation. Furthermore, it stems from an overall lower segment-reconstruction efficiency in the CSC relative to simulation predictions. Since tracks with $\pt$ below $4$~\GeV{} often have insufficient residual energy to reach the second station of MS precision chambers, efficiency losses in the innermost MS station have a direct impact on the overall reconstruction efficiency.
 
\begin{figure}
\centering
\includegraphics[width=0.59\textwidth]{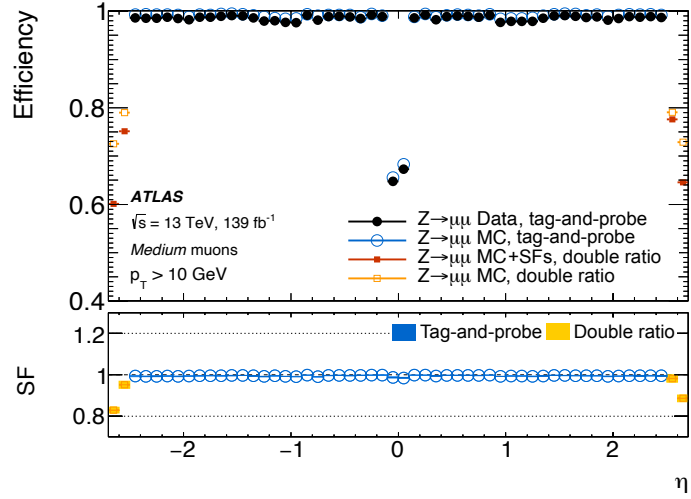}
\caption{Muon reconstruction and identification efficiencies for the \medium{} criteria measured in $Z\rightarrow\mu\mu$ events as a function of $\eta$ for muons with $\pt>10$~\GeV{}. Circular markers show the results obtained using the tag-and-probe method for $|\eta|<2.5$, while square markers represent the MC simulation efficiencies for $|\eta|>2.5$ before and after the SFs computed with the double-ratio method are applied. The predicted efficiencies are depicted as open markers, while filled markers illustrate the efficiencies resulting from a direct measurement in collision data ($|\eta|<2.5$), or from the application of the measured SFs ($|\eta|>2.5$). The data efficiencies are not shown for $|\eta|>2.5$ as the double-ratio method allows only the SFs to be measured. The panel at the bottom shows the measured SFs. The statistical and systematic uncertainties are smaller than the size of the markers, and thus not displayed.}
\label{fig:Zeff_CentralForward}
\end{figure}

Figure~\ref{fig:Zeff_CentralForward} summarises the efficiencies and SFs for \medium\ muons in the pseudorapidity range of $2.5<|\eta|<2.7$ as measured with the double-ratio method described in Section~\ref{subsec:forward}, and compares them with those obtained with the tag-and-probe technique for $|\eta|<2.5$. The observed decrease in reconstruction and identification efficiency for $|\eta|>2.5$ muons stems from the different reconstruction strategy and the more stringent selection criteria applied to tracks in a region where the ID coverage is partial or absent, and reconstruction is mainly based on the MS information. The measured SFs in the forward regions deviate significantly from unity, and account for an observed degradation of reconstruction and identification efficiencies in $\phi$ sectors of the MS with overlapping precision chambers, which is only partially reproduced in simulation. For this reason, the SFs for $|\eta|>2.5$ muons used in physics analyses are computed as a function of $\eta$ and $\phi$.
 
\subsection{Vertex association efficiencies}
\begin{figure}
\centering
\includegraphics[width=0.49\textwidth]{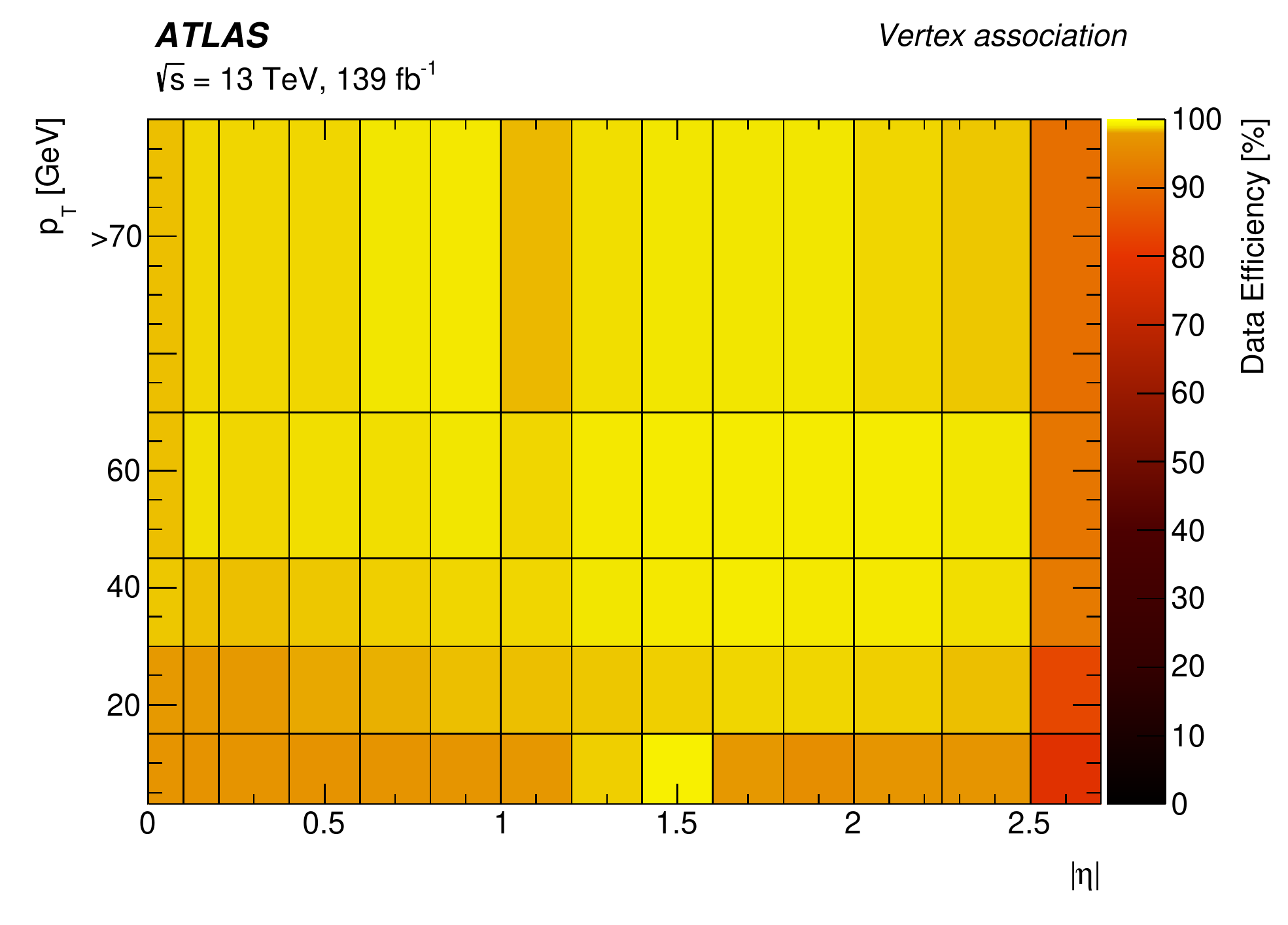}
\includegraphics[width=0.49\textwidth]{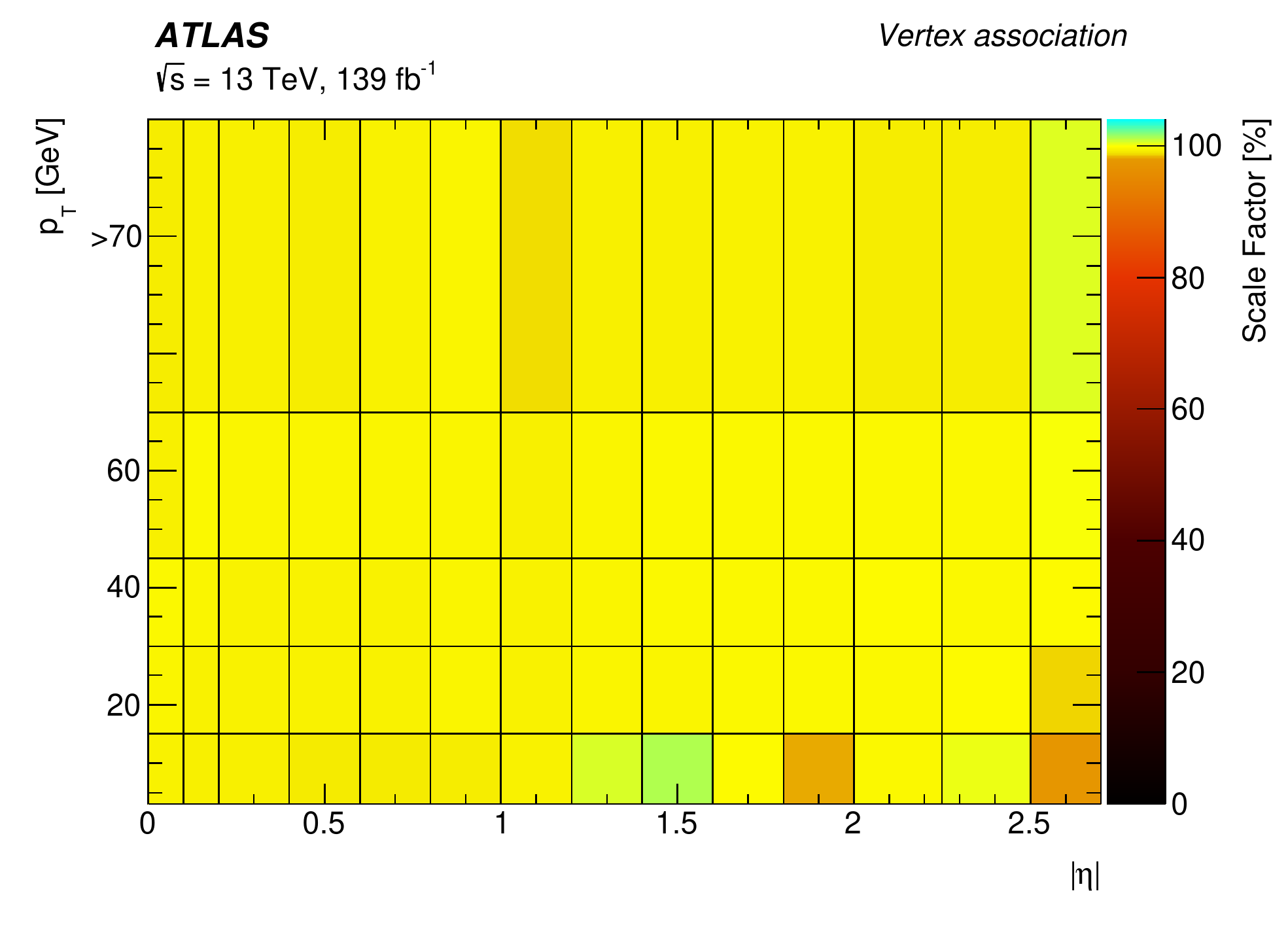}
\caption{Efficiency of the vertex association criteria measured in data (left), and the data/MC efficiency scale factor (right) as a function of $\pt$ and $|\eta|$ for muons with $\pt>3$~\GeV{} in $Z\rightarrow\mu\mu$ events.}
\label{fig:TTVASF}
\end{figure}

Figure~\ref{fig:TTVASF} shows the muon vertex association efficiency and SFs as a function of the muon $\pt$ and $|\eta|$. With the exception of the $|\eta|>2.5$ region, where tracks fall outside the acceptance of the ID and can therefore only be reconstructed with limited impact parameter resolution, the vertex association efficiency is observed to always exceed 97\%, approaching 99\% when $\pt$ is greater than $20$~\GeV{}. In the lowest $\pt$ bins, the poorer impact parameter resolution due to multiple interactions with the detector material leads to lower efficiency. Excellent agreement between collision data and detector simulation is found everywhere, with the largest deviation within the ID coverage being of the order of 2\% for low-$\pt$ tracks near the edge of the TRT detector acceptance around $|\eta|$ of 1.9.
 
\subsection{Isolation efficiencies}
 
\begin{figure}[p]
\centering
\includegraphics[width=0.49\textwidth]{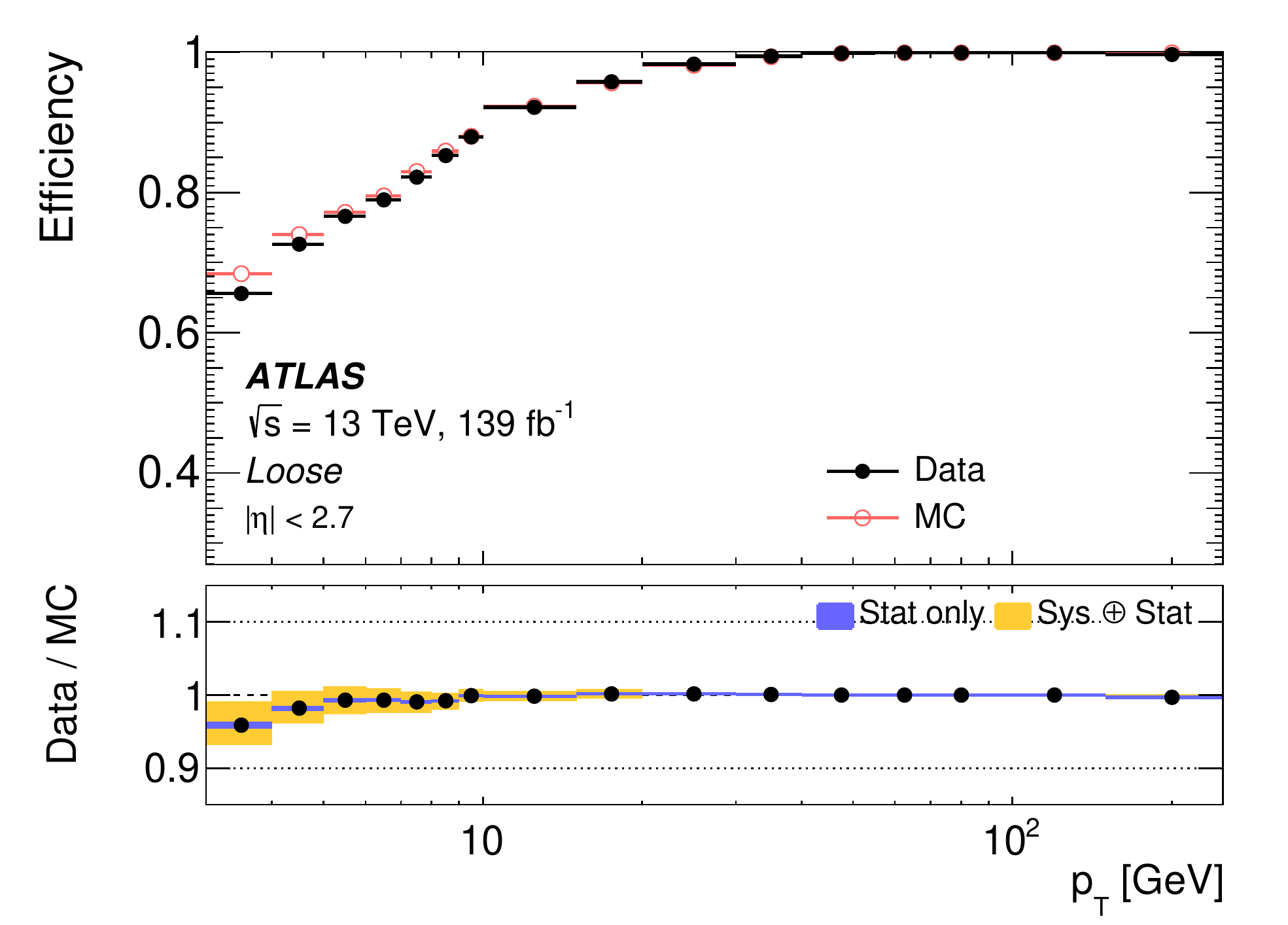}
\includegraphics[width=0.49\textwidth]{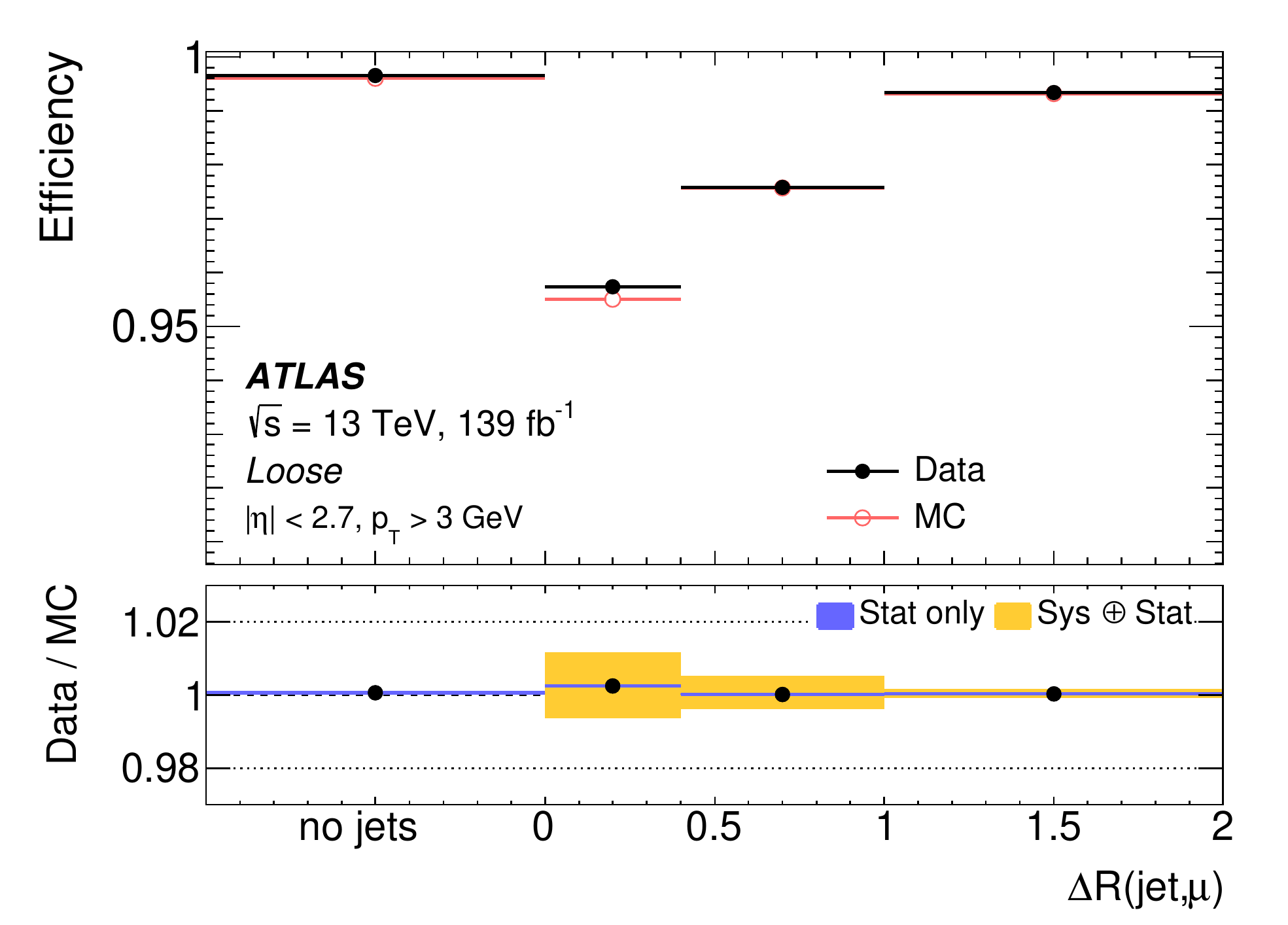}
\includegraphics[width=0.49\textwidth]{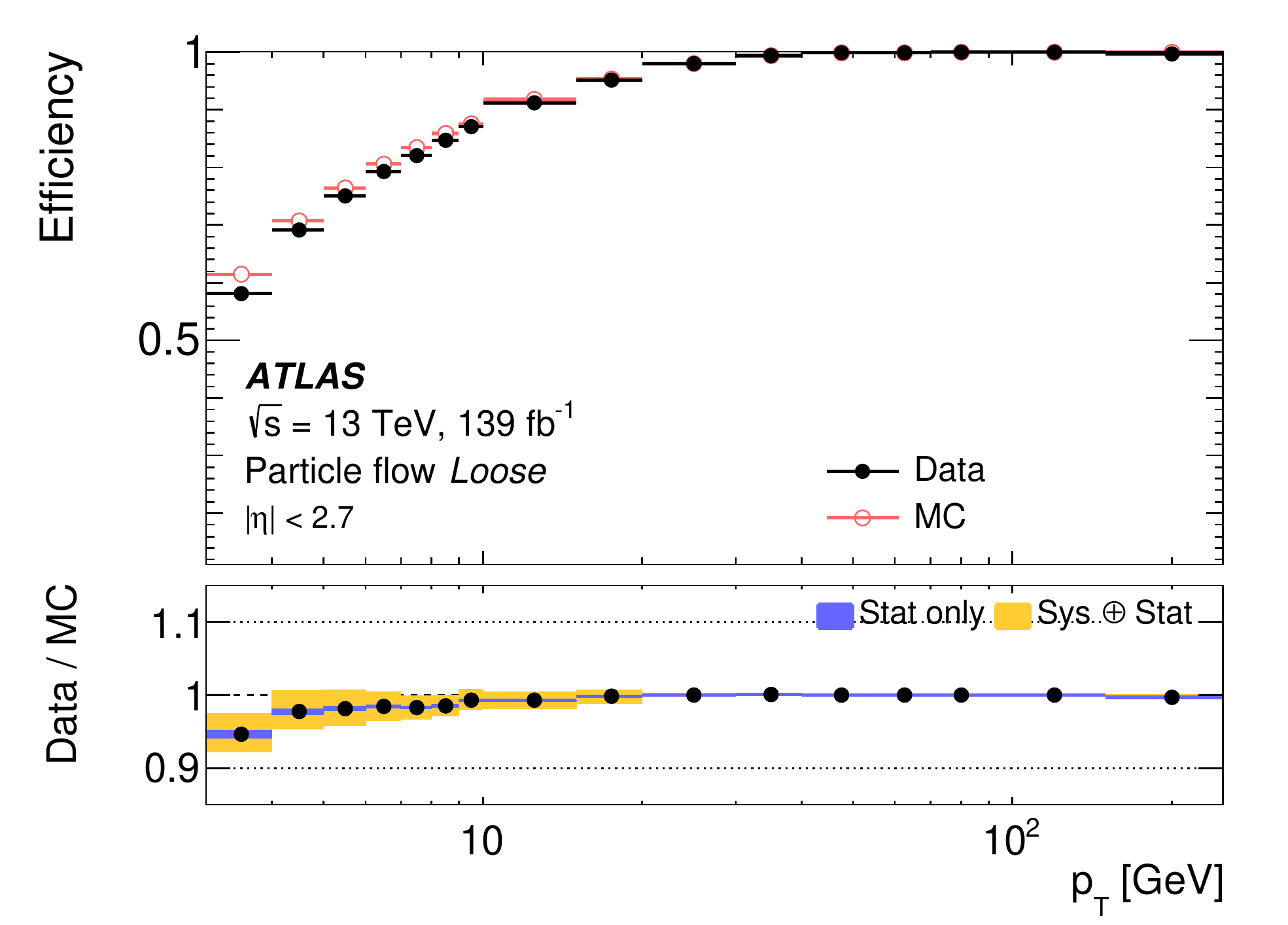}
\includegraphics[width=0.49\textwidth]{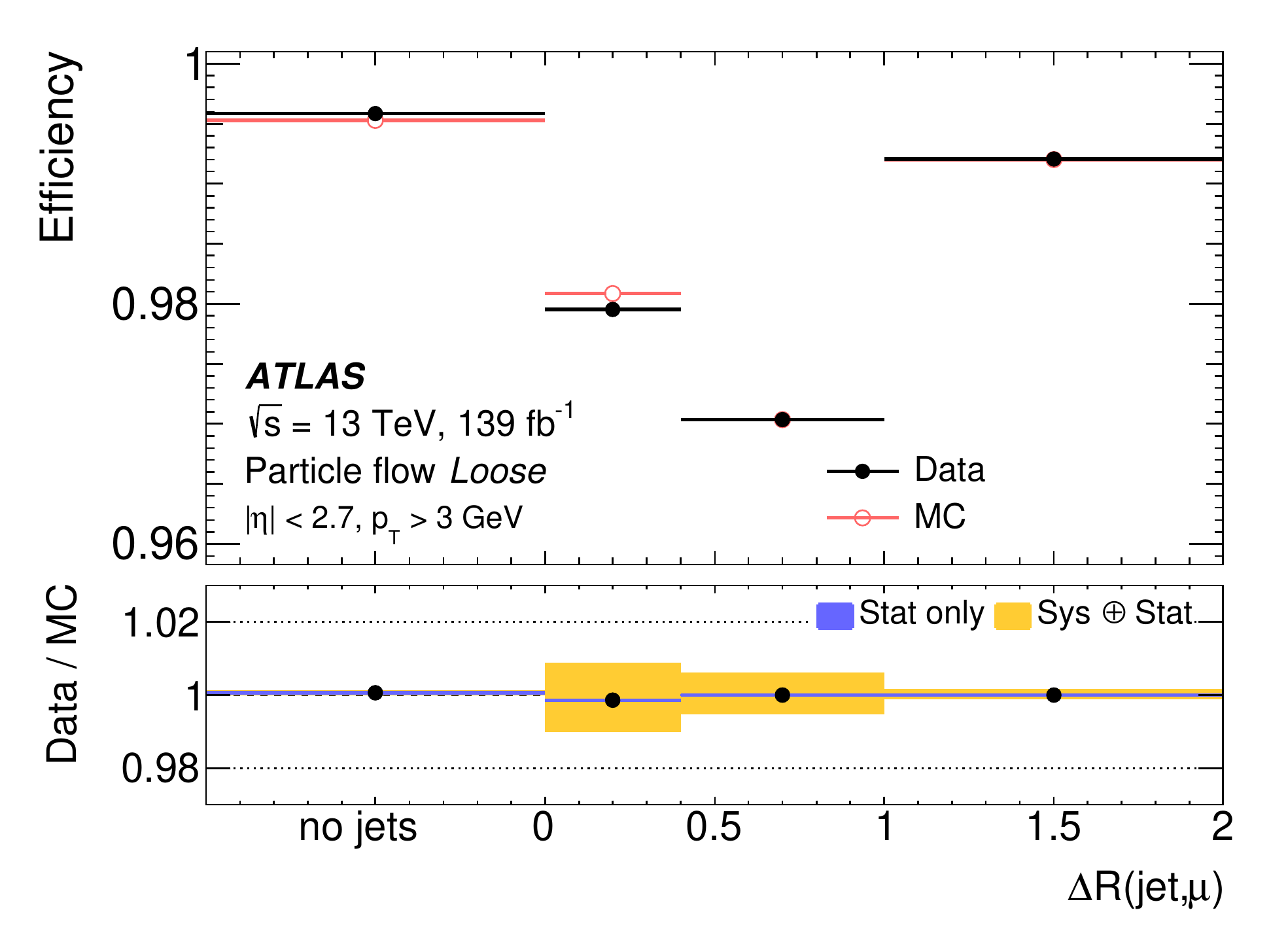}
\includegraphics[width=0.49\textwidth]{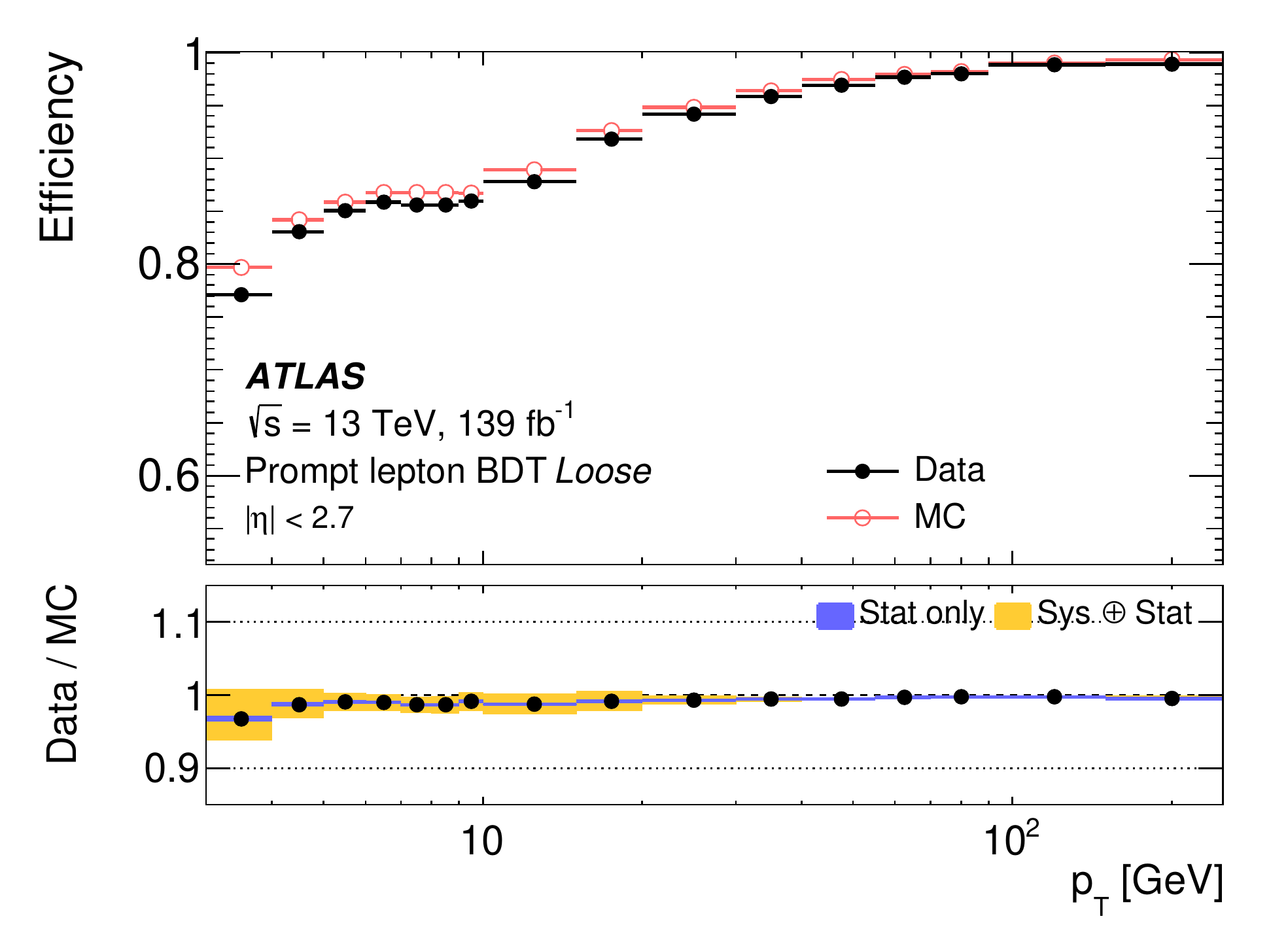}
\includegraphics[width=0.49\textwidth]{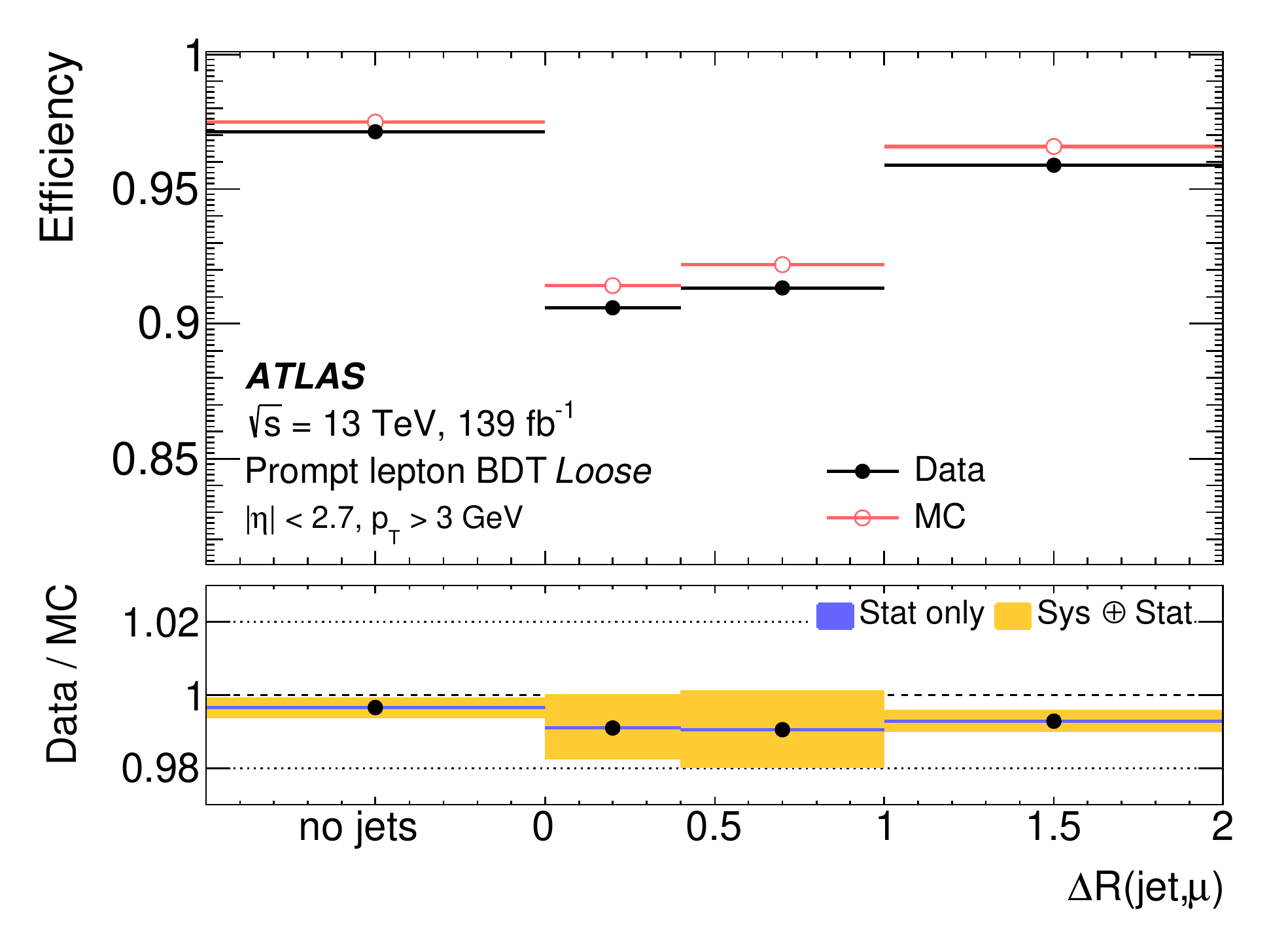}
\caption{Muon isolation efficiency measured in $Z\rightarrow\mu\mu$ events for the \emph{Loose} (top plots), \emph{PflowLoose} (middle plots), and \emph{PLBDTLoose} (bottom plots) criteria, as a function of $\pt$ (left plots) and \deltarmuj{} (right plots) for muons with $\pt>3$~\GeV{}. The statistical uncertainty in the efficiency measurement is smaller than the size of the markers, and thus not displayed. The panel at the bottom shows the ratio of the measured to predicted efficiencies, with statistical and systematic uncertainties.}
\label{fig:IsoEffLoose}
\end{figure}

\begin{figure}[p]
\centering
\includegraphics[width=0.49\textwidth]{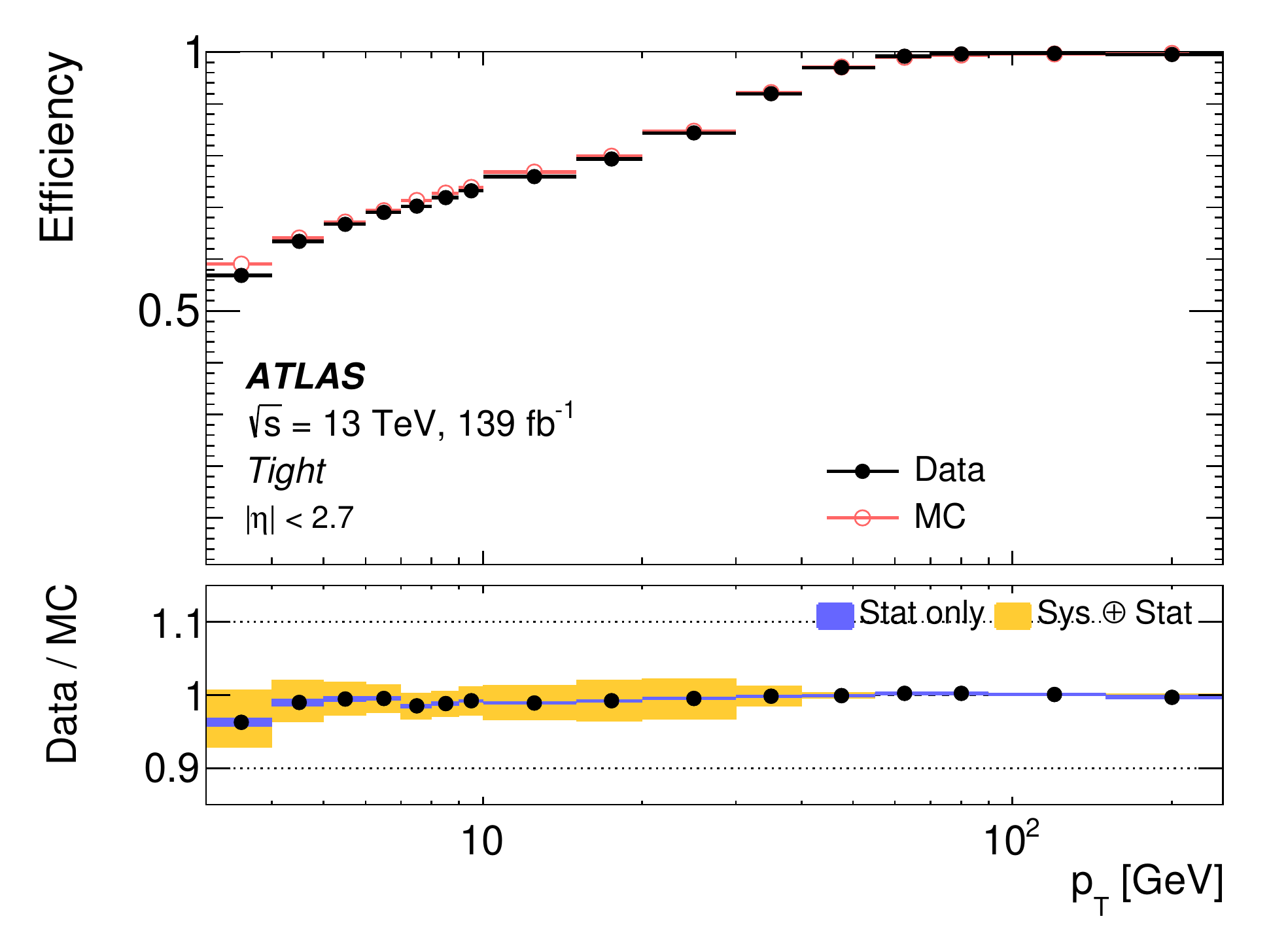}
\includegraphics[width=0.49\textwidth]{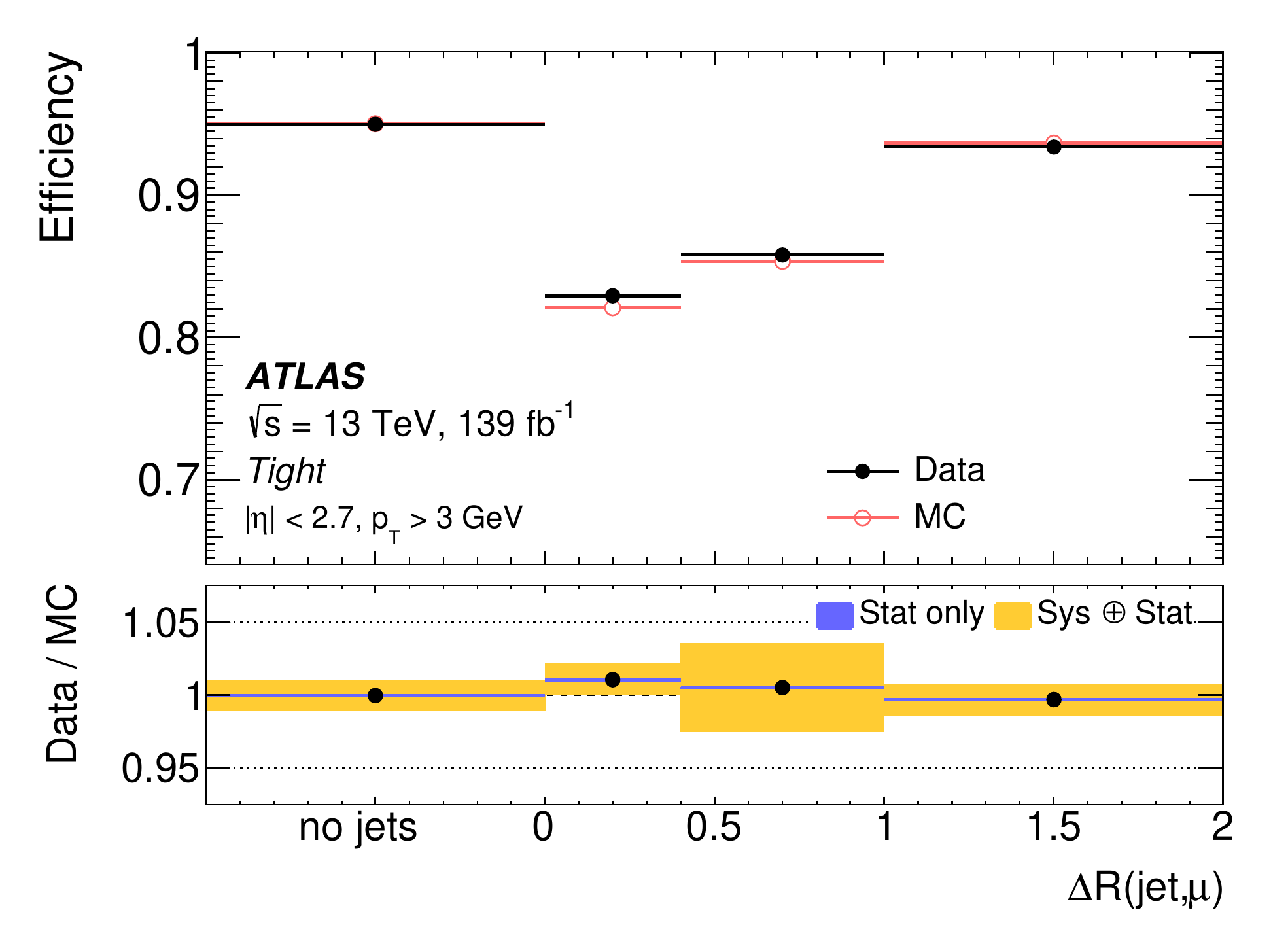}
\includegraphics[width=0.49\textwidth]{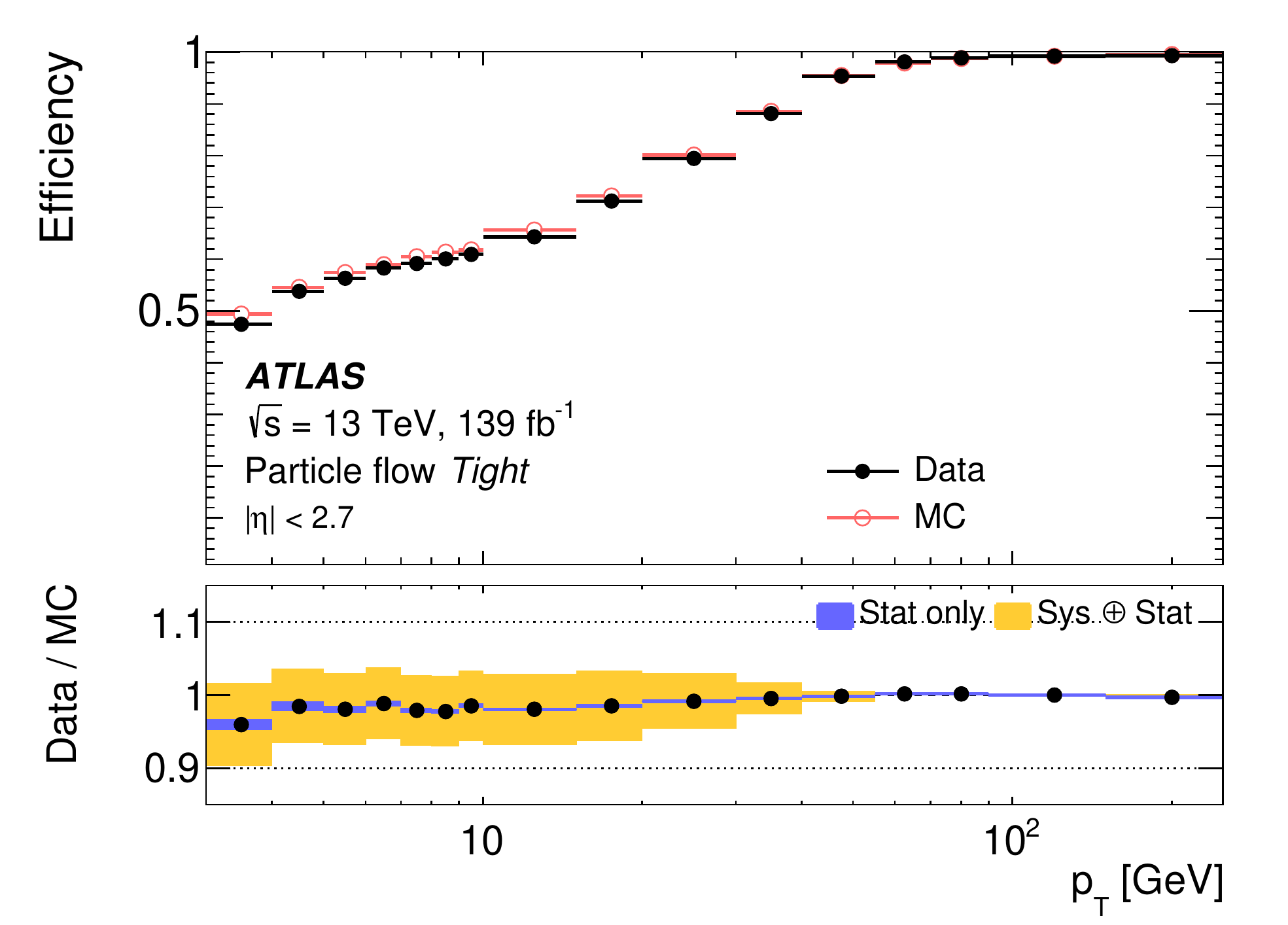}
\includegraphics[width=0.49\textwidth]{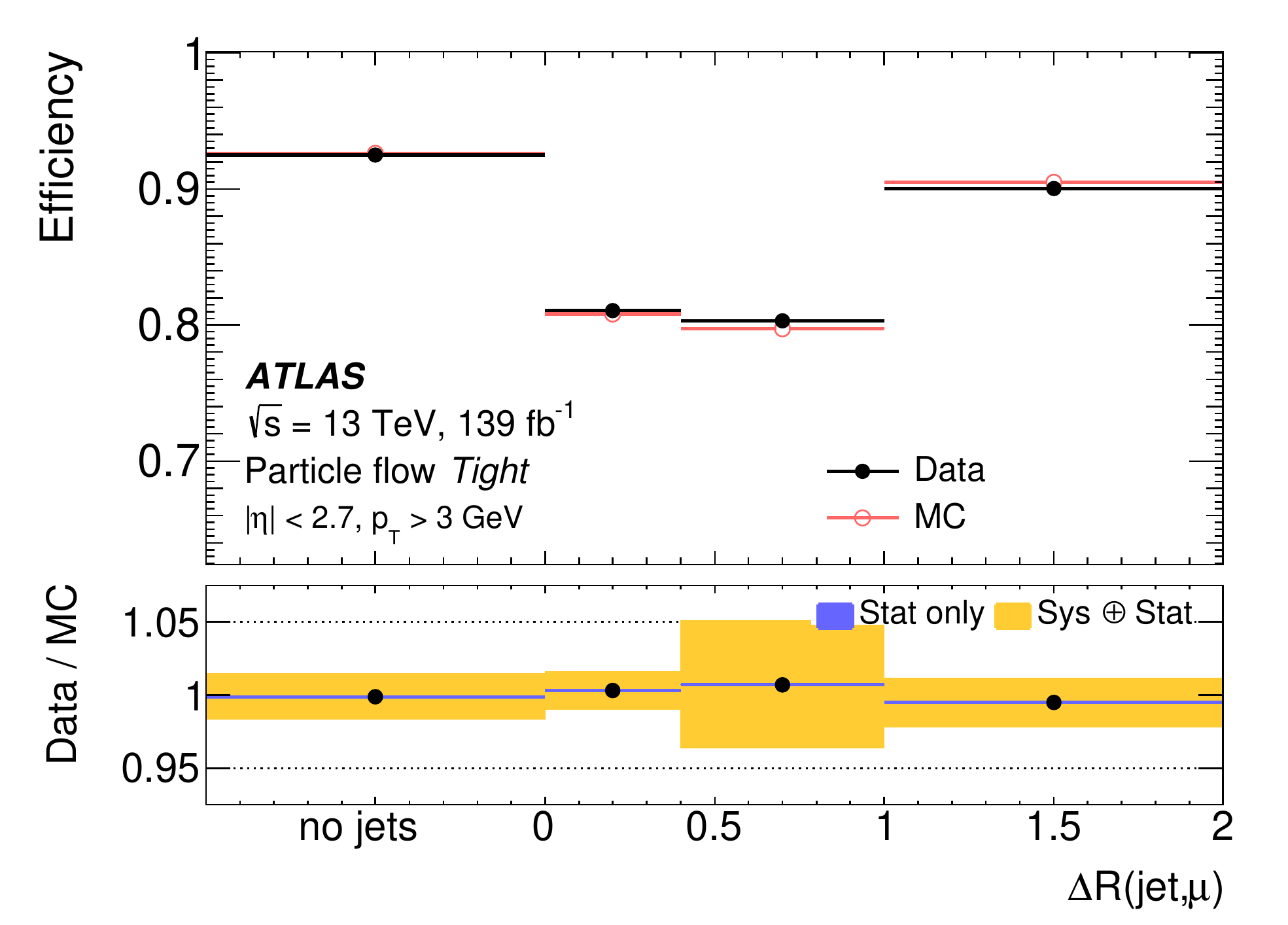}
\includegraphics[width=0.49\textwidth]{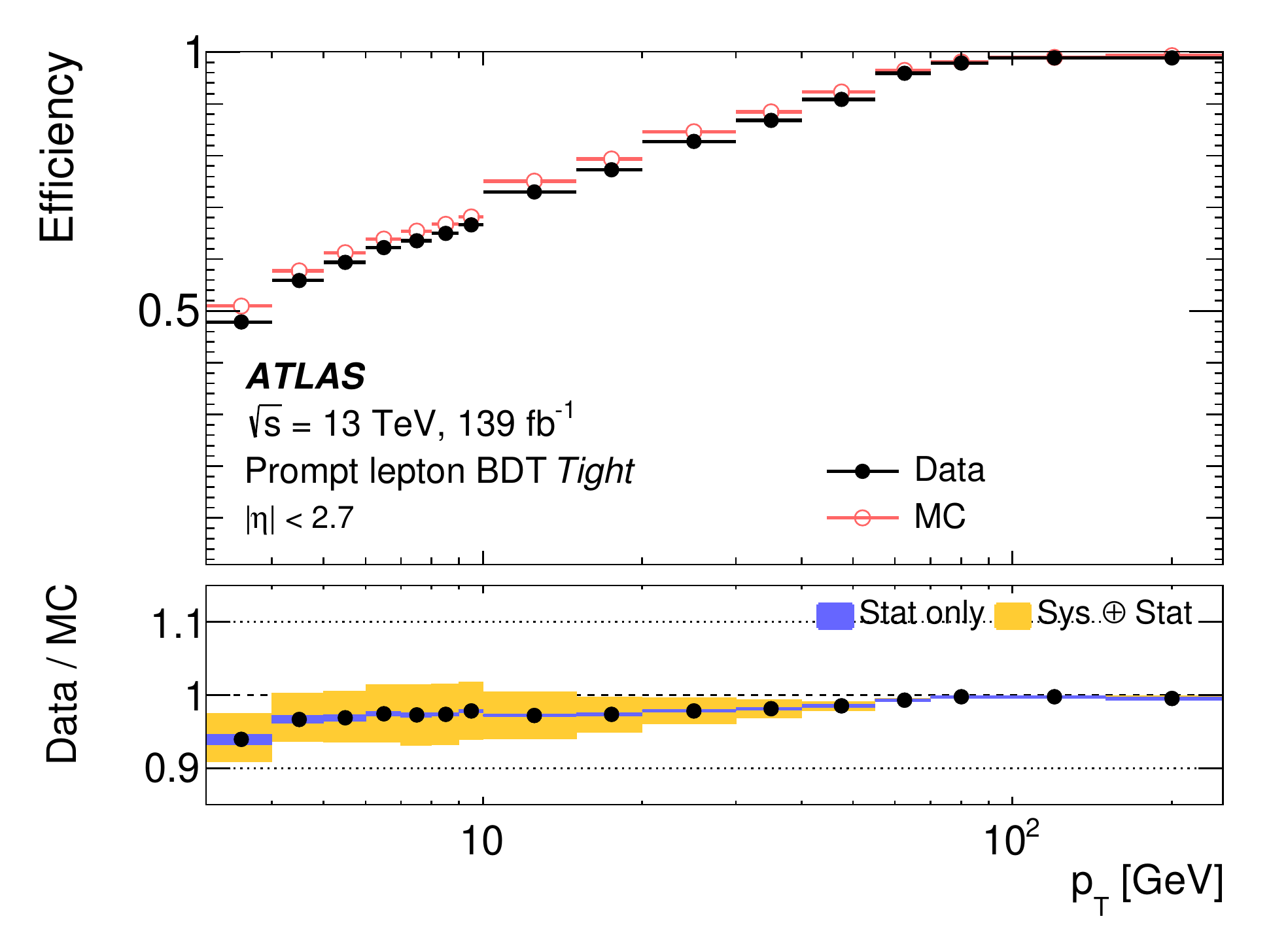}
\includegraphics[width=0.49\textwidth]{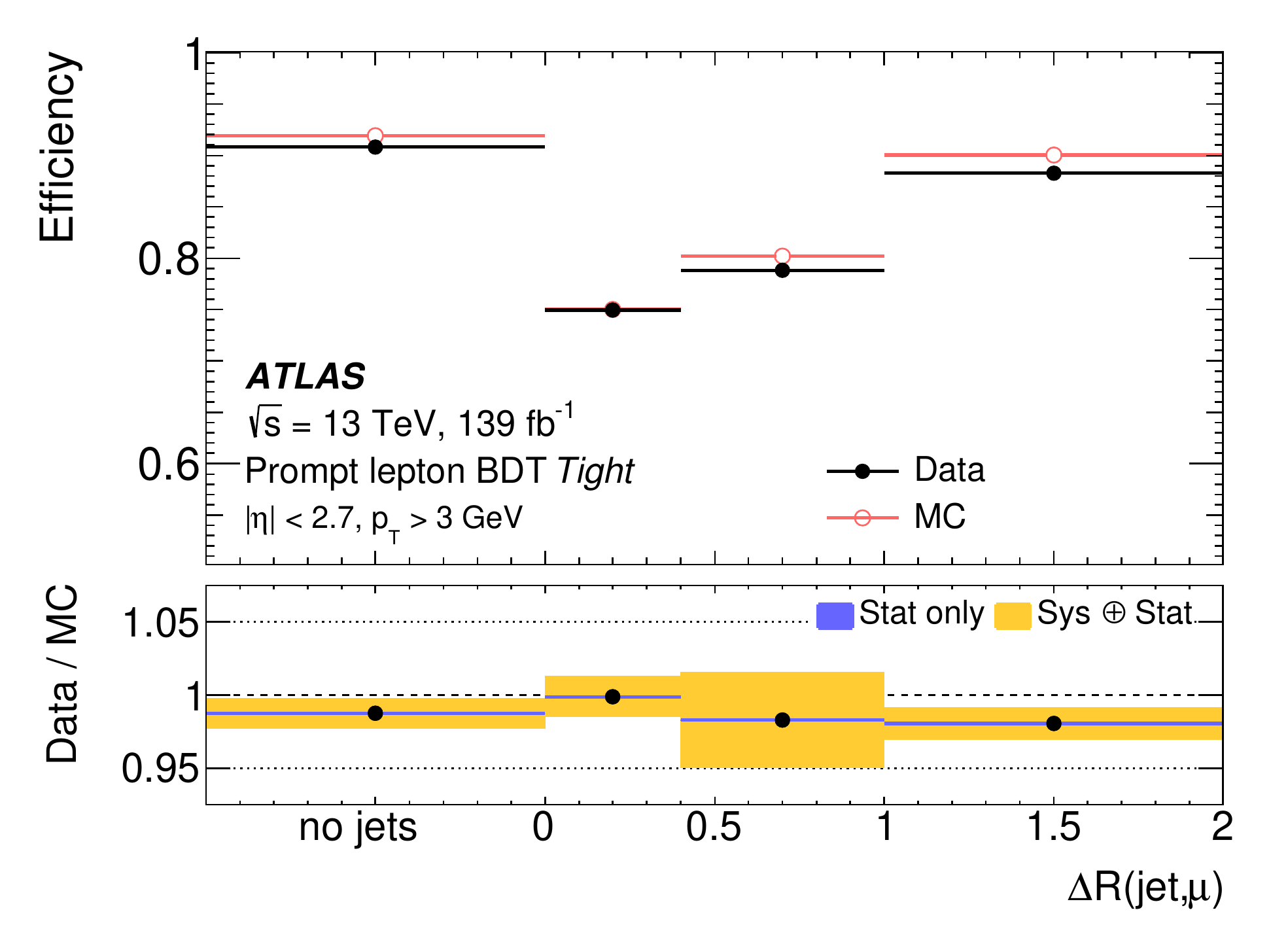}
\caption{Muon isolation efficiency measured in $Z\rightarrow\mu\mu$ events for the \emph{Tight} (top plots), \emph{PflowTight} (middle plots), and \emph{PLBDTTight} (bottom plots) criteria, as a function of $\pt$ (left plots) and \deltarmuj{} (right plots) for muons with $\pt>3$~\GeV{}. The statistical uncertainty in the efficiency measurement is smaller than the size of the markers, and thus not displayed. The panel at the bottom shows the ratio of the measured to predicted efficiencies, with statistical and systematic uncertainties.}
\label{fig:IsoEffTight}
\end{figure}

Figures~\ref{fig:IsoEffLoose} and~\ref{fig:IsoEffTight} display one-dimensional projections of the measured isolation efficiencies for the \emph{Loose}, \emph{PflowLoose}, and\emph{PLBDTLoose} selections, and the \emph{Tight}, \emph{PflowTight}, and \emph{PLBDTTight} selections, respectively, along the muon $\pt$ and the angular distance \deltarmuj{} from the closest jet. Muons well separated from jets and with $\pt>20$~\GeV{} show SFs very close to unity for all selections, with uncertainties at the per-mille level. At very low transverse momentum and near or within jets, the uncertainties increase to approximately 5\%, and are dominated by the \emph{Jet modelling} uncertainty discussed in Section~\ref{subsubsec:ztp_syst}. Reflecting the plots on the right of Figures~\ref{fig:IsoEffLoose} and~\ref{fig:IsoEffTight}, the SFs provided to physics analyses are computed as a function of $\pt$ in four wide \deltarmuj{} bins: $\deltarmuj{}<0$ corresponding to the case of no jets in the event, $0<\deltarmuj{}<0.4$ corresponding to muons within an anti-$k_t$ jet with $R=0.4$, $0.4<\deltarmuj{}<1.0$ for muons near an $R=0.4$ jet or within a large radius jet, and $\deltarmuj{}>1.0$ for muons far from any jet.
 
\subsection{Stability throughout data taking}
 
\begin{figure}
\centering
\includegraphics[width=0.59\textwidth]{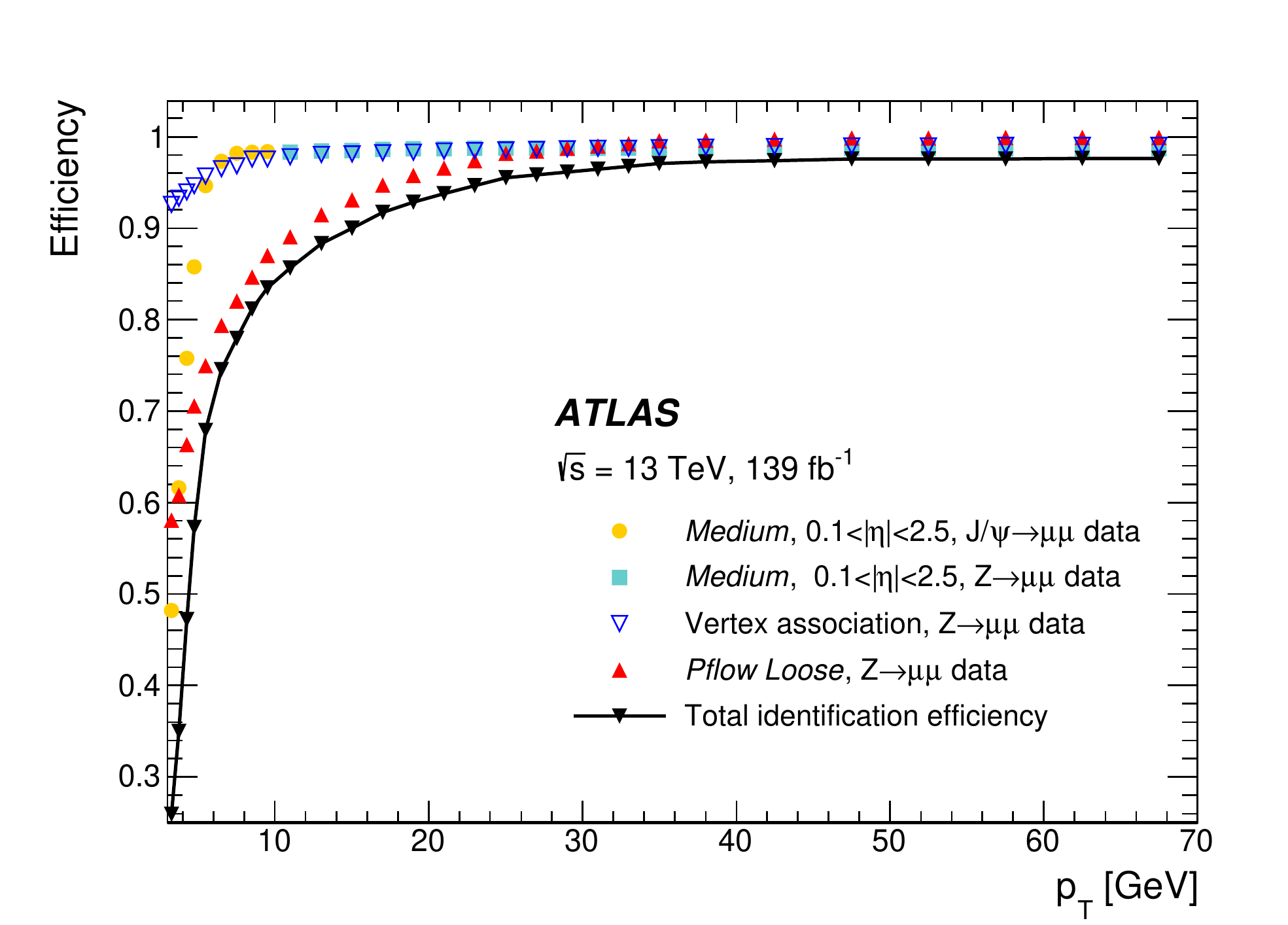}
\caption{Overall reconstruction and identification efficiency measured in data with $Z\to\mu\mu$ and $J/\psi\to\mu\mu$ decays for prompt muons with $\pt>3$~\GeV{}. The total identification efficiency for satisfying simultaneously the \medium, \emph{PflowLoose} isolation and vertex association criteria (black line) is shown together with its separate components (coloured markers).}
\label{fig:UltimateSummaryPlot}
\end{figure}

The overall reconstruction and identification efficiency for \medium\ muons fulfilling vertex association and \emph{PflowLoose} isolation criteria is summarised in Figure~\ref{fig:UltimateSummaryPlot}.
 
\begin{figure}
\centering
\includegraphics[width=0.49\textwidth]{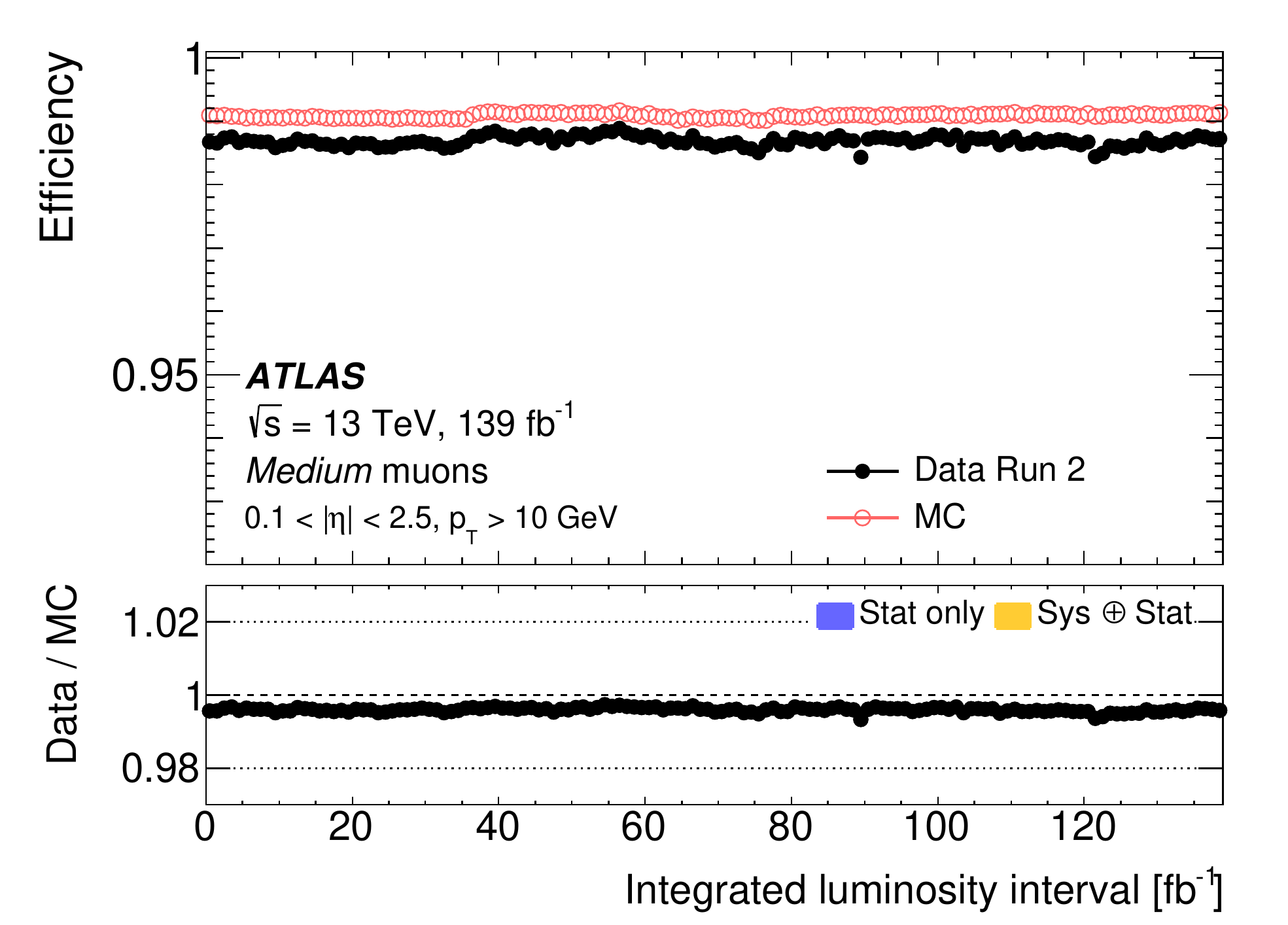}
\includegraphics[width=0.49\textwidth]{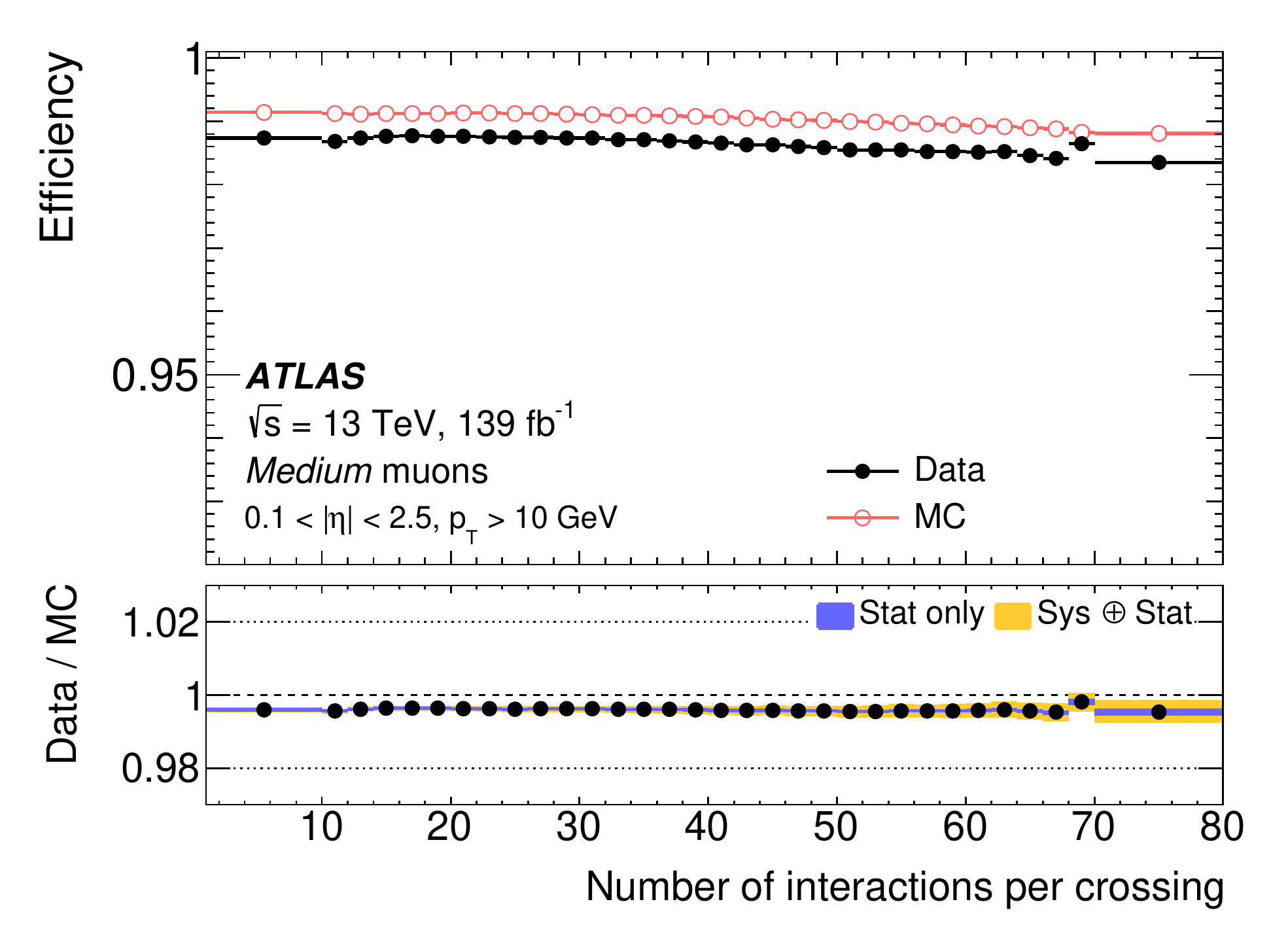}
\caption{Muon reconstruction and identification efficiency for the \medium\ identification criteria measured in $Z\rightarrow\mu\mu$ events as a function of the integrated luminosity interval (left) and the actual number of interactions per bunch crossing (right) for muons with $\pt>10$~\GeV{} and $0.1<|\eta|<2.5$. The statistical uncertainty in the efficiency measurement is smaller than the size of the markers, and thus not displayed. In the left plot each data point corresponds to 1~\ifb\ of collected data in Run~2. The panel at the bottom shows the ratio of the measured to predicted efficiencies, with statistical and systematic uncertainties.}
\label{fig:EffVsTime_Medium}
\end{figure}

In Figure~\ref{fig:EffVsTime_Medium} the reconstruction and identification efficiency for \medium\ muons is studied as a function of the delivered integrated luminosity during Run~2, and as a function of the number of simultaneous interactions per bunch crossing. Thanks to the high standards maintained in the operation of the detector and to the robustness of muon reconstruction, no significant drops in efficiency are observed within Run~2. Furthermore, the performance of the reconstruction and identification algorithms remained insensitive to the harshening of the pile-up conditions.
 
\begin{figure}
\centering
\includegraphics[width=0.49\textwidth]{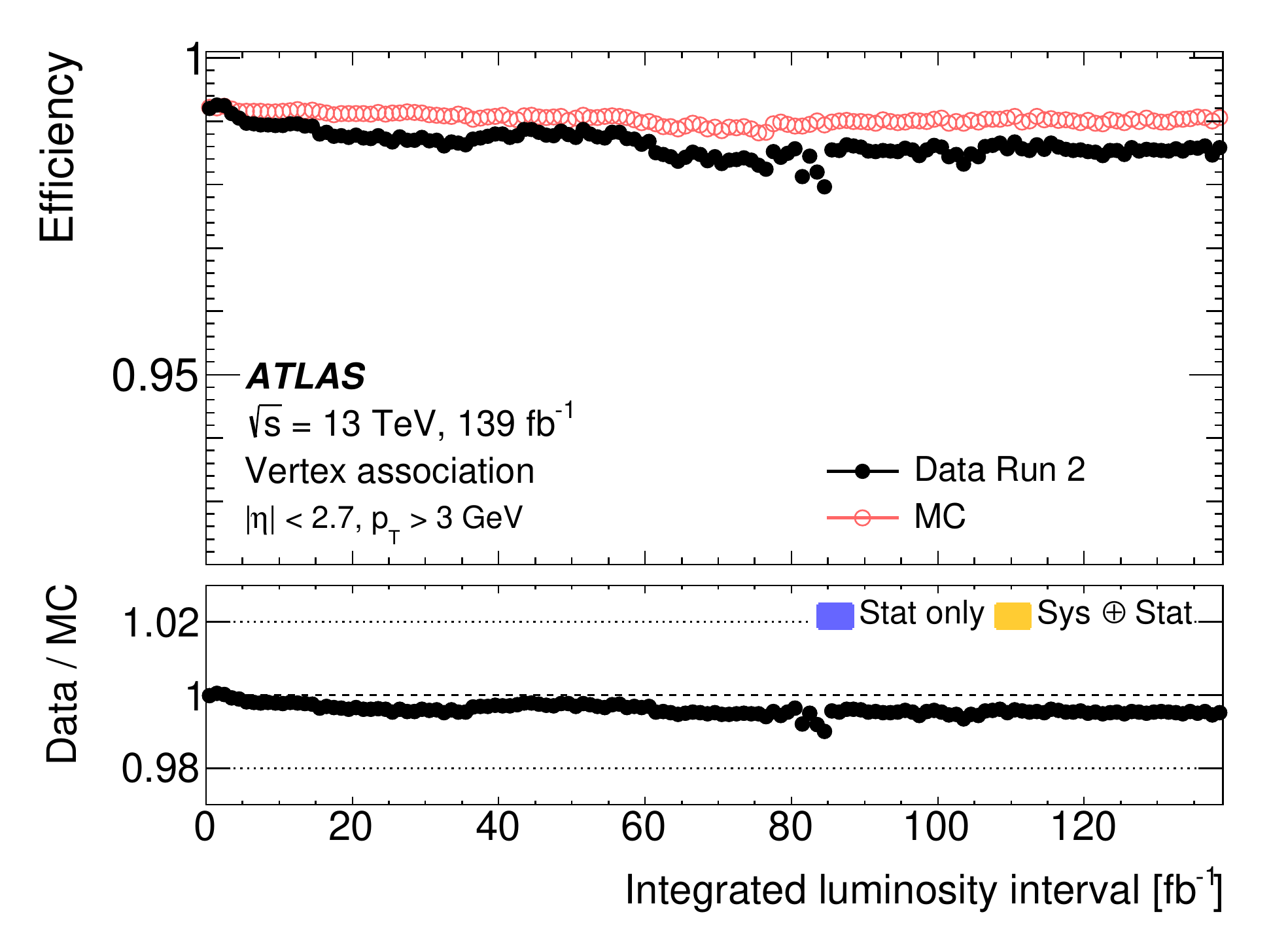}
\includegraphics[width=0.49\textwidth]{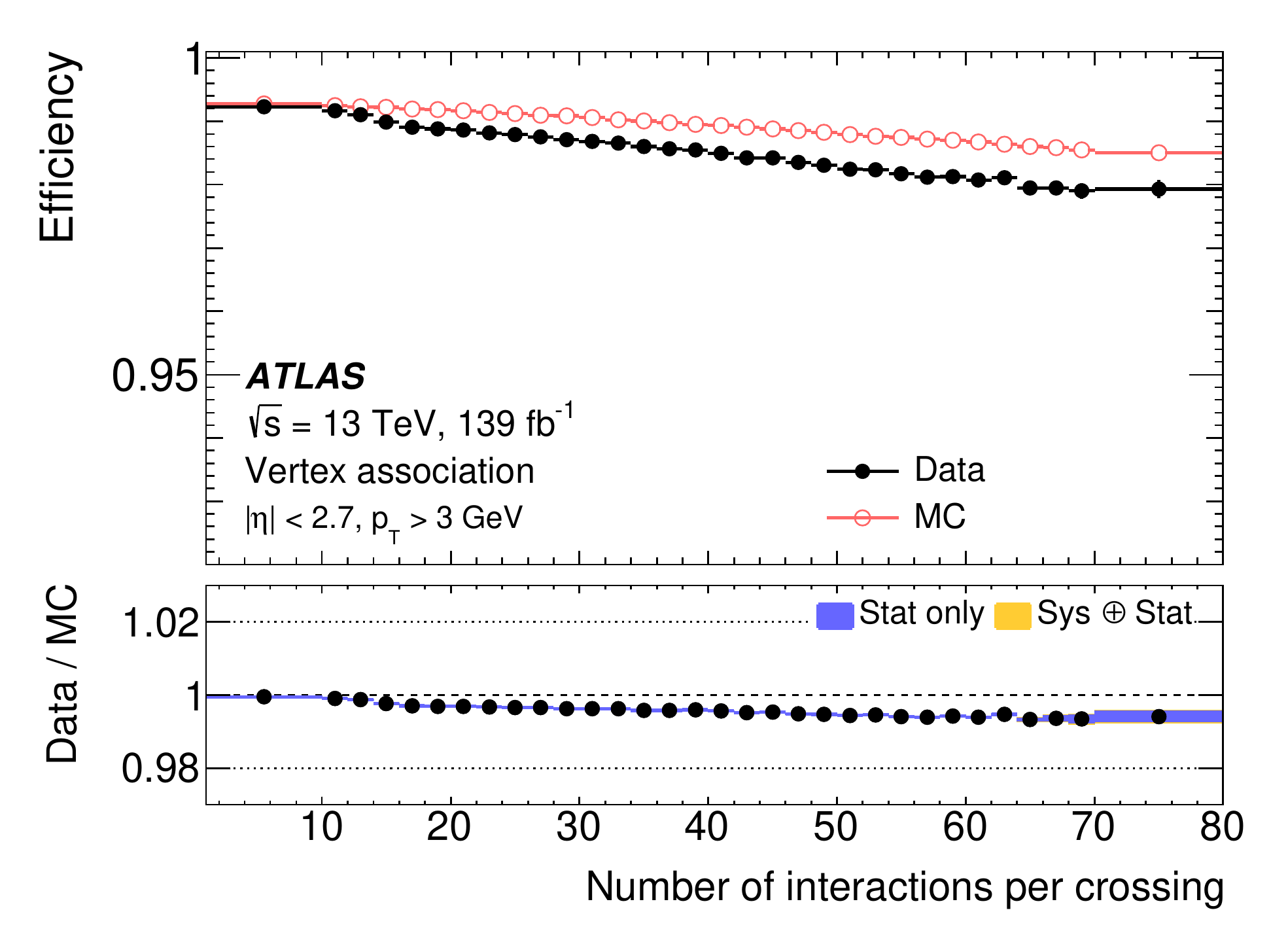}
\caption{Efficiency for the vertex association criteria measured in $Z\rightarrow\mu\mu$ events as a function of the integrated luminosity interval (left) and the actual number of interactions per bunch crossing (right) for muons with $\pt>3$~\GeV{}. In the left plot each data point corresponds to 1~\ifb\ of collected data in Run~2. The statistical uncertainty in the efficiency measurement is smaller than the size of the markers, and thus not displayed. The panel at the bottom shows the ratio of the measured to predicted efficiencies, with statistical and systematic uncertainties. Stable efficiencies are reached during 2017, after 60~\ifb\ of collected data, and maintained until the end of Run~2, with some fluctuations at the beginning of 2018 corresponding to collision runs with high pile-up.}
\label{fig:EffVsTime_TTVA}
\end{figure}

The stability of the vertex association efficiency is illustrated in Figure~\ref{fig:EffVsTime_TTVA}. A progressive but small decrease in the measured efficiency throughout data taking is observed, and corresponds to a gradual deterioration of the impact parameter resolution related to the worse pile-up conditions. Stability is reached during the second half of the 2017 data taking and maintained until the end of Run 2, with some fluctuations at the beginning of 2018 corresponding to collision runs with high pile-up.
 
\begin{figure}
\centering
\includegraphics[width=0.49\textwidth]{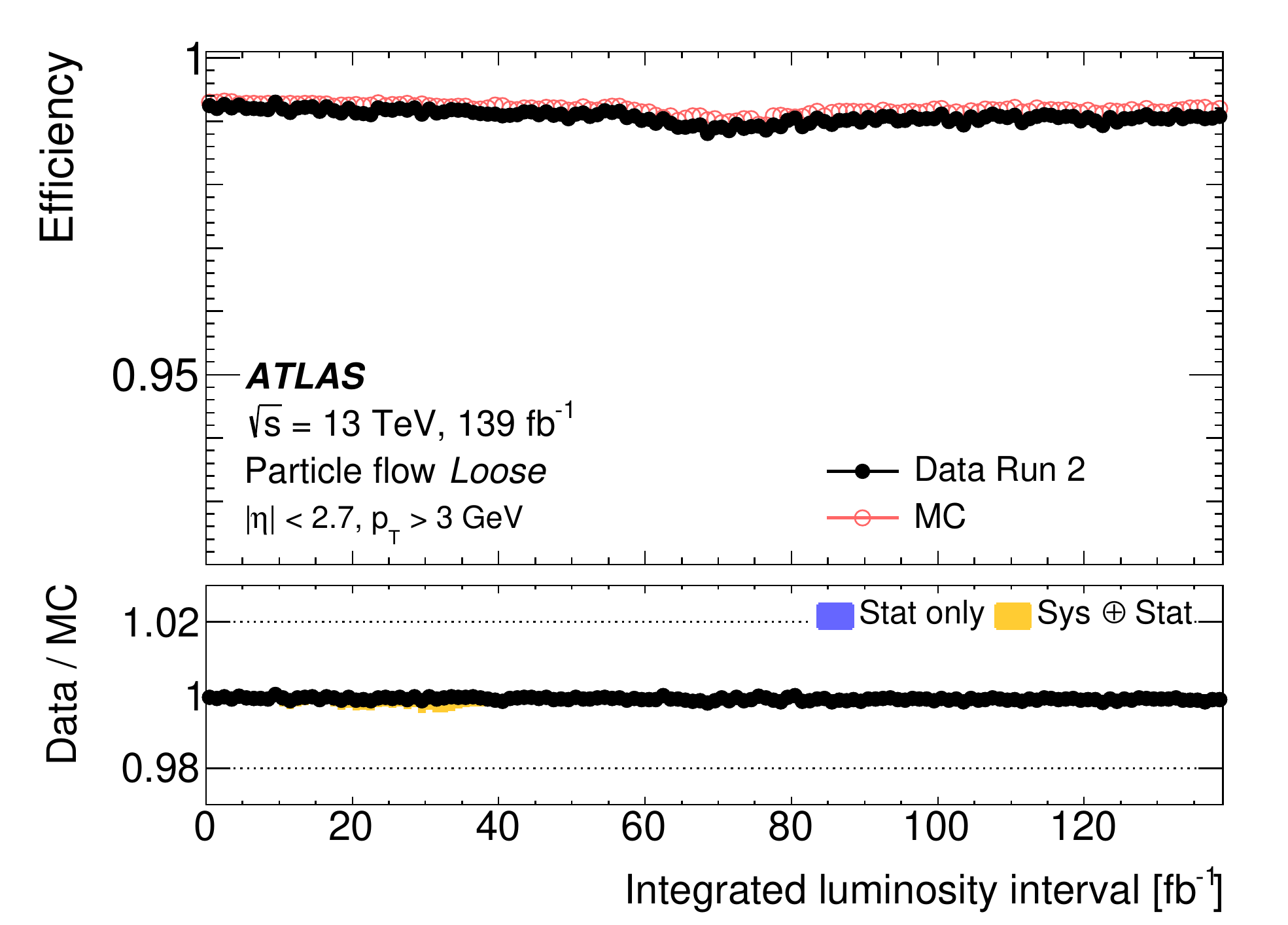}
\includegraphics[width=0.49\textwidth]{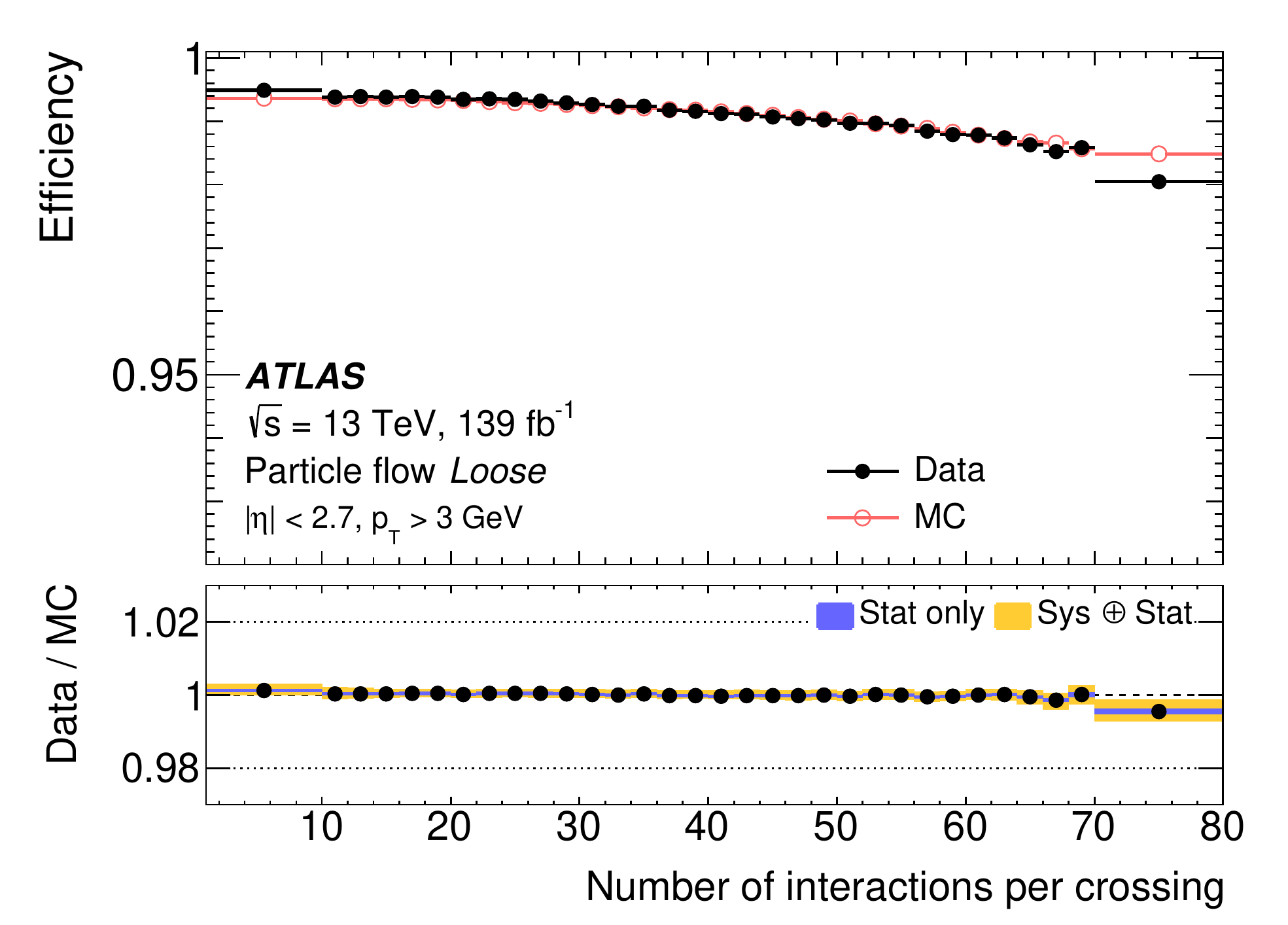}
\caption{Efficiency for the \emph{PflowLoose} isolation criteria measured in $Z\rightarrow\mu\mu$ events as a function of the integrated luminosity interval (left) and the actual number of interactions per bunch crossing (right) for muons with $\pt>3$~\GeV{}. In the left plot each data point corresponds to 1~\ifb\ of collected data in Run~2. The statistical uncertainty in the efficiency measurement is smaller than the size of the markers, and thus not displayed. The panel at the bottom shows the ratio of the measured to predicted efficiencies, with statistical and systematic uncertainties.}
\label{fig:EffVsTime_PflowLoose}
\end{figure}

Finally, Figure~\ref{fig:EffVsTime_PflowLoose} shows the isolation efficiency for the \emph{PflowLoose} criteria throughout Run~2 and as a function of pile-up, for muons with $\pt$ greater than $10$~\GeV{}. In spite of the very different pile-up conditions reached within the various data-taking periods, which have a direct but overall small impact on the efficiency of the isolation selections, agreement between data and simulation remained excellent and stable.
 
\FloatBarrier
\section{Conclusions}
\label{sec:conclusions}
The ATLAS muon reconstruction, identification, vertex association, and isolation efficiencies have been measured using 139~$\ifb{}$ of $pp$ collision data at $\sqrt{s}=13~\TeV$ recorded between 2015 and 2018 at the LHC. The measured efficiencies have been compared with the predictions from simulation over the full acceptance of $|\eta|<2.7$ and over the transverse momentum range of $3~\GeV<\pt<250$~\GeV{}, deploying large MC samples of $Z\to\mu\mu$ and $J/\psi\to\mu\mu$ decays consisting of more than 210 and 45 million events, respectively. In the efficiency and SF measurements, the available phase space was subdivided into well-populated regions, choosing a granularity suitable for most of the physics analyses in the ATLAS experiment.
 
The $Z\to\mu\mu$ sample allows the reconstruction and identification efficiencies to be measured with a precision better than the per-mille level for muons with $\pt$ above $10$~\GeV{} in most of the detector regions. The $J/\psi\to\mu\mu$ sample extends the measurement down to $\pt=3$~\GeV{}, with a precision better than 1\% in the 5--20~\GeV{} range.
 
The efficiency of the muon vertex association criteria has been measured with a precision better than 0.2\% in the entire transverse momentum range considered, and not exceeding 0.01\% for muons with $\pt$ above $20$~\GeV{}. Excellent agreement with the MC simulation was found. Similarly, the measured efficiencies for the eight isolation working points have been found to agree well with the predictions from simulation, with calibration factors very close to unity and uncertainties at the per-mille level for muons with $\pt$ above $20$~\GeV{} and well separated from jets.
 
These results have been used to correct the MC simulation to improve the data--simulation agreement and to minimise the uncertainties in physics analyses.
 
\section*{Acknowledgements}
 

We thank CERN for the very successful operation of the LHC, as well as the
support staff from our institutions without whom ATLAS could not be
operated efficiently.
 
We acknowledge the support of ANPCyT, Argentina; YerPhI, Armenia; ARC, Australia; BMWFW and FWF, Austria; ANAS, Azerbaijan; SSTC, Belarus; CNPq and FAPESP, Brazil; NSERC, NRC and CFI, Canada; CERN; ANID, Chile; CAS, MOST and NSFC, China; COLCIENCIAS, Colombia; MSMT CR, MPO CR and VSC CR, Czech Republic; DNRF and DNSRC, Denmark; IN2P3-CNRS and CEA-DRF/IRFU, France; SRNSFG, Georgia; BMBF, HGF and MPG, Germany; GSRT, Greece; RGC and Hong Kong SAR, China; ISF and Benoziyo Center, Israel; INFN, Italy; MEXT and JSPS, Japan; CNRST, Morocco; NWO, Netherlands; RCN, Norway; MNiSW and NCN, Poland; FCT, Portugal; MNE/IFA, Romania; JINR; MES of Russia and NRC KI, Russian Federation; MESTD, Serbia; MSSR, Slovakia; ARRS and MIZ\v{S}, Slovenia; DST/NRF, South Africa; MICINN, Spain; SRC and Wallenberg Foundation, Sweden; SERI, SNSF and Cantons of Bern and Geneva, Switzerland; MOST, Taiwan; TAEK, Turkey; STFC, United Kingdom; DOE and NSF, United States of America. In addition, individual groups and members have received support from BCKDF, CANARIE, Compute Canada, CRC and IVADO, Canada; Beijing Municipal Science \& Technology Commission, China; COST, ERC, ERDF, Horizon 2020 and Marie Sk{\l}odowska-Curie Actions, European Union; Investissements d'Avenir Labex, Investissements d'Avenir Idex and ANR, France; DFG and AvH Foundation, Germany; Herakleitos, Thales and Aristeia programmes co-financed by EU-ESF and the Greek NSRF, Greece; BSF-NSF and GIF, Israel; La Caixa Banking Foundation, CERCA Programme Generalitat de Catalunya and PROMETEO and GenT Programmes Generalitat Valenciana, Spain; G\"{o}ran Gustafssons Stiftelse, Sweden; The Royal Society and Leverhulme Trust, United Kingdom.
 
The crucial computing support from all WLCG partners is acknowledged gratefully, in particular from CERN, the ATLAS Tier-1 facilities at TRIUMF (Canada), NDGF (Denmark, Norway, Sweden), CC-IN2P3 (France), KIT/GridKA (Germany), INFN-CNAF (Italy), NL-T1 (Netherlands), PIC (Spain), ASGC (Taiwan), RAL (UK) and BNL (USA), the Tier-2 facilities worldwide and large non-WLCG resource providers. Major contributors of computing resources are listed in Ref.~\cite{ATL-SOFT-PUB-2020-001}.
 

\clearpage

\printbibliography
 
\clearpage \input{atlas_authlist}

\end{document}

%% file: atlas_authlist.tex
 
\begin{flushleft}
\hypersetup{urlcolor=black}
{\Large The ATLAS Collaboration}

\bigskip

\AtlasOrcid[0000-0002-6665-4934]{G.~Aad}$^\textrm{\scriptsize 102}$,    
\AtlasOrcid[0000-0002-5888-2734]{B.~Abbott}$^\textrm{\scriptsize 128}$,    
\AtlasOrcid[0000-0002-7248-3203]{D.C.~Abbott}$^\textrm{\scriptsize 103}$,    
\AtlasOrcid[0000-0002-2788-3822]{A.~Abed~Abud}$^\textrm{\scriptsize 36}$,    
\AtlasOrcid[0000-0002-1002-1652]{K.~Abeling}$^\textrm{\scriptsize 53}$,    
\AtlasOrcid[0000-0002-2987-4006]{D.K.~Abhayasinghe}$^\textrm{\scriptsize 94}$,    
\AtlasOrcid[0000-0002-8496-9294]{S.H.~Abidi}$^\textrm{\scriptsize 167}$,    
\AtlasOrcid[0000-0002-8279-9324]{O.S.~AbouZeid}$^\textrm{\scriptsize 40}$,    
\AtlasOrcid{N.L.~Abraham}$^\textrm{\scriptsize 156}$,    
\AtlasOrcid[0000-0001-5329-6640]{H.~Abramowicz}$^\textrm{\scriptsize 161}$
% ...
\bigskip
\\

% ...
$^{36}$CERN, Geneva; Switzerland.\\
% ...
$^{40}$Niels Bohr Institute, University of Copenhagen, Copenhagen; Denmark.\\
% ...
$^{53}$II. Physikalisches Institut, Georg-August-Universit\"{a}t G\"ottingen, G\"ottingen; Germany.\\
% ...
$^{94}$Department of Physics, Royal Holloway University of London, Egham; United Kingdom.\\
% ...
$^{102}$CPPM, Aix-Marseille Universit\'e, CNRS/IN2P3, Marseille; France.\\
$^{103}$Department of Physics, University of Massachusetts, Amherst MA; United States of America.\\
% ...
$^{128}$Homer L. Dodge Department of Physics and Astronomy, University of Oklahoma, Norman OK; United States of America.\\
% ...
$^{156}$Department of Physics and Astronomy, University of Sussex, Brighton; United Kingdom.\\
% ...
$^{161}$Raymond and Beverly Sackler School of Physics and Astronomy, Tel Aviv University, Tel Aviv; Israel.\\
% ...
$^{167}$Department of Physics, University of Toronto, Toronto ON; Canada.\\
% ...

\end{flushleft}
